\documentclass[pra,superscriptaddress,showpacs,longbibliography,reprint]{revtex4-2} 

\usepackage[colorlinks=true,linkcolor=blue,urlcolor=blue,citecolor=blue,pdfusetitle]{hyperref}
\usepackage{color}
\usepackage[utf8]{inputenc}
\usepackage[usenames,dvipsnames]{xcolor}
\usepackage{amsmath}
\usepackage{amssymb}
\usepackage{graphicx}
\graphicspath{{graphics/}}
\usepackage{epsfig}
\usepackage{dcolumn}
\usepackage{bm}
\usepackage{mathrsfs}
\usepackage{multirow}
\usepackage[all]{xy}
\usepackage{pbox}
\usepackage{lipsum}
\usepackage{verbatim}
\usepackage{braket}
\usepackage{dsfont}
\usepackage{array}
\usepackage{makecell}
\usepackage{tabularx}
\usepackage{isotope}
\usepackage{xr}
\usepackage{float}

\begin{document}
\raggedbottom

\title{Entangled time-crystal phase in an open quantum light-matter system
}

\newcommand{\tubingen}{Institut f\"{u}r Theoretische Physik,  Universit\"{a}t T\"{u}bingen, Auf der Morgenstelle 14, 72076 T\"{u}bingen, Germany}

\author{Robert Mattes}
\affiliation{\tubingen}

\author{Igor Lesanovsky}
\affiliation{\tubingen}
\affiliation{School of Physics and Astronomy and Centre for the Mathematics and Theoretical Physics of Quantum Non-Equilibrium Systems, The University of Nottingham, Nottingham, NG7 2RD, United Kingdom}

\author{Federico Carollo}
\affiliation{\tubingen}

\date{\today}

\begin{abstract}
Time-crystals are nonequilibrium many-body phases in which the state of the system dynamically approaches a limit cycle. While these phases are recently in the focus of intensive research, it is still far from clear whether they can host quantum correlations. In fact, mostly classical correlations have been observed so far and time-crystals appear to be effectively classical high-entropy phases. Here, we consider the nonequilibrium behavior of an open quantum light-matter system, realizable in current experiments, which maps onto a paradigmatic time-crystal model after an adiabatic elimination of the light field. The system displays a bistable regime, with coexistent time-crystal and stationary phases, terminating at a tricritical point from which a second-order phase transition line departs. 
While light and matter are uncorrelated in the stationary phase, the time-crystal phase features bipartite correlations, both of quantum and classical nature. Our work unveils that time-crystal phases in collective open quantum systems can sustain quantum correlations, including entanglement, and are thus more than effectively classical many-body phases. 
\end{abstract}

\maketitle

\section{Introduction}
Interacting light-matter systems can feature intriguing collective behavior and phase transitions. An example is given by transitions into superradiant phases, i.e., phases with a ``macroscopically" excited light field  \cite{dicke1954,hepp1973superradiant,hepp1973equilibrium,wang1973,hioe1973,carmichael1973,gross1982,emary2003,emary2003b,lambert2005,bastidas2012}. In the presence of Markovian dissipation, these systems generically approach, at long times, a stationary state. However, under certain conditions, genuine dynamical regimes may occur \cite{keeling2010}, as it happens, for instance, in the case of lasing \cite{kirton2019} or {\it counter-lasing} regimes \cite{zhiqiang2017,zhiqiang2018,stitely2020,shchadilova2020,stitely2022}. The emergence of non-stationary many-body behavior \cite{iemini2018,buca2019,buca2019b,chiacchio2019,booker2020}, with the system undergoing persistent oscillatory dynamics, witnesses the breaking of the continuous time-translation symmetry of the dynamical generator and the concomitant formation of a crystalline structure in time. Because of this reason, these nonequilibrium phases are referred to as time-crystals (see, e.g., Refs.~\cite{wilczek2012,shapere2012,bruno2013,watanabe2015,sacha2015,khemani2016,else2016,yao2017,sacha2018,gambetta2019,else2020,booker2020,taheri2022,kongkhambut2022,nie2023}).

A minimal Markovian open quantum system displaying non-stationary behavior is the so-called {\it boundary time-crystal} \cite{iemini2018}, generalized in Refs.~\cite{sanchezmunoz2019,piccitto2021,prazeres2021,passarelli2022} and experimentally realized in Ref.~\cite{ferioli2022}. It consists of a collective many-body spin model which allows for both efficient numerical simulations  \cite{chase2008,baragiola2010,kirton2017,kirton2018,shammah2018,huybrechts2020} and exact analytical solutions \cite{carmichael1980,alicki1983,benatti2018,sanchezmunoz2019,buonaiuto2021,carollo2022,boneberg2022}. This model features quantum correlations  between spins  in its stationary phase --- witnessed by non-zero spin-squeezing and two-qubit entanglement --- but only classical correlations in the time-crystal regime \cite{hannukainen2018,sanchezmunoz2019,buonaiuto2021,carollo2022,lourenco2022,montenegro2023,pavlov2023}, which is described by highly mixed and effectively classical states \cite{piccitto2021,buonaiuto2021,carollo2022,cabot2022}. 
It thus remains an open question whether (boundary) time-crystal phases can host quantum effects or whether these phases are essentially purely {\it classical} dynamical regimes. 
\begin{figure}[t!]
    \centering
    \includegraphics[width=\columnwidth]{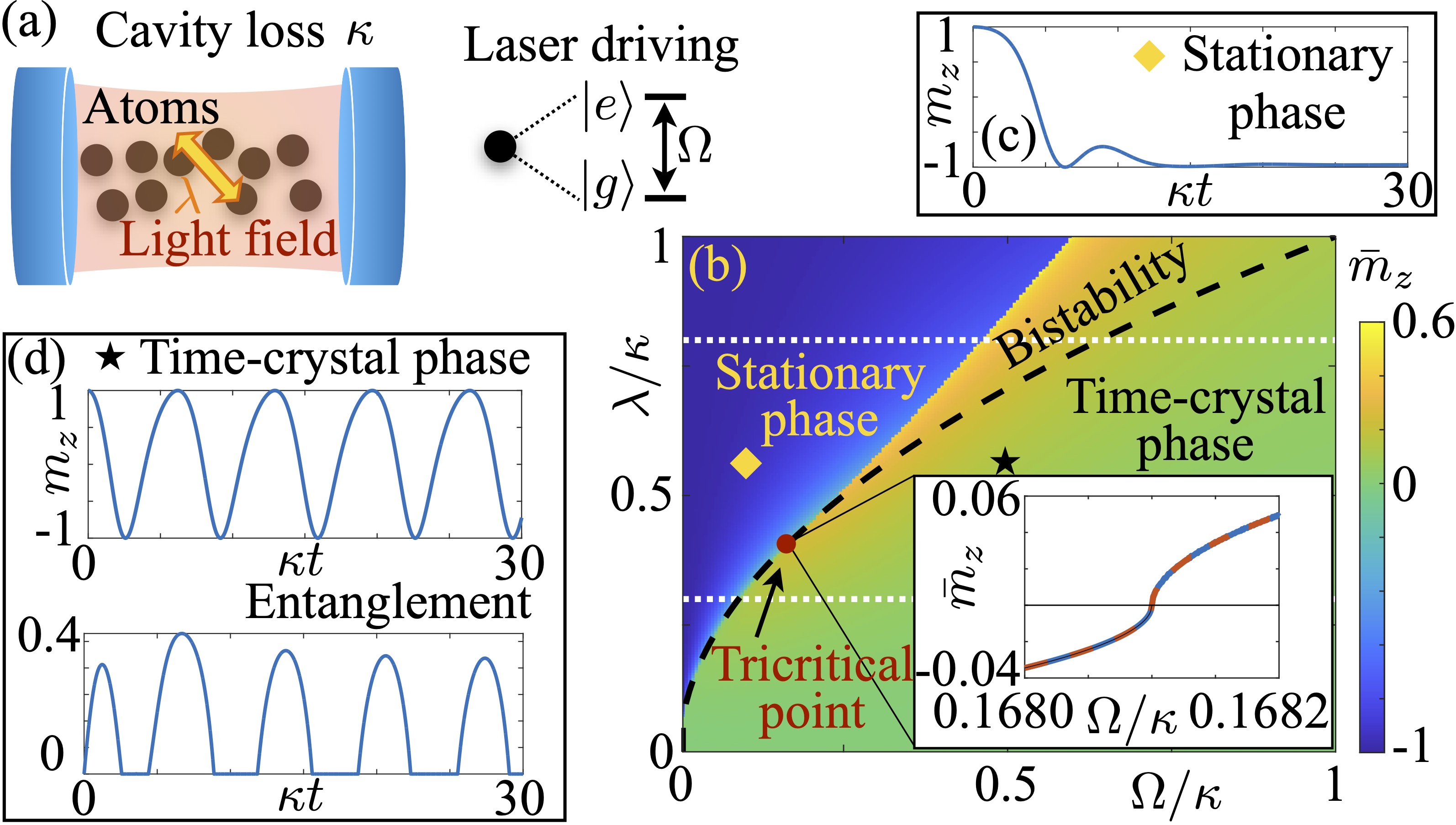}
    \caption{{\bf System and nonequilibrium phase diagram.} (a) An ensemble of two-level atoms, with ground state $\ket{g}$ and excited state $\ket{e}$, is driven by a laser with Rabi-frequency $\Omega$ and interacts via exchange of excitations with the light field of a cavity (coupling constant $\lambda$). The cavity is subject to photon loss at rate $\kappa$. (b) Phase diagram in terms of the time-averaged {\it magnetization} $\bar{m}_z$ as a function of $\Omega$ and $\lambda$. It features a bistable regime terminating at a tricritical point, $(\Omega/\kappa,\lambda/\kappa)\approx (0.17,0.41)$. Here, the transition becomes of second order (see inset). Below the dashed line, the system does not possess a well-defined stationary state. (c) In the stationary phase the magnetization $m_z(t)$ approaches a constant value. (d) In the time-crystal phase, $m_z(t)$ undergoes persistent oscillations and the atoms and the light field are {\it collectively} entangled.  }
    \label{fig:Fig1}
\end{figure}

In this paper, we consider a paradigmatic model describing atoms coupled to a light field via an excitation-exchange interaction \cite{tavis1967,tavis1969,carollo2020,paulino2022}, see sketch in Fig.~\ref{fig:Fig1}(a). This system can be realized in current cavity-atom experiments \cite{ritsch2013,feng2015,piazza2015,norcia2016,rose2017,norcia2018,dogra2019,schaffer2020,mivehvar2021} and realizes a boundary time-crystal  \cite{agarwal1997,gopalakrishnan2011,damanet2019,carollo2020} when adiabatically eliminating the light field. It features both a stationary superradiant and a time-crystal phase  \cite{paulino2022}. Within the parameter regime in which one may expect the adiabatic elimination to hold, the nonequilibrium transition between the two phases is a second-order one. However, in contrast to the boundary time-crystal, the system features a bistable regime, characterized by the coexistence of a limit cycle and a stationary phase. This signals that the phase transition eventually becomes of first order [cf.~Fig.~\ref{fig:Fig1}(b)]. 
Explicitly taking into account the light field further allows us --- to best of our knowledge for the first time --- to observe the emergence of quantum correlations, including entanglement, in a time-crystal regime. The existence of these  correlations may motivate the development of alternative strategies for exploiting these phases for enhanced  metrological applications \cite{montenegro2023,pavlov2023}.
\section{The model}
We consider a driven-dissipative version of the so-called Tavis-Cummings spin-boson model \cite{tavis1967,tavis1969,kirton2019,mivehvar2021}. For concreteness, we focus on a realization of the model in a cavity setup, as depicted in Fig.~\ref{fig:Fig1}(a). The spins describe $N$ two-level atoms with ground state $\ket{g}$, excited state $\ket{e}$, and energy splitting $\omega_{\rm at}$. The bosonic operators $a,a^\dagger$ are associated with the light field inside the cavity (frequency $\omega_{\rm cav}=\omega_{\rm at}$). For later convenience, we define the quadrature operators $q=i(a-a^\dagger)/\sqrt{2}$ and $p=(a+a^\dagger)/\sqrt{2}$, such that $[q,p]=i$. 

The atoms are resonantly driven by a laser and, in the rotating frame, the system Hamiltonian is given by 
\begin{equation}
    H=\Omega (S_++S_-) +\frac{\lambda}{\sqrt{N}}\left(a^\dagger S_-+aS_+\right)\, ,
    \label{Hamiltonian}
\end{equation}
with $\Omega$ being the laser Rabi-frequency and $\lambda$ the coupling constant providing the ``rate" of the coherent exchange of excitations. The collective atom operators $S_\pm$ are defined as $S_-=\sum_{k=1}^N\sigma_-^{(k)}$, with $\sigma_-=\ket{g}\!\bra{e}$, $S_+=S_-^\dagger$. The factor $1/\sqrt{N}$ in front of the coupling term ensures a well-defined thermodynamic limit \cite{kirton2019,carollo2021}. Photon losses, at rate $\kappa$, are described by the dissipator \cite{lindblad1976,gorini1976} 
\begin{equation}
    \mathcal{L}[X]=\kappa \left(a X a^\dagger -\frac{1}{2}\left\{a^\dagger a, X\right\}\right)\, .
    \label{dissipator}
\end{equation}
The full quantum state of the system, $\rho$, thus evolves according to the quantum master equation $\dot{\rho}_t=-i[H,\rho_t]+\mathcal{L}[\rho_t]$ and allows for the calculation of the expectation value of any operator $O$,  as $\langle O\rangle_t:={\rm Tr}(\rho_t O)$. 

While the Tavis-Cummings model \cite{tavis1967,tavis1969} has been considered in several works \cite{bogoliubov1996,lopez2007,wood2014,genway2014,feng2015,rose2017,lamata2017,trivedi2019,andolina2019,shapiro2020,restrepo2021,baum2022,blaha2022,valenciatortora2022,kelly2022,stitely2023}, the setting analyzed here appears to be not much explored \cite{paulino2022}, and even less for what concerns the analysis of quantum correlations (see related studies in Refs.~\cite{hebenstreit2017,plankensteiner2017,plankensteiner2019,reitz2022} also for related models in Refs.~\cite{boneberg2022,dimer2007,zhou2023}), mostly investigated in the few-atom case \cite{tessier2003,retzker2007,Guo2008,zhang2009,man2009,youssef2010,mohamed2012,hu2013,torres2014,fan2014,restrepo2016,rundle2021,carnio2021}.
\section{Time-crystal phase transition}
To demonstrate the emergence of a phase characterized by non-stationary asymptotic dynamics, we analyze the long-time behavior of our system, in the thermodynamic limit ($N\to\infty$). To this end, we introduce the average ``magnetization" operators  $m^N_{r}=\sum_{k=1}^N\sigma_r^{(k)}/N$ for the atoms, with $\sigma_r$ being the Pauli matrices constructed from the states $\ket{g}$ and $\ket{e}$. For the light field, we consider the rescaled quadratures $m_q^N=q/\sqrt{N}$ and $m_p^N=p/\sqrt{N}$. In the thermodynamic limit, both the atom and the light-field operators $m_r=\lim_{N\to\infty} m_r^N$ describe average properties of the system  \cite{carollo2021,carollo2022} and provide suitable order parameters. 
\subsection{Mean-field equations and fixed points}
Since we are interested in the long-time regime, we derive the evolution equations for the average operators.
We focus on physically-relevant initial states of the system \cite{strocchi2005}, i.e., states with sufficiently short-range correlations.
For such initial states, following the derivation put forward in Ref.~\cite{carollo2021}, it is possible to show that, in the thermodynamic limit, the order-parameter dynamics is exactly captured by a set of nonlinear differential equations \cite{carollo2021,fiorelli2023}. \vphantom{\cite{strogatz,cabot2022quantum,adesso2010,giorda2010,simon2000,adesso2014}}
These equations are the so-called mean-field equations and for the model considered they are given by 
\begin{align*}
  \dot{m}_x(t) &= \sqrt{2}\lambda m_q(t)m_z(t)\, ,\\\
  \dot{m}_y(t) &= -2\Omega m_z(t) - \sqrt{2}\lambda m_p(t)m_z(t)\, ,\\\
  \dot{m}_z(t) &= 2\Omega m_y(t) + \sqrt{2}\lambda m_p(t)m_y(t) -\sqrt{2}\lambda m_q(t)m_x(t)\, ,\\\
  \dot{m}_q(t) &= \frac{\lambda}{\sqrt{2}} m_x(t)- \frac{\kappa}{2}m_q(t)\, ,\\\
  \dot{m}_p(t) &= -\frac{\lambda}{\sqrt{2}} m_y(t)- \frac{\kappa}{2}m_p(t)\, .
\end{align*}
The latter show that $m_x^2+m_y^2+m_z^2$ is a constant of motion, which we set to one, and that assuming an initial state for which $m_x(0)=m_q(0)=0$, results in having $m_x(t)=m_q(t)=0$, for all times $t>0$.
The remaining operators evolve via the equations
\begin{equation}
\begin{split}
\dot{m}_y(t)&=-2\Omega m_z(t)-\sqrt{2}\lambda m_p(t)m_z(t)\, ,\\
\dot{m}_z(t)&=2\Omega m_y(t) +\sqrt{2}\lambda m_p(t) m_y(t)\, ,\\
\dot{m}_p(t)&= -\frac{\lambda}{\sqrt{2}} m_y(t)-\frac{\kappa}{2}m_p(t)\, .
\end{split}
    \label{mean-field-eqs}
\end{equation}
We note that, adiabatically eliminating $m_p(t)$, by setting the last of the equations above to zero and substituting the result in the other two equations, leads to the equations of motion for the boundary time-crystal model \cite{iemini2018,carollo2022}. A similar mapping holds also at an operatorial level \cite{carollo2020}.
By setting the derivatives in the above equations to zero and using the constant of motion ($m_x^2+m_y^2+m_z^2=1$), we find the stationary solutions to the mean-field equations, given by 
\begin{align}
m_y^*= \frac{\Omega \kappa}{\lambda^2}\, , \, m_z^*=\pm \sqrt{1-\left(\frac{\Omega \kappa}{\lambda^2}\right)^2}\, , \, m_p^*=-\frac{\sqrt{2}\Omega}{\lambda}\, .
    \label{stat-mean-field}
\end{align}
The stability of the stationary solutions can be analyzed by looking at the Jacobian matrix $J$, obtained by linearizing the mean-field equations around the stationary values. This matrix can be obtained by writing $m_c(t) \approx m_c^*+\delta m_c$, with $\delta m_c$ being small, and considering perturbations only up to first-order.
The linearized Jacobian matrix takes the form
\begin{align*}
  J = \left(
    \begin{array}{ccc}
      0&0&-\sqrt{2}\lambda m_z^*\\
      0&0&\sqrt{2}\lambda m_y^*\\
      -\frac{\lambda}{\sqrt{2}}&0&-\frac{\kappa}{2}
    \end{array}
  \right)\, .
\end{align*}
A stationary solution is stable if the real part of all the eigenvalues of the matrix $J$ is smaller or at most equal to zero.
The eigenvalues $\mu_i$ of the matrix $J$ are given by
\begin{align*}
  \mu_1 = 0 &&\text{and} && \mu_{2,3} = -\frac{\kappa}{4}\left(1\pm \sqrt{1 + \frac{4\lambda^2m_z^*}{\left(\frac{\kappa}{2}\right)^2}}\right)\, ,
\end{align*}
which  immediately shows that the stationary state with positive $m_z^*$ is unstable. The only stable stationary mean-field solution is the one with negative $m_z^*$ [see also Fig.~\ref{fig:Fig1}(c)]. Such a stationary solution is physical only when $\Omega \le \lambda^2/\kappa$. Here, the light field becomes macroscopically occupied, $\langle a^\dagger a \rangle\propto N (m_p^*)^2 $, denoting the superradiant character of the phase \cite{kirton2019}. For $\Omega>\lambda^2/\kappa$, no stationary solution exists  (within the sector identified by the choice of the conserved quantities) and the system belongs to a time-crystal phase, as shown in Fig.~\ref{fig:Fig1}(b-d).

\subsection{Proof of existence of the limit cycle}\label{existence_LC}
The non-stationary behavior of the system in the time-crystal phase is, as we will show analytically by closely following the derivation in Section 8.5 of Ref.~\cite{strogatz}, the result of an emergent limit-cycle dynamics. To show the existence of limit cycles for the mean-field equations [cf.~Eq.~\eqref{mean-field-eqs}], we first bring them into a more convenient form. We make use of the fact that $m_y^2+m_z^2=1$ is a conserved quantity and thus  Eq.~\eqref{mean-field-eqs} describes an evolution taking place on the surface of a cylinder. The dynamics of the system is then captured by
\begin{align*}
    \dot{\theta}(t) &= -2\Omega - \sqrt{2}\lambda m_p(t)\, ,\\\
     \dot{m}_p(t) &= -\frac{\lambda}{\sqrt{2}} m_y(t)- \frac{\kappa}{2}m_p(t)\, ,
\end{align*}
which can be obtained by exploiting the ansatz
\begin{equation}\label{eq:ansatz}
    \begin{split}
    m_y(t)&=\cos\theta(t)m_y(0) + \sin\theta(t)m_z(0)\, ,\\\
    m_z(t)&=\cos\theta(t)m_z(0) - \sin\theta(t)m_y(0)\, ,
\end{split}
\end{equation}
obeying  $\dot{m}_{y(z)}(t) = +(-) \dot{\theta}(t) m_{z(y)}(t)$.
Secondly, we perform the substitution $Y=-2\Omega - \sqrt{2}\lambda m_p$, yielding  
\begin{align}\label{eq:mod_josephson}
    \dot{\theta} = Y && \mathrm{and} && \dot{Y} = -\kappa\Omega + \lambda^2 m_y - \frac{\kappa}{2} Y\, .
\end{align}
The above equation is closely related to the differential equations for the dynamics of the phase difference across a Josephson junction  (see Section 8.5 of Ref.~\cite{strogatz}).

With the restriction $|m_y|\leq 1$, we find again that the stationary solutions of Eq.~(\ref{eq:mod_josephson}) only exist for $\Omega\kappa<\lambda^2$. Above the critical value $\Omega=\lambda^2/\kappa$, we find persistent oscillations of the system witnessing a stable limit cycle to which all trajectories are attracted.
To analyze the long-time behavior of the system in this parameter regime, we consider the nullcline $Y=\frac{2\lambda^2}{\kappa}m_y -2\Omega$, with $|m_y|\leq 1$, which defines a regime with vanishing derivative $\dot{Y}=0$. For smaller (larger) values of $Y$, the derivative $\dot{Y}$ is positive (negative) so that for long times all trajectories end up in a regime restricted to the strip $y_1 \le Y \le y_2$, for all  $y_1<-\frac{2\lambda^2}{\kappa}-2\Omega$ and $y_2>+\frac{2\lambda^2}{\kappa} -2\Omega$ \cite{strogatz}.

Given the periodicity of $m_y$ [cf. Eq.~(\ref{eq:ansatz})], it is sufficient to consider values $0\leq \theta \leq 2\pi$. For $\Omega > \lambda^2/\kappa$, we can fix $y_2<0$, such that the derivative of the angle $\dot{\theta} = Y < 0 $ does not change its sign inside the  strip. Thus, in the long-time limit a periodic solution can only exist within this strip.
A limit cycle is a trajectory that starts at a point $Y^*$ and ends after one period at the same point $P(Y^*) = Y^*$, where $P$ is called Poincaré map \cite{strogatz}. In order to show the existence of such a point inside the strip, we  make use of the fact that $P(y_1)>y_1\,, \forall\,  y_1 < -\frac{2\lambda^2}{\kappa}-2\Omega $, which is due to the fact that the derivative $\dot{Y}$ is strictly positive for $y_1 < -\frac{2\lambda^2}{\kappa}-2\Omega $ and thus $Y$ cannot go back to the value $y_1$ \cite{strogatz}.
Similarly, we have $P(y_2)<y_2\,, \forall\, y_2 < +\frac{2\lambda^2}{\kappa}-2\Omega $. Since the Poincaré map $P$ is continuous and monotonic, there must thus exist a value $Y^*$ such that $P(Y^*)=Y^*$, implying the existence of the limit cycle \cite{strogatz}. It is also possible to show that the closed orbit is unique (for details we refer to Section 8.5 of Ref.~\cite{strogatz}).

In Appendix~\ref{appendix_existence_TC} we further demonstrate that the emergent limit-cycle dynamics is associated with the spontaneous breaking of continuous time-translation symmetry. This shows that the considered system  features a proper time-crystal phase \cite{kongkhambut2022}.

\begin{figure}[t]
    \centering
    \includegraphics[width=\columnwidth]{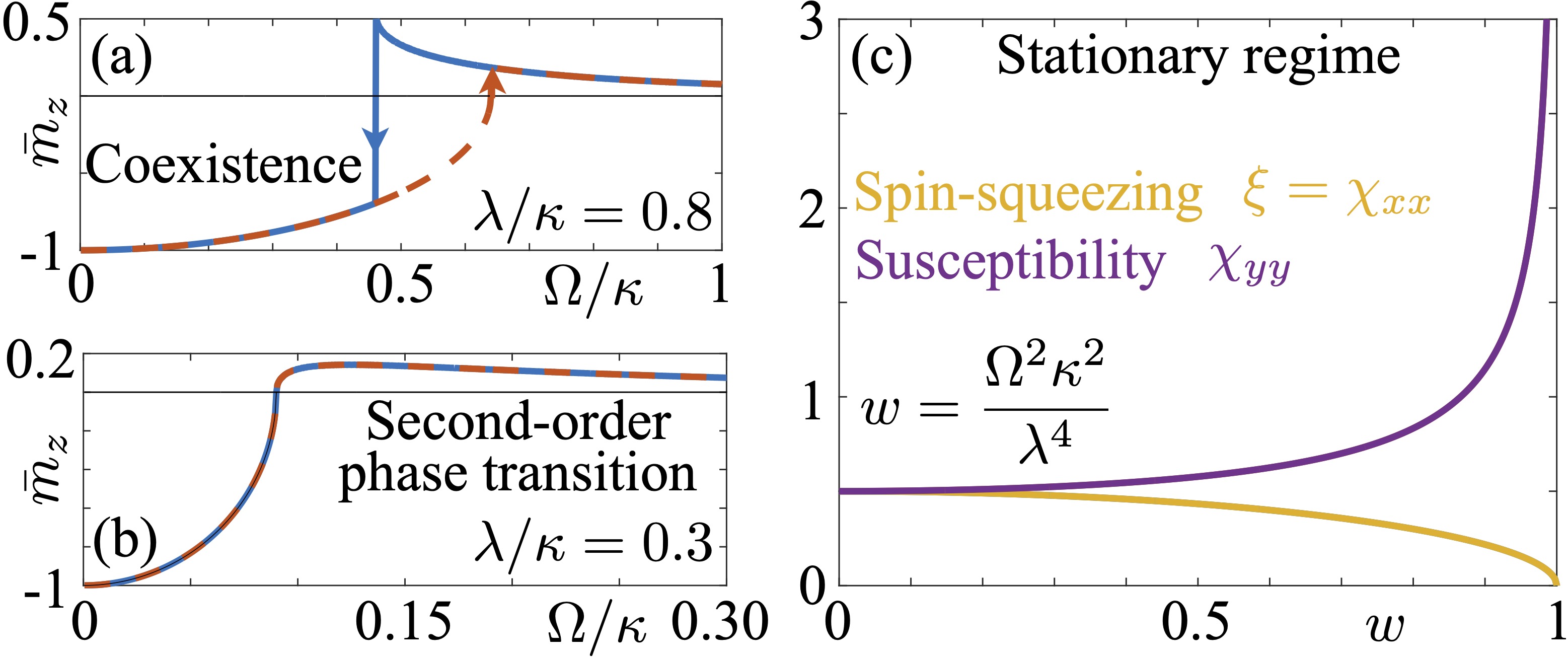}
    \caption{{\bf Coexistence and critical behavior.} 
    (a) Time-averaged $\bar{m}_z$ as a function of $\Omega$, for $\lambda/\kappa=0.8$ [upper dotted line in Fig.~\ref{fig:Fig1}(b)]. The solid (dashed) curve is obtained by starting from the time-crystal (stationary) phase and moving ``adiabatically" $\Omega$ towards the stationary (time-crystal) one. (b) Time-averaged $\bar{m}_z$ as a function of $\Omega$, for $\lambda/\kappa=0.3$ [lower dotted line in Fig.~\ref{fig:Fig1}(b)]. Here, the phase transition is of second order [see also inset of Fig.~\ref{fig:Fig1}(b)]. (c) Critical behavior of the spin-squeezing and of the  susceptibility in the stationary regime, as a function of $(\Omega \kappa/\lambda^2)^2$. The coordinates $x,y$ refers to the frame in which the $z$-axis is aligned with the direction of the vector identified by the stable stationary values in Eq.~\eqref{stat-mean-field}.}
    \label{fig:Fig2}
\end{figure}
\subsection{Phase diagram and bistability}
Having established the existence of a non-stationary regime, we now analyze in detail the nonequilibrium phase diagram of the system. We observe that, also within the parameter regime in which the stable stationary state of Eq.~\eqref{stat-mean-field} is well-defined, the system can approach a limit cycle. This implies the existence of a region where the stationary phase [cf.~Eq.~\eqref{stat-mean-field}] and the time-crystal one coexist. Such bistable regimes usually occur for stationary phases (see, e.g., Refs.~\cite{carr2013,marcuzzi2014}) and are characterized through a stability analysis. However, in our case one of the two asymptotic solutions is a limit cycle. As such, to fully explore the bistable region we take an approach which exploits the coexistence between the two phases. To treat the latter on an equal footing, we will focus on the time-averaged order-parameter $\bar{m}_z=\frac{1}{t} \int_0^t {\rm d}u \, m_z(u)$, which converges to the stable value in Eq.~\eqref{stat-mean-field} within the stationary regime while it gives an average over the oscillations in the time-crystal phase. 

When $\Omega > \lambda^2/\kappa$, the system can only be found in the time-crystal phase. The curve $\Omega=\lambda^2/\kappa$ thus provides one of the boundaries of the bistability region. To find the other boundary, i.e., the line separating the bistable regime from the stationary phase [cf.~Fig.~\ref{fig:Fig1}(b)], we probe coexistence behavior. The idea is as follows. We start at a point $(\Omega,\lambda)$ in parameter space, with $\Omega\gg\lambda$ where only the time-crystal phase is stable [cf.~Fig.~\ref{fig:Fig1}(b)]. We initialize the system in the state $\ket{\psi}$, with all atoms in the excited state $\ket{e}$ and the light field in the vacuum, and let it relax towards the asymptotic limit cycle. We then increase $\lambda$, in small discrete steps, in an {\it adiabatically slow} fashion, i.e., always giving the system sufficient time to accommodate into the new limit cycle. In this way, we can enter the bistable regime lying within the basin of attraction of the time-crystal phase. For sufficiently large $\lambda$, only the stationary phase is eventually stable. As shown in Fig.~\ref{fig:Fig1}(b), this makes the second (upper) spinodal line emerge as the line where $\bar{m}_z$ jumps from positive values, attained in the time-crystal phase, to the negative ones given by Eq.~\eqref{stat-mean-field}. A similar sweep through the phase diagram can be done by fixing $\lambda$. This procedure also shows coexistence of the two phases as apparent from Fig.~\ref{fig:Fig2}(a). 
The two spinodal lines meet at a tricritical point, highlighted in Fig.~\ref{fig:Fig1}(b). Beyond this point, the phase transition does not switch to a crossover as it usually happens, but it rather changes nature and becomes a second-order one, see Fig.~\ref{fig:Fig2}(b). Note that the curve in Fig.~\ref{fig:Fig2}(b) displays  a proper phase transition since i) the stable stationary value $m_z^*$ approaches the critical point with an infinite derivative [cf.~Eq.~\eqref{stat-mean-field}] and ii) as we calculate below and anticipate in Fig.~\ref{fig:Fig2}(c), approaching the critical point from the stationary regime the system features a diverging susceptibility.
{\color{black}
\subsection{Characterization of the phase transitions in terms of bifurcations}
\begin{figure}[t]
    \centering
    \includegraphics[width=0.6\columnwidth]{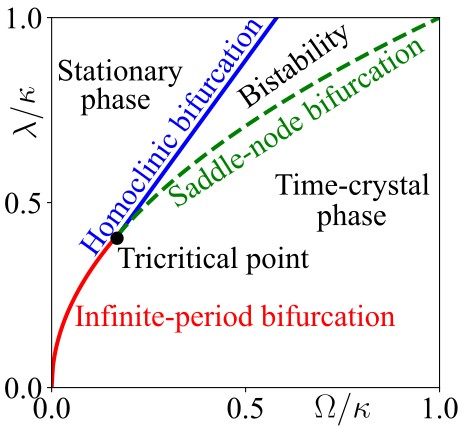}
    \caption{{\bf Phase diagram and bifurcations.} 
    Sketch of the phase diagram of the model specifying the types of bifurcation occurring at the critical and at the spinodal lines.}
    \label{fig:Fig_Bifurcation}
\end{figure}
The different phase-transition behavior can be  related to the different types of bifurcations \cite{strogatz} occurring at the transition lines, see also animations provided as Supplemental Material \cite{SM}.
 
In the regime where the adiabatic elimination is valid [lower left corner of the phase diagram in Fig.~\ref{fig:Fig1}(b)] the system undergoes a phase transition at the critical line $\Omega=\lambda^2/\kappa$. Crossing the latter from the time-crystal phase the periodic solution is disrupted by the emergence of a pair of fixed points, a saddle and a node \cite{SM}. Here, an infinite-period bifurcation [see also Fig.~\ref{fig:Fig_Bifurcation}] occurs and the  behavior of the system is analogous to that of the  boundary time-crystal \cite{cabot2022quantum}.

Above the tricritical point, the system can be  found in a bistable regime, in which the stable stationary solution and the stable limit cycle coexist. When starting from the time-crystal  phase and moving adiabatically slow inside and within the bistable regime, it is possible to remain within the basin of attraction of the time-crystal phase. In this case, when approaching the upper spinodal line [the line  separating the bistable regime from the stationary one in Fig.~\ref{fig:Fig_Bifurcation}] the limit cycle eventually hits the unstable (saddle) stationary solution. Here, a homoclinic bifurcation takes place and the system ``jumps" into the stable stationary solution \cite{SM}.
On the other hand, coming from the stationary phase and increasing the parameters adiabatically slow, the system stays in the basin of attraction of the stable stationary solution, even within  the coexistence regime. In this case, approaching the lower spinodal line [the line separating the bistable regime and the time-crystal phase in Fig.~\ref{fig:Fig_Bifurcation}], stable and unstable stationary solutions coalesce (saddle-node bifurcation). Beyond this line, the only attractor is the limit cycle.

The presence of different types of bifurcations (infinite-period bifurcation below the tricritical point \cite{cabot2022quantum}, saddle-node and homoclinic ones above) explains the appearance of different phase-transition behavior [cf.~Fig.~\ref{fig:Fig2}(a-b)].  Approaching the critical line below the tricritical point, the limit cycle acquires an infinite period and spends an infinite amount of time close to where the stable solution emerges. In this way, when passing from the limit cycle to the stationary solution the time-averaged magnetizations change continuously, i.e., the system undergoes a second-order phase transition. On the other hand, above the tricritical point, when passing from one phase to the other, the system experiences sudden jumps between two already existing solutions, which live in different regions of the ``phase space". This fact gives rise to a first-order phase transition with the associated jump of the order parameters.} 

\section{Dynamics of quantum fluctuations} Average operators converge, in the thermodynamic limit, to multiples of the identity \cite{landford1969} and thus cannot carry information about correlations. The natural next step is thus to consider suitable {\it susceptibility} parameters. In analogy with classical central limit theorems, for the atoms we introduce the quantum fluctuation operators \cite{goderis1989,goderis1990,narnhofer2002,benatti2014,benatti2016,benatti2016b,benatti2018}\begin{equation}
F_r^N =\frac{1}{\sqrt{2N}}\left(S_r-\langle S_r\rangle\right)\, ,
\label{quantum_fluct}
\end{equation}
whose variance, $\chi_{rr}=\langle F_r^2\rangle$, provides the fluctuations of the order parameter $m_r^N$, that is, its susceptibility. The operators in Eq.~\eqref{quantum_fluct} retain a quantum character in the thermodynamic limit. To understand this, let us consider the state with all atoms in $\ket{e}$. The commutator $[F_x^N,F_y^N]=im_z^N$ is proportional to an average operator and, thus, converges in the thermodynamic limit to the multiple of the identity $i m_z$, with $m_z=1$, due to our choice of the state. This commutation relation identifies the limiting fluctuation operators, $q_A=\lim_{N\to\infty}F_x^N$ and $p_A=\lim_{N\to\infty}F_y^N$, as two (bosonic) quadrature operators such that $[q_A,p_A]=i$. Together with these atom fluctuations, we consider the light-field fluctuation operators $q_L=q-\langle q\rangle$ and $p_L=p-\langle p\rangle$ \cite{boneberg2022}. The emergent two-mode bosonic description formed by the fluctuation operators $R=(q_A,p_A,q_L,p_L)^T$ can be used to analyze correlations between the atoms and the light field \cite{boneberg2022}. 
\begin{figure*}[t]
    \centering
    \includegraphics[width=\textwidth]{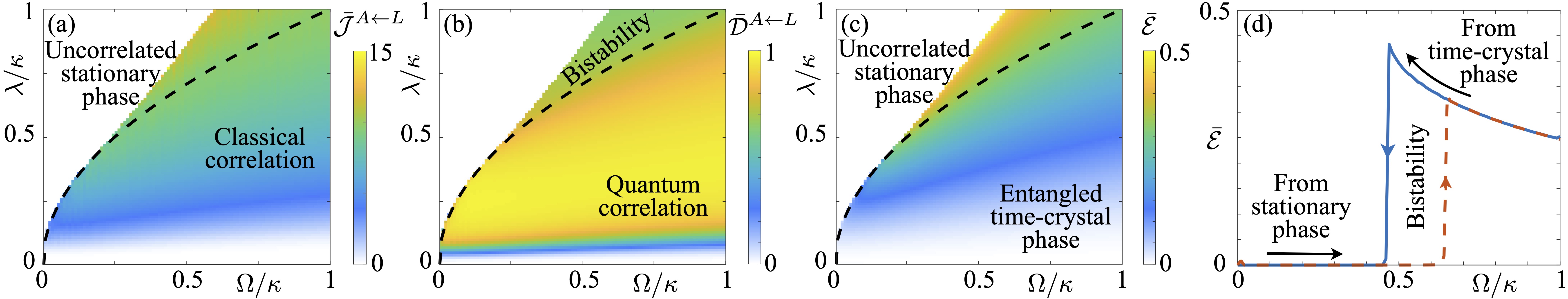}
    \caption{{\bf Quantum and classical correlations.} (a) Time-averaged classical correlation $\bar{\mathcal{J}}^{A\leftarrow L}$, as a function of $\lambda$ and $\Omega$. For each value of $\Omega$, the data are obtained by initializing the system in state $\ket{\psi}$, evolving it with the smallest value of $\lambda$, and then adiabatically increasing the interaction parameter $\lambda$ in discrete steps up to the largest values. The evolution time for each value of $\lambda$ is $\kappa t=5000$ and coincides with the averaging window for the correlation measure. (b) Same as in panel (a) for the quantum discord  $\bar{\mathcal{D}}^{A\leftarrow L}$. The latter shows that the time-crystal phase features quantum correlations. (c) Same as in panels (a) and (b) but for the logarithmic negativity quantifying the amount of entanglement between the atoms and the light field.  (d) Coexistence of different bipartite entanglement behavior, as measured by the time-averaged logarithmic negativity $\bar{\mathcal{E}}$, for $\lambda/\kappa=0.8$ [cf.~upper dotted line in Fig.~\ref{fig:Fig1}(b)], and different values of $\Omega$ slowly varied in discrete steps both starting from the stationary phase and the time-crystal phase. }
    \label{fig:Fig3}
\end{figure*}

To this end, we introduce the covariance matrix $\Sigma_{uv}=\langle \{R_u,R_v\}\rangle/2$ and investigate its time evolution. 
Due to the dynamics of average operators, the commutation relation between the fluctuation operators associated with the atoms generically depends on time \cite{benatti2018}. To ``remove" this dependence, we move to the frame rotating with the time-evolving average operators. Here, we can derive the Lindblad generator for the dynamics of the two-mode bosonic system related to quantum fluctuations (see Appendix~\ref{appendix_dynamics_q_fluc}).
{\color{black} 
The time-dependent Lindblad generator is of the form
$$
\mathcal{W}^*_{A-L}(t)[O]=i[H_{A-L}(t),O]+\mathcal{L}_L^*[O]\, ,
$$
with the Hamiltonian
$$
H_{A-L}(t)=\sum_{i,j=1}^4 h_{ij}(t)R_iR_j\, ,
$$
where 
$$
h(t)=\frac{\lambda}{2}
\begin{pmatrix}
0&0&0&1\\
0&0&m_z(t)&0\\
0&m_z(t)&0&0\\
1&0&0&0
\end{pmatrix}\,.
$$
In the generator above, the map $\mathcal{L}_L^*$ is the dual of the map
\begin{equation}
\mathcal{L}_L[X]=\kappa \left(a_L X a_L^\dagger-\frac{1}{2}\left\{a_L^\dagger a_L,X\right\}\right)\, ,
\label{L_L}
\end{equation}
which is analogous to the one in   Eq.~\eqref{dissipator} but with jump operator $a_L=(p_L-iq_L)/\sqrt{2}$.

Under this dynamics the covariance matrix evolves according to the differential equation
$$
\dot{\Sigma}(t)=[2sh(t)+sb]\Sigma(t)+\Sigma(t)[2sh(t)+sb]^T +scs^T\, ,
$$
with $s$ being the symplectic matrix of a two mode bosonic system \cite{benatti2018} 
\begin{align*}
        s = 
  \left(
    \begin{array}{cccc}
      0&1&0&0\\
      -1&0&0&0\\
      0&0&0&1\\
      0&0&-1&0
    \end{array}
  \right)\, ,
 \end{align*}
and the matrices 
\begin{align*}
        c = 
  \frac{\kappa}{2}\left(
    \begin{array}{cccc}
      0&0&0&0\\
      0&0&0&0\\
      0&0&1&0\\
      0&0&0&1
    \end{array}
  \right)\, ,
  \quad 
      b = 
  \frac{\kappa}{2}\left(
    \begin{array}{cccc}
      0&0&0&0\\
      0&0&0&0\\
      0&0&0&1\\
      0&0&-1&0
    \end{array}
  \right)\, ,
\end{align*}
encoding the dissipative dynamics of the system.
}\\

The emergent two-mode Hamiltonian, which can also be written as
\begin{equation}
H_{A-L}(t)=\lambda\left[q_Ap_L+m_z(t)q_Lp_A\right]\, ,
\label{Hamiltonian_fluct}
\end{equation}
is time-dependent as a consequence of the time-dependence of the instantaneous magnetization $m_z(t)$ and encodes 
{\color{black}
both an excitation-exchange and a two-mode squeezing process. To show this, we represent the fluctuation operators $q_A,p_A,q_L,p_L$ in terms of bosonic creation and annihilation operators. Due to the definition of the original quadrature operators of the light field, we write 
\begin{align}
    q_L=\frac{i}{\sqrt{2}}\left(a_L-a_L^\dagger\right)\, , \qquad p_L=\frac{1}{\sqrt{2}}(a_L+a_L^\dagger)\, .
\end{align}
For the atoms, we instead recall that $q_A$ is the limit of $F_x^N$ and $p_A$ the one of $F_y^N$ and, in order to associate the annihilation operator with $S_-$, we write
$$
q_A=\frac{1}{\sqrt{2}}\left(a_A+a_A^\dagger\right)\, , \qquad p_A=\frac{i}{\sqrt{2}}\left(a_A^\dagger -a_A\right)\, .
$$
Substituting these definitions into the Hamiltonian in Eq.~\eqref{Hamiltonian_fluct}, we find 
\begin{align*}
    H_{A-L}(t)=\frac{\lambda}{2}&\Big[\left(a_A a_L+a_A^\dagger a_L^\dagger \right)\left[1+m_z(t)\right]\\\
    &+\left(a_L^\dagger a_A+a_A^\dagger a_L\right)\left[1-m_z(t)\right]\Big]\, ,
\end{align*}
}which makes apparent that the Hamiltonian can be decomposed into an excitation-exchange term, proportional to $1-m_z(t)$, and a two-mode squeezing term, proportional to $1+m_z(t)$. These contributions provide the only coupling between the atoms and the light field and, as we show below, can generate quantum correlations between the two subsystems. In contrast to the boundary time-crystal model \cite{carollo2022}, the dynamics of fluctuations is not fully dissipative due to the collective Hamiltonian in Eq.~\eqref{Hamiltonian}. Since the emergent dynamical generator is quadratic, quantum fluctuations remain Gaussian  \cite{heinosaari2010}.

\section{Quantum correlations and entanglement} From the time evolution of the covariance matrix $\Sigma$, we can calculate classical correlation, quantum discord \cite{henderson2001,ollivier2001,adesso2010,giorda2010,isar2014}, as well as bipartite (collective) entanglement between the atoms and the light field \cite{simon2000,adesso2014}, in the thermodynamic limit. Within the stationary phase, the asymptotic covariance matrix can be computed exactly as 
\begin{equation}
    \Sigma=\frac{1}{2}{\rm diag}(-m_z^*,-(m_z^*)^{-1},1,1)\, ,
    \label{stat-Sigma}
\end{equation}
with {\color{black}the stable $m_z^*$ [cf. Eq.~\eqref{stat-mean-field}}]. This expression shows that the light field (described by the operators $q_L,p_L$) is in the vacuum state, while the collective atom operators $q_A,p_A$ are in a squeezed state. Eq.~\eqref{stat-Sigma} shows no correlations between the atoms and the light field in the stationary phase. Yet, the atoms display spin-squeezing, with a squeezing parameter, $\xi=|m_z^*|$, which diverges (to zero) on the spinodal line separating the bistable regime from the pure time-crystal phase. The divergence (to infinity) of $\Sigma_{22}\propto |1/m_z^*|$ is related to the divergence of the susceptibility close to the second-order phase transition [cf.~Fig.~\ref{fig:Fig2}(c)].  Since fluctuations are in the frame aligned with the direction of the stable state in Eq.~\eqref{stat-mean-field}, $\Sigma_{22}$, in the stationary regime and close to the phase transition line, is essentially the susceptibility of the order-parameter $m_{z}$. 

We now turn to the time-crystal phase. Here, there is no significant spin-squeezing in the atom ensemble. Moreover, it can be shown that the determinant of the covariance matrix increases indefinitely with time, which indicates that the state of the system becomes more and more mixed. Nonetheless, in this regime the atoms and the light field are correlated. This can be seen, for instance, through the one-way classical correlation. This quantity is a measure of the maximal information about one of the two subsystems, let us say the atoms, that can be gained by performing measurements on the other subsystem, in our case the light field. This one-way classical correlation, denoted as $\mathcal{J}^{A\leftarrow L}$, is shown in Fig.~\ref{fig:Fig3}(a) and demonstrates the existence of correlations in the time-crystal phase. Even more interestingly, also correlations of genuine quantum nature can be observed in this regime, as measured by the (one-way) quantum discord $\mathcal{D}^{A\leftarrow L}=\mathcal{I}-\mathcal{J}^{A\leftarrow L}$, with $\mathcal{I}$ being the mutual information between the atoms and light field. The quantum discord quantifies the amount of correlations which are not of classical nature. In Fig.~\ref{fig:Fig3}(b), we show that in the time-crystal phase the quantum discord is non-zero throughout. (We report results for $\mathcal{J}^{A\rightarrow L},\mathcal{D}^{A\rightarrow L}$ in Appendix~\ref{appendix_corr}.) Remarkably, a fraction of these quantum correlations is related to bipartite entanglement between the atom ensemble and the light field, which can be quantified through the logarithmic negativity $\mathcal{E}$ shown in Fig.~\ref{fig:Fig3}(c). Both classical and quantum correlations display coexistence behavior, as for instance shown in Fig.~\ref{fig:Fig3}(d), due to the coexistence of the uncorrelated stationary phase and the correlated time-crystal. 

{\color{black} To conclude we note that Fig.~\ref{fig:Fig3}(a-c) clearly shows that increasing the coupling strength $\lambda$ between the atoms and the light field does not always lead to increased correlations. Indeed, for fixed $\Omega$ and $\kappa$, a too large coupling strength $\lambda$  brings the system into the stationary uncorrelated phase.  }

\section{Discussion} 
The system we have investigated is related to the well-known boundary time-crystal model \cite{iemini2018} through an adiabatic elimination of the light field \cite{agarwal1997,gopalakrishnan2011,damanet2019,carollo2020,paulino2022}. For what concerns the atoms, it shows features which are similar to those of the boundary time-crystal. That is, we observe spin-squeezing in the stationary regime, and absence of quantum correlations among the atoms in the oscillatory phase \cite{hannukainen2018,sanchezmunoz2019,buonaiuto2021,carollo2022,lourenco2022,montenegro2023,pavlov2023}. However, explicitly considering the light field allowed us to uncover the existence of genuine quantum correlations in the time-crystal regime, even though the latter is characterized by a mixed state, established between atoms and light field. From a fundamental perspective our results demonstrate that time-crystal phases can display quantum correlations and are thus certainly not classical.  {\color{black} Given the Gaussian character of the quantum state of the atoms and the cavity mode, the correlations we have investigated here may be accessed experimentally via measurements of two-point correlation functions. } Our findings are valid in the thermodynamic limit. For a finite system, they are accurate up to a timescale $t^*$ (diverging for $N\to\infty$). Beyond this timescale, the oscillations in  single realizations of the dynamics dephase \cite{cabot2022quantum}. The average state thus consists of the sum over all possible dephased limit-cycles and becomes asymptotically time invariant. This phenomenology is related, in the thermodynamic limit, to mode-softening and phase diffusion in time-crystals \cite{nie2023,chan2015,benlloch2017}.

Finally, we note that the time-crystal phase appears to be related to lasing --- since there is an inversion of population signalled by a positive magnetization $\bar{m}_z>0$ \cite{kirton2019} --- even though the model does not possess a $U(1)$ symmetry. This is due to the fact that the ``pumping" is not incoherent but rather is implemented through external laser driving. However, the oscillations established are not harmonic [see Fig.~\ref{fig:Fig1}(c)] and, deep in the time-crystal phase, $|\Omega| \gg|\lambda|$, the time-averaged magnetization $\bar{m}_z$ tends to zero, i.e., there is no inversion of population \cite{kirton2019}.

\section*{Acknowledgments}
We thank Albert Cabot for useful discussions. We are grateful for financing from the Baden-W\"urttemberg Stiftung through Project No.~BWST\_ISF2019-23. We also acknowledge funding from the Deutsche Forschungsgemeinschaft (DFG, German Research Foundation) under Project No. 435696605 and through the Research Unit FOR 5413/1, Grant No. 465199066. This project has also received funding from the European Union’s Horizon Europe research and innovation program under Grant Agreement No. 101046968 (BRISQ), and from EPSRC under Grant No. EP/V031201/1. FC~is indebted to the Baden-W\"urttemberg Stiftung for the financial support of this research project by the Eliteprogramme for Postdocs.
\appendix
\section{Time-translation symmetry breaking}\label{appendix_existence_TC}
\begin{figure}[t]
    \centering
    \includegraphics[width=0.7\columnwidth]{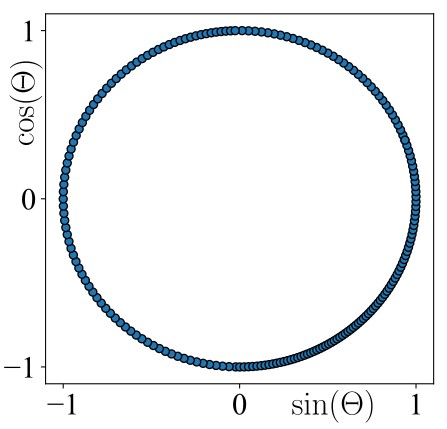}
    \caption{{\bf Continuous symmetry-breaking in the time-domain.} Distribution of the phase angle $\Theta$ in the time-crystal phase for fixed parameters $(\Omega/\kappa,\lambda/\kappa)=(0.5,0.5)$ and 200 random initial conditions encoded in the angle $\alpha$. The time at which $\Theta$ is evaluated is fixed for all initial values. As shown in the plot, $\Theta$ can assume all values between $0$ and $2\pi$, witnessing a continuous-time symmetry breaking.}
    \label{fig:FigSM0}
\end{figure}
In this section, we discuss the spontaneous breaking of the continuous time-translation symmetry associated with the observed time-crystal phase \cite{kongkhambut2022}.

As discussed in Ref.~\cite{kongkhambut2022}, a continuous time-crystal is characterized by emergent persistent oscillations of the system and the spontaneous breaking of continuous time-translation symmetry. After showing the former in Sec.~\ref{existence_LC} through the existence of the limit cycle we will now focus on the latter. With our ansatz in Eq.~\eqref{eq:ansatz} and an initial angle $\alpha$, where $m_y(0)=\sin\alpha$ and $m_z(0)=\cos\alpha$, we find $m_y(t)=\sin\Theta(t)$ and $m_z(t)=\cos\Theta(t)$, with the phase angle $\Theta(t)=\theta(t)+\alpha$. For randomly sampled initial conditions the system can assume all phases in the limit cycle [see Fig.~\ref{fig:FigSM0}]. Similarly to the discussion in Ref.~\cite{kongkhambut2022}, this witnesses the breaking of  continuous symmetry in the time domain. Additionally, this also shows that the system always approaches the time-crystal phase demonstrating its robustness against varying initial conditions.

\section{Dynamics of quantum fluctuations}\label{appendix_dynamics_q_fluc}
In this Appendix, we give details on the derivation of the evolution of the covariance matrix for fluctuation operators, as well as on the transformation to the frame which rotates solidly with the main direction of the atom average operators. We then explicitly derive the dynamical generator for quantum fluctuations in such rotating frame.

\subsection{Time evolution of the covariance matrix of quantum fluctuations}
The derivation of the time evolution of the covariance matrix for fluctuation operators follows closely the one presented in Ref.~\cite{boneberg2022}. We start by introducing the vector of fluctuation operators $\tilde{R}^N=(F^N_x,F^N_y,F^N_z,q_L,p_L)^T$ in the time-independent frame. The covariance matrix of these fluctuation operators can be written as $\tilde{\Sigma} = \underset{N\to\infty}{\lim} \left(K^N+(K^N)^T\right)/2$, where we have defined $K_{uv}^N = \left\langle \tilde{R}_u^ N\tilde{R}_v^N \right\rangle$. Here, the expectation $\langle\cdot\rangle$ denotes expectation with respect to the state at time $t$. 

We now consider the time evolution of this correlation function $K_{uv}^N$. First, we note that 
$$
\dot{F^N_u}=-\frac{1}{\sqrt{2N}}\dot{\langle S_u\rangle} \, , \mbox{ for } u=x,y,z\, ,
$$
as well as $\dot{q_L}=-\dot{\langle q\rangle}$ and $\dot{p_L}=-\dot{\langle p\rangle}$, and that $\langle \tilde{R}_u^N\rangle=0$.  
Taking the time derivative of $K^N_{uv}$ then leads to 
\begin{align*}
  \dot{K}_{uv}^N = &\left\langle i\left[H,\tilde{R}_u^N\right]\tilde{R}_v^N \right\rangle + \left\langle i\tilde{R}_u^N\left[H,\tilde{R}_v^N\right] \right\rangle \\\
  &+ \left\langle \mathcal{L}^*\left[\tilde{R}_u^N \tilde{R}_v^N\right] \right\rangle\, ,
\end{align*}
where $\mathcal{L}^*$ is the dissipator in the Heisenberg picture, i.e., the map dual to $\mathcal{L}$. 
To proceed, we compute the commutators in the above equations which can then be rewritten in terms of fluctuation operators exploiting again that $\langle \tilde{R}_u^N\rangle=0$. A similar calculation also applies to the dissipative part in the above equation. As in Ref.~\cite{boneberg2022}, this gives rise to products of fluctuation operators and average operators. Making use of the fact that, in the thermodynamic limit, $\underset{N\to\infty}{\lim}\left\langle \tilde{R}_r^N m_u^N \tilde{R}_v^N \right\rangle = m_u(t)\left\langle \tilde{R}_r \tilde{R}_v \right\rangle$, and recalling the relation between $K^N$ and the covariance matrix, we find that 
\begin{align*}
  \dot{\tilde{\Sigma}}(t) =\tilde{W}(t)\tilde{\Sigma}(t) + \tilde{\Sigma}(t) \tilde{W}^T(t) + \tilde{S}(t){C}\tilde{S}^T(t)\, ,
\end{align*}
with $\tilde{W}(t) = \tilde{P}(t) +\tilde{S}(t)B$.
Here, we have defined the symplectic matrix 
\begin{align*}
    \tilde{S}(t) =
  \left(
    \begin{array}{ccccc}
      0&m_z(t)&-m_y(t)&0&0\\
      -m_z(t)&0&m_x(t)&0&0\\
      m_y(t)&-m_x(t)&0&0&0\\
      0&0&0&0&1\\
      0&0&0&-1&0\\
    \end{array}
  \right)\, ,
\end{align*}
encoding the commutation relation between fluctuation operators. We further have 
\begin{align*}
    D = 
  \frac{\kappa}{2}\left(
    \begin{array}{ccccc}
      0&0&0&0&0\\
      0&0&0&0&0\\
      0&0&0&0&0\\
      0&0&0&1&i\\
      0&0&0&-i&1
    \end{array}
  \right)\, ,
\end{align*}
through which we can define $C=(D+D^T)/2$ and $B=(D-D^T)/(2i)$, and 
\begin{widetext}
\begin{align*}
  \tilde{P}(t) = \left(
    \begin{array}{ccccc}
      0&0&\sqrt{2}\lambda m_q(t)&\lambda m_z(t)&0\\
      0&0&-2\Omega-\sqrt{2}\lambda m_p(t) &0&-\lambda m_z(t)\\
      -\sqrt{2}\lambda m_q(t)&2\Omega+\sqrt{2}\lambda m_p(t)&0&-\lambda m_x(t)&\lambda m_y(t)\\
      \lambda&0&0&0&0\\
      0&-\lambda&0&0&0
    \end{array}
  \right)\, .
\end{align*}
\end{widetext}
The evolution for the case considered in the main text is obtained by setting $m_x=m_q=0$. 

\subsection{Covariance matrix in the rotating frame}
We now focus on the case in which the system is initialized in the state with all atoms in $\ket{e}$ and the light field in the vacuum. This gives $m_x(t)=m_q(t)=0$ and $m_y^2(t)+m_z^2(t)=1$. This is the initial state considered for producing the plots in the main text. Our task is now to find the time evolution of the covariance matrix in the frame which rotates solidly with the direction identified by the average operators. 
To rotate the reference frame of the atom operators back to the initial one, we need to find the rotation matrix which maps the instantaneous vector of the average operators $m=[0,m_y(t),m_z(t),m_q(t),0]^T$ into the one $m=[0,0,1,m_q(t),0]^T$. Exploiting the conservation law $m_y^2(t)+m_z^2(t)=1$, this matrix can be found to be the matrix 
\begin{align*}
    U(t)=
  \left(
    \begin{array}{ccccc}
      1 &0&0&0&0\\
      0&m_z(t)&-m_y(t)&0&0\\
      0&m_y(t)&m_z(t)&0&0\\
      0&0&0&1&0\\
      0&0&0&0&1\\
    \end{array}
  \right)\, .
\end{align*}
Under this transformation, the symplectic matrix becomes 
\begin{align*}
    S=U(t)\tilde{S}(t) U^T(t)=  \left(
    \begin{array}{ccccc}
      0&1&0&0&0\\
      -1&0&0&0&0\\
      0&0&0&0&0\\
      0&0&0&0&1\\
      0&0&0&-1&0\\
    \end{array}
  \right)\, .
\end{align*}
The time evolution of the covariance matrix in the rotating frame can be calculated by taking the derivative of $\hat{\Sigma}=U(t)\tilde{\Sigma}(t) U^T(t)$, which gives 
\begin{align}
\label{eq:cov_rot_deriv}
  \dot{{\hat{\Sigma}}} (t) = Q(t) {\hat{\Sigma}}(t) + {\hat{\Sigma}}(t)Q^T(t) + SCS^T\, ,
 \end{align}
where
\begin{align*}
 Q(t) = \left(
    \begin{array}{ccccc}
      0&0&0&\lambda m_z(t)&0\\
      0&0&0&0&-\lambda\\
      0&0&0&0&0\\
      \lambda&0&0&-\frac{\kappa}{2}&0\\
      0&-\lambda m_z(t)&-\lambda m_y(t)&0&-\frac{\kappa}{2}
    \end{array}
  \right)\, .
\end{align*}
\\

For the considered initial state, the covariance matrix is given by the diagonal matrix $\hat{\Sigma}(0)=1/2\, {\rm diag}(1,1,0,1,1)$. Starting from this covariance matrix, it is possible to see that the third row and the third column of the covariance matrix are not coupled with the remainder of the matrix. We thus define ${\Sigma}$ to be the covariance matrix of the fluctuation operator $q_A$ (which is the limiting operator of the fluctuation $F_x^N$ in the rotating frame), $p_A$ (which is the limiting operator of the fluctuation $F_y^N$ in the rotating frame) coupled to the fluctuations $q_L$,$p_L$ (see also main text). 

For such a matrix, the time evolution is given by the equation 
\begin{equation}
\dot{\Sigma}(t)=X(t)\Sigma(t)+\Sigma(t)X^T(t)+s c s^T\, ,
\label{cov_mat_rot_frame}
\end{equation}
where $X,s,c$ are the $4\times4$ matrices obtained by removing the third row and third column in $Q,S,C$ respectively. 
\subsection{Dynamical generator for the quantum fluctuation dynamics in the rotating frame}
We now want to find the generator for the dynamics of the two-mode bosonic system described by the vector of bosonic operators $R=(q_A,p_A,q_L,p_L)^T$. As done in Ref.~\cite{carollo2022}, to this end we consider a time-dependent Lindblad generator on bosonic operators of the form 
$$
\mathcal{W}^*_{A-L}(t)[O]=i[H_{A-L}(t),O]+\mathcal{L}_L^*[O]\, ,
$$
with an ansatz for the Hamiltonian given by 
$$
H_{A-L}(t)=\sum_{i,j=1}^4 h_{ij}(t)R_iR_j\, .
$$
The dissipative part of the generator is essentially equivalent to the one of the original system, except that it now features the ``rescaled" fluctuation operators of the light [cf.~Eq.~\eqref{L_L} in the main text]. Using the generator $\mathcal{W}_{A-L}^*(t)$ to calculate the time evolution of the covariance matrix one finds  
$$
\dot{\Sigma}(t)=[2sh(t)+sb]\Sigma(t)+\Sigma(t)[2sh(t)+sb]^T +scs^T\, .
$$
Here, we have that the $4\times4$ matrix $b$ is obtained by removing the third row and the third column from the matrix $B$ introduced above. Comparing the above equation with Eq.~\eqref{cov_mat_rot_frame} gives that the generator correctly captures the dynamics of the covariance matrix if the relation 
$$
2sh(t)=\begin{pmatrix}
0&0&\lambda m_z(t)&0\\
0&0&0&-\lambda\\
\lambda & 0&0&0\\
0 &-\lambda m_z(t)&0&0
\end{pmatrix}\, 
$$
is satisfied.  Exploiting that $s^2=-\mathbb{I}$, we can invert the relation to find the Hamiltonian reported in the main text.
\section{Quantum and classical correlations}\label{appendix_corr}
In this Section, we describe how to calculate the correlation measures that we analyze in our work and we further present additional results on these. For details on the derivation of these measures, we refer to Refs.~\cite{adesso2010,giorda2010,simon2000,adesso2014}.\\ 
\begin{figure*}[t]
    \centering
    \includegraphics[width=0.75\textwidth]{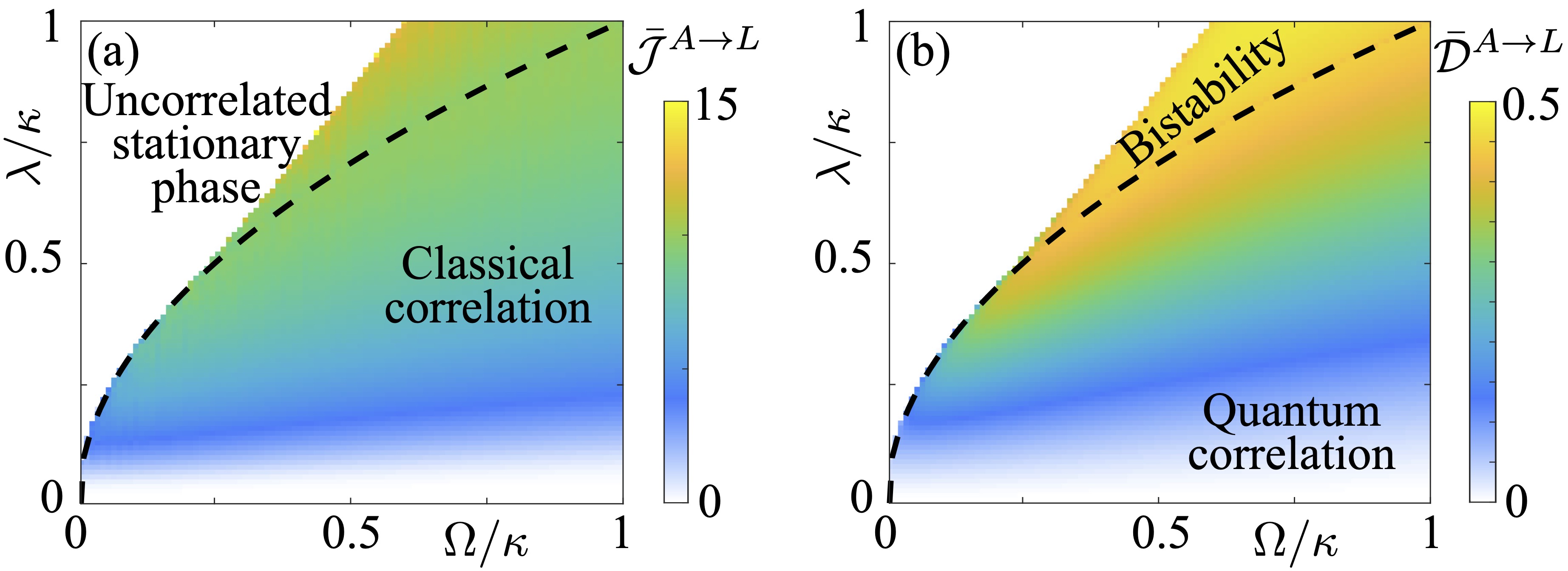}
    \caption{{\bf Additional results on quantum and classical correlations.} (a) Time-averaged classical correlation $\bar{\mathcal{J}}^{A\rightarrow L}$, as a function of $\lambda$ and $\Omega$. For each value of $\Omega$, the data are obtained by initializing the system in state $\ket{\psi}$, evolving it with the smallest value of $\lambda$, and then adiabatically increasing the interaction parameter $\lambda$ in discrete steps up to the largest values. The evolution time for each value of $\lambda$ is $\kappa t=5000$ and coincides with the averaging window for the correlation measure. (b) Same as in panel (a) for the quantum discord  $\bar{\mathcal{D}}^{A\rightarrow L}$. The latter shows that the time-crystal phase features quantum correlations. }
    \label{fig:FigSM}
\end{figure*}

Given the two-mode covariance matrix $\Sigma(t)$, we now show how to compute measures for  the classical correlation, for the quantum discord, and for the logarithmic negativity. To start, we identify the relevant $2\times2$ minors of $\Sigma$ as the matrices $\alpha$, $\beta$ and $\gamma$, such that 
$$
2\Sigma(t)=\begin{pmatrix}
\alpha&\gamma\\
\gamma& \beta\, .
\end{pmatrix}
$$
Here, the matrix $\alpha$ contains the variances of the atom fluctuations, $\beta$ those of the light-field fluctuations, while $\gamma$ contains the covariances between the atoms and the light field. We now define the quantities 
$$
c_\alpha={\rm det}(\alpha)\, , \, c_\beta={\rm det}(\beta)\, ,\, c_\gamma={\rm det}(\gamma)\, , \, c_\delta={\rm det}(2\Sigma)\, ,
$$
as well as the function 
\begin{align*}
  f(x) = \left(\frac{x+1}{2}\right)\log\left(\frac{x+1}{2}\right)-\left(\frac{x-1}{2}\right)\log\left(\frac{x-1}{2}\right)\, .
\end{align*}

For a two-mode Gaussian state an expression for the one-way classical correlation, quantifying the information on the first mode obtained by measurements performed on the second mode, is given by 
\begin{align}
  \mathcal{J}^{A\leftarrow L} = f(\sqrt{c_\alpha}) - f(\sqrt{E_{\mathrm{min} } } )\, ,
  \label{eq:class_corr_2}
\end{align}
while the quantum discord is 
\begin{align}
  \mathcal{D}^{A\leftarrow L} = f(\sqrt{c_\beta}) - f(\nu_-) - f(\nu_+) + f(\sqrt{E_{\mathrm{min} } } )\, ,
  \label{eq:q_discord_2}
\end{align}
where $E_{\mathrm{min} }$ is defined as
\begin{widetext}
\begin{align*}
  E_{\mathrm{min} } = \left\{ \begin{array}{cc}
    \frac{2c_\gamma^2+(c_\beta-1)(c_\delta-c_\alpha)+2|c_\gamma|\sqrt{c_\gamma^2+(c_\beta-1)(c_\delta-c_\alpha)}}{(c_\beta-1)^2} & \textrm{for}\, (c_\delta-c_\alpha c_\beta)^2 \leq (1+c_\beta)c_\gamma^2(c_\alpha+c_\delta)\, ,\\
    \frac{c_\alpha c_\beta-c_\gamma^2+c_\delta-\sqrt{c_\gamma^4+(c_\delta-c_\alpha c_\beta)^2-2c_\gamma^2(c_\alpha c_\beta+c_\delta)}}{2c_\beta} & \textrm{otherwise}\, .
  \end{array}\right.
\end{align*}
\end{widetext}
The quantities $\nu_-$ and $\nu_+$ are the symplectic eigenvalues of the matrix $2\Sigma$, with $\nu_- < \nu_+$. These are found as the positive eigenvalues of the matrix $2is\Sigma$. 
To compute the correlations $\mathcal{J}^{A\rightarrow L}$ and $\mathcal{D}^{A\rightarrow L}$, quantifying the information about the light field that can be obtained from a measurement on the atoms, one can exploit the same definitions as above but exchanging the roles of $\alpha$ and $\beta$ in all of the above relations. 

In order to quantify the amount of bipartite entanglement between the atoms and the light field, we compute the logarithmic negativity. This is defined as 
\begin{align*}
  \mathcal{E}= \text{max}\left(0, -\log\left(\tilde{\nu}_-\right)\right)\, ,
\end{align*}
where $\tilde{\nu}_-$ is the smallest symplectic eigenvalue of the partially transposed covariance $\Sigma^{PT}=\Lambda \Sigma \Lambda$, where $\Lambda={\rm diag}(1,1,1,-1)$. The latter is computed as the smallest positive eigenvalues of the matrix $2is\Sigma^{PT}$.
\bibliography{refs}

\begin{thebibliography}{135}%
\makeatletter
\providecommand \@ifxundefined [1]{%
 \@ifx{#1\undefined}
}%
\providecommand \@ifnum [1]{%
 \ifnum #1\expandafter \@firstoftwo
 \else \expandafter \@secondoftwo
 \fi
}%
\providecommand \@ifx [1]{%
 \ifx #1\expandafter \@firstoftwo
 \else \expandafter \@secondoftwo
 \fi
}%
\providecommand \natexlab [1]{#1}%
\providecommand \enquote  [1]{``#1''}%
\providecommand \bibnamefont  [1]{#1}%
\providecommand \bibfnamefont [1]{#1}%
\providecommand \citenamefont [1]{#1}%
\providecommand \href@noop [0]{\@secondoftwo}%
\providecommand \href [0]{\begingroup \@sanitize@url \@href}%
\providecommand \@href[1]{\@@startlink{#1}\@@href}%
\providecommand \@@href[1]{\endgroup#1\@@endlink}%
\providecommand \@sanitize@url [0]{\catcode `\\12\catcode `\$12\catcode
  `\&12\catcode `\#12\catcode `\^12\catcode `\_12\catcode `\%12\relax}%
\providecommand \@@startlink[1]{}%
\providecommand \@@endlink[0]{}%
\providecommand \url  [0]{\begingroup\@sanitize@url \@url }%
\providecommand \@url [1]{\endgroup\@href {#1}{\urlprefix }}%
\providecommand \urlprefix  [0]{URL }%
\providecommand \Eprint [0]{\href }%
\providecommand \doibase [0]{https://doi.org/}%
\providecommand \selectlanguage [0]{\@gobble}%
\providecommand \bibinfo  [0]{\@secondoftwo}%
\providecommand \bibfield  [0]{\@secondoftwo}%
\providecommand \translation [1]{[#1]}%
\providecommand \BibitemOpen [0]{}%
\providecommand \bibitemStop [0]{}%
\providecommand \bibitemNoStop [0]{.\EOS\space}%
\providecommand \EOS [0]{\spacefactor3000\relax}%
\providecommand \BibitemShut  [1]{\csname bibitem#1\endcsname}%
\let\auto@bib@innerbib\@empty
\bibitem [{\citenamefont {Dicke}(1954)}]{dicke1954}%
  \BibitemOpen
  \bibfield  {author} {\bibinfo {author} {\bibfnamefont {R.~H.}\ \bibnamefont
  {Dicke}},\ }\bibfield  {title} {\bibinfo {title} {{Coherence in Spontaneous
  Radiation Processes}},\ }\href {https://doi.org/10.1103/PhysRev.93.99}
  {\bibfield  {journal} {\bibinfo  {journal} {Phys. Rev.}\ }\textbf {\bibinfo
  {volume} {93}},\ \bibinfo {pages} {99} (\bibinfo {year} {1954})}\BibitemShut
  {NoStop}%
\bibitem [{\citenamefont {Hepp}\ and\ \citenamefont
  {Lieb}(1973{\natexlab{a}})}]{hepp1973superradiant}%
  \BibitemOpen
  \bibfield  {author} {\bibinfo {author} {\bibfnamefont {K.}~\bibnamefont
  {Hepp}}\ and\ \bibinfo {author} {\bibfnamefont {E.~H.}\ \bibnamefont
  {Lieb}},\ }\bibfield  {title} {\bibinfo {title} {{On the superradiant phase
  transition for molecules in a quantized radiation field: the Dicke maser
  model}},\ }\href
  {https://doi.org/https://doi.org/10.1016/0003-4916(73)90039-0} {\bibfield
  {journal} {\bibinfo  {journal} {Ann. Phys.}\ }\textbf {\bibinfo {volume}
  {76}},\ \bibinfo {pages} {360} (\bibinfo {year}
  {1973}{\natexlab{a}})}\BibitemShut {NoStop}%
\bibitem [{\citenamefont {Hepp}\ and\ \citenamefont
  {Lieb}(1973{\natexlab{b}})}]{hepp1973equilibrium}%
  \BibitemOpen
  \bibfield  {author} {\bibinfo {author} {\bibfnamefont {K.}~\bibnamefont
  {Hepp}}\ and\ \bibinfo {author} {\bibfnamefont {E.~H.}\ \bibnamefont
  {Lieb}},\ }\bibfield  {title} {\bibinfo {title} {{Equilibrium statistical
  mechanics of matter interacting with the quantized radiation field}},\ }\href
  {https://doi.org/https://doi.org/10.1103/PhysRevA.8.2517} {\bibfield
  {journal} {\bibinfo  {journal} {Phys. Rev. A}\ }\textbf {\bibinfo {volume}
  {8}},\ \bibinfo {pages} {2517} (\bibinfo {year}
  {1973}{\natexlab{b}})}\BibitemShut {NoStop}%
\bibitem [{\citenamefont {Wang}\ and\ \citenamefont {Hioe}(1973)}]{wang1973}%
  \BibitemOpen
  \bibfield  {author} {\bibinfo {author} {\bibfnamefont {Y.~K.}\ \bibnamefont
  {Wang}}\ and\ \bibinfo {author} {\bibfnamefont {F.~T.}\ \bibnamefont
  {Hioe}},\ }\bibfield  {title} {\bibinfo {title} {{Phase Transition in the
  Dicke Model of Superradiance}},\ }\href
  {https://doi.org/10.1103/PhysRevA.7.831} {\bibfield  {journal} {\bibinfo
  {journal} {Phys. Rev. A}\ }\textbf {\bibinfo {volume} {7}},\ \bibinfo {pages}
  {831} (\bibinfo {year} {1973})}\BibitemShut {NoStop}%
\bibitem [{\citenamefont {Hioe}(1973)}]{hioe1973}%
  \BibitemOpen
  \bibfield  {author} {\bibinfo {author} {\bibfnamefont {F.~T.}\ \bibnamefont
  {Hioe}},\ }\bibfield  {title} {\bibinfo {title} {{Phase Transitions in Some
  Generalized Dicke Models of Superradiance}},\ }\href
  {https://doi.org/10.1103/PhysRevA.8.1440} {\bibfield  {journal} {\bibinfo
  {journal} {Phys. Rev. A}\ }\textbf {\bibinfo {volume} {8}},\ \bibinfo {pages}
  {1440} (\bibinfo {year} {1973})}\BibitemShut {NoStop}%
\bibitem [{\citenamefont {Carmichael}\ \emph {et~al.}(1973)\citenamefont
  {Carmichael}, \citenamefont {Gardiner},\ and\ \citenamefont
  {Walls}}]{carmichael1973}%
  \BibitemOpen
  \bibfield  {author} {\bibinfo {author} {\bibfnamefont {H.}~\bibnamefont
  {Carmichael}}, \bibinfo {author} {\bibfnamefont {C.}~\bibnamefont
  {Gardiner}},\ and\ \bibinfo {author} {\bibfnamefont {D.}~\bibnamefont
  {Walls}},\ }\bibfield  {title} {\bibinfo {title} {{Higher order corrections
  to the Dicke superradiant phase transition}},\ }\href
  {https://doi.org/https://doi.org/10.1016/0375-9601(73)90679-8} {\bibfield
  {journal} {\bibinfo  {journal} {Phys. Lett. A}\ }\textbf {\bibinfo {volume}
  {46}},\ \bibinfo {pages} {47} (\bibinfo {year} {1973})}\BibitemShut {NoStop}%
\bibitem [{\citenamefont {Gross}\ and\ \citenamefont
  {Haroche}(1982)}]{gross1982}%
  \BibitemOpen
  \bibfield  {author} {\bibinfo {author} {\bibfnamefont {M.}~\bibnamefont
  {Gross}}\ and\ \bibinfo {author} {\bibfnamefont {S.}~\bibnamefont
  {Haroche}},\ }\bibfield  {title} {\bibinfo {title} {{Superradiance: An essay
  on the theory of collective spontaneous emission}},\ }\href
  {https://doi.org/https://doi.org/10.1016/0370-1573(82)90102-8} {\bibfield
  {journal} {\bibinfo  {journal} {Phys. Rep.}\ }\textbf {\bibinfo {volume}
  {93}},\ \bibinfo {pages} {301} (\bibinfo {year} {1982})}\BibitemShut
  {NoStop}%
\bibitem [{\citenamefont {Emary}\ and\ \citenamefont
  {Brandes}(2003{\natexlab{a}})}]{emary2003}%
  \BibitemOpen
  \bibfield  {author} {\bibinfo {author} {\bibfnamefont {C.}~\bibnamefont
  {Emary}}\ and\ \bibinfo {author} {\bibfnamefont {T.}~\bibnamefont
  {Brandes}},\ }\bibfield  {title} {\bibinfo {title} {{Chaos and the quantum
  phase transition in the Dicke model}},\ }\href
  {https://doi.org/10.1103/PhysRevE.67.066203} {\bibfield  {journal} {\bibinfo
  {journal} {Phys. Rev. E}\ }\textbf {\bibinfo {volume} {67}},\ \bibinfo
  {pages} {066203} (\bibinfo {year} {2003}{\natexlab{a}})}\BibitemShut
  {NoStop}%
\bibitem [{\citenamefont {Emary}\ and\ \citenamefont
  {Brandes}(2003{\natexlab{b}})}]{emary2003b}%
  \BibitemOpen
  \bibfield  {author} {\bibinfo {author} {\bibfnamefont {C.}~\bibnamefont
  {Emary}}\ and\ \bibinfo {author} {\bibfnamefont {T.}~\bibnamefont
  {Brandes}},\ }\bibfield  {title} {\bibinfo {title} {{Quantum Chaos Triggered
  by Precursors of a Quantum Phase Transition: The Dicke Model}},\ }\href
  {https://doi.org/10.1103/PhysRevLett.90.044101} {\bibfield  {journal}
  {\bibinfo  {journal} {Phys. Rev. Lett.}\ }\textbf {\bibinfo {volume} {90}},\
  \bibinfo {pages} {044101} (\bibinfo {year} {2003}{\natexlab{b}})}\BibitemShut
  {NoStop}%
\bibitem [{\citenamefont {Lambert}\ \emph {et~al.}(2005)\citenamefont
  {Lambert}, \citenamefont {Emary},\ and\ \citenamefont
  {Brandes}}]{lambert2005}%
  \BibitemOpen
  \bibfield  {author} {\bibinfo {author} {\bibfnamefont {N.}~\bibnamefont
  {Lambert}}, \bibinfo {author} {\bibfnamefont {C.}~\bibnamefont {Emary}},\
  and\ \bibinfo {author} {\bibfnamefont {T.}~\bibnamefont {Brandes}},\
  }\bibfield  {title} {\bibinfo {title} {{Entanglement and entropy in a
  spin-boson quantum phase transition}},\ }\href
  {https://doi.org/10.1103/PhysRevA.71.053804} {\bibfield  {journal} {\bibinfo
  {journal} {Phys. Rev. A}\ }\textbf {\bibinfo {volume} {71}},\ \bibinfo
  {pages} {053804} (\bibinfo {year} {2005})}\BibitemShut {NoStop}%
\bibitem [{\citenamefont {Bastidas}\ \emph {et~al.}(2012)\citenamefont
  {Bastidas}, \citenamefont {Emary}, \citenamefont {Regler},\ and\
  \citenamefont {Brandes}}]{bastidas2012}%
  \BibitemOpen
  \bibfield  {author} {\bibinfo {author} {\bibfnamefont {V.~M.}\ \bibnamefont
  {Bastidas}}, \bibinfo {author} {\bibfnamefont {C.}~\bibnamefont {Emary}},
  \bibinfo {author} {\bibfnamefont {B.}~\bibnamefont {Regler}},\ and\ \bibinfo
  {author} {\bibfnamefont {T.}~\bibnamefont {Brandes}},\ }\bibfield  {title}
  {\bibinfo {title} {{Nonequilibrium Quantum Phase Transitions in the Dicke
  Model}},\ }\href {https://doi.org/10.1103/PhysRevLett.108.043003} {\bibfield
  {journal} {\bibinfo  {journal} {Phys. Rev. Lett.}\ }\textbf {\bibinfo
  {volume} {108}},\ \bibinfo {pages} {043003} (\bibinfo {year}
  {2012})}\BibitemShut {NoStop}%
\bibitem [{\citenamefont {Keeling}\ \emph {et~al.}(2010)\citenamefont
  {Keeling}, \citenamefont {Bhaseen},\ and\ \citenamefont
  {Simons}}]{keeling2010}%
  \BibitemOpen
  \bibfield  {author} {\bibinfo {author} {\bibfnamefont {J.}~\bibnamefont
  {Keeling}}, \bibinfo {author} {\bibfnamefont {M.~J.}\ \bibnamefont
  {Bhaseen}},\ and\ \bibinfo {author} {\bibfnamefont {B.~D.}\ \bibnamefont
  {Simons}},\ }\bibfield  {title} {\bibinfo {title} {{Collective Dynamics of
  Bose-Einstein Condensates in Optical Cavities}},\ }\href
  {https://doi.org/10.1103/PhysRevLett.105.043001} {\bibfield  {journal}
  {\bibinfo  {journal} {Phys. Rev. Lett.}\ }\textbf {\bibinfo {volume} {105}},\
  \bibinfo {pages} {043001} (\bibinfo {year} {2010})}\BibitemShut {NoStop}%
\bibitem [{\citenamefont {Kirton}\ \emph {et~al.}(2019)\citenamefont {Kirton},
  \citenamefont {Roses}, \citenamefont {Keeling},\ and\ \citenamefont
  {Dalla~Torre}}]{kirton2019}%
  \BibitemOpen
  \bibfield  {author} {\bibinfo {author} {\bibfnamefont {P.}~\bibnamefont
  {Kirton}}, \bibinfo {author} {\bibfnamefont {M.~M.}\ \bibnamefont {Roses}},
  \bibinfo {author} {\bibfnamefont {J.}~\bibnamefont {Keeling}},\ and\ \bibinfo
  {author} {\bibfnamefont {E.~G.}\ \bibnamefont {Dalla~Torre}},\ }\bibfield
  {title} {\bibinfo {title} {{Introduction to the Dicke Model: From Equilibrium
  to Nonequilibrium, and Vice Versa}},\ }\href
  {https://doi.org/https://doi.org/10.1002/qute.201800043} {\bibfield
  {journal} {\bibinfo  {journal} {Adv. Quantum Technol.}\ }\textbf {\bibinfo
  {volume} {2}},\ \bibinfo {pages} {1800043} (\bibinfo {year}
  {2019})}\BibitemShut {NoStop}%
\bibitem [{\citenamefont {Zhiqiang}\ \emph {et~al.}(2017)\citenamefont
  {Zhiqiang}, \citenamefont {Lee}, \citenamefont {Kumar}, \citenamefont
  {Arnold}, \citenamefont {Masson}, \citenamefont {Parkins},\ and\
  \citenamefont {Barrett}}]{zhiqiang2017}%
  \BibitemOpen
  \bibfield  {author} {\bibinfo {author} {\bibfnamefont {Z.}~\bibnamefont
  {Zhiqiang}}, \bibinfo {author} {\bibfnamefont {C.~H.}\ \bibnamefont {Lee}},
  \bibinfo {author} {\bibfnamefont {R.}~\bibnamefont {Kumar}}, \bibinfo
  {author} {\bibfnamefont {K.~J.}\ \bibnamefont {Arnold}}, \bibinfo {author}
  {\bibfnamefont {S.~J.}\ \bibnamefont {Masson}}, \bibinfo {author}
  {\bibfnamefont {A.~S.}\ \bibnamefont {Parkins}},\ and\ \bibinfo {author}
  {\bibfnamefont {M.~D.}\ \bibnamefont {Barrett}},\ }\bibfield  {title}
  {\bibinfo {title} {{Nonequilibrium phase transition in a spin-1 Dicke
  model}},\ }\href {https://doi.org/10.1364/OPTICA.4.000424} {\bibfield
  {journal} {\bibinfo  {journal} {Optica}\ }\textbf {\bibinfo {volume} {4}},\
  \bibinfo {pages} {424} (\bibinfo {year} {2017})}\BibitemShut {NoStop}%
\bibitem [{\citenamefont {Zhiqiang}\ \emph {et~al.}(2018)\citenamefont
  {Zhiqiang}, \citenamefont {Lee}, \citenamefont {Kumar}, \citenamefont
  {Arnold}, \citenamefont {Masson}, \citenamefont {Grimsmo}, \citenamefont
  {Parkins},\ and\ \citenamefont {Barrett}}]{zhiqiang2018}%
  \BibitemOpen
  \bibfield  {author} {\bibinfo {author} {\bibfnamefont {Z.}~\bibnamefont
  {Zhiqiang}}, \bibinfo {author} {\bibfnamefont {C.~H.}\ \bibnamefont {Lee}},
  \bibinfo {author} {\bibfnamefont {R.}~\bibnamefont {Kumar}}, \bibinfo
  {author} {\bibfnamefont {K.~J.}\ \bibnamefont {Arnold}}, \bibinfo {author}
  {\bibfnamefont {S.~J.}\ \bibnamefont {Masson}}, \bibinfo {author}
  {\bibfnamefont {A.~L.}\ \bibnamefont {Grimsmo}}, \bibinfo {author}
  {\bibfnamefont {A.~S.}\ \bibnamefont {Parkins}},\ and\ \bibinfo {author}
  {\bibfnamefont {M.~D.}\ \bibnamefont {Barrett}},\ }\bibfield  {title}
  {\bibinfo {title} {{Dicke-model simulation via cavity-assisted Raman
  transitions}},\ }\href {https://doi.org/10.1103/PhysRevA.97.043858}
  {\bibfield  {journal} {\bibinfo  {journal} {Phys. Rev. A}\ }\textbf {\bibinfo
  {volume} {97}},\ \bibinfo {pages} {043858} (\bibinfo {year}
  {2018})}\BibitemShut {NoStop}%
\bibitem [{\citenamefont {Stitely}\ \emph {et~al.}(2020)\citenamefont
  {Stitely}, \citenamefont {Masson}, \citenamefont {Giraldo}, \citenamefont
  {Krauskopf},\ and\ \citenamefont {Parkins}}]{stitely2020}%
  \BibitemOpen
  \bibfield  {author} {\bibinfo {author} {\bibfnamefont {K.~C.}\ \bibnamefont
  {Stitely}}, \bibinfo {author} {\bibfnamefont {S.~J.}\ \bibnamefont {Masson}},
  \bibinfo {author} {\bibfnamefont {A.}~\bibnamefont {Giraldo}}, \bibinfo
  {author} {\bibfnamefont {B.}~\bibnamefont {Krauskopf}},\ and\ \bibinfo
  {author} {\bibfnamefont {S.}~\bibnamefont {Parkins}},\ }\bibfield  {title}
  {\bibinfo {title} {{Superradiant switching, quantum hysteresis, and
  oscillations in a generalized Dicke model}},\ }\href
  {https://doi.org/10.1103/PhysRevA.102.063702} {\bibfield  {journal} {\bibinfo
   {journal} {Phys. Rev. A}\ }\textbf {\bibinfo {volume} {102}},\ \bibinfo
  {pages} {063702} (\bibinfo {year} {2020})}\BibitemShut {NoStop}%
\bibitem [{\citenamefont {Shchadilova}\ \emph {et~al.}(2020)\citenamefont
  {Shchadilova}, \citenamefont {Roses}, \citenamefont {Dalla~Torre},
  \citenamefont {Lukin},\ and\ \citenamefont {Demler}}]{shchadilova2020}%
  \BibitemOpen
  \bibfield  {author} {\bibinfo {author} {\bibfnamefont {Y.}~\bibnamefont
  {Shchadilova}}, \bibinfo {author} {\bibfnamefont {M.~M.}\ \bibnamefont
  {Roses}}, \bibinfo {author} {\bibfnamefont {E.~G.}\ \bibnamefont
  {Dalla~Torre}}, \bibinfo {author} {\bibfnamefont {M.~D.}\ \bibnamefont
  {Lukin}},\ and\ \bibinfo {author} {\bibfnamefont {E.}~\bibnamefont
  {Demler}},\ }\bibfield  {title} {\bibinfo {title} {{Fermionic formalism for
  driven-dissipative multilevel systems}},\ }\href
  {https://doi.org/10.1103/PhysRevA.101.013817} {\bibfield  {journal} {\bibinfo
   {journal} {Phys. Rev. A}\ }\textbf {\bibinfo {volume} {101}},\ \bibinfo
  {pages} {013817} (\bibinfo {year} {2020})}\BibitemShut {NoStop}%
\bibitem [{\citenamefont {Stitely}\ \emph {et~al.}(2022)\citenamefont
  {Stitely}, \citenamefont {Giraldo}, \citenamefont {Krauskopf},\ and\
  \citenamefont {Parkins}}]{stitely2022}%
  \BibitemOpen
  \bibfield  {author} {\bibinfo {author} {\bibfnamefont {K.~C.}\ \bibnamefont
  {Stitely}}, \bibinfo {author} {\bibfnamefont {A.}~\bibnamefont {Giraldo}},
  \bibinfo {author} {\bibfnamefont {B.}~\bibnamefont {Krauskopf}},\ and\
  \bibinfo {author} {\bibfnamefont {S.}~\bibnamefont {Parkins}},\ }\bibfield
  {title} {\bibinfo {title} {{Lasing and counter-lasing phase transitions in a
  cavity-QED system}},\ }\href
  {https://doi.org/10.1103/PhysRevResearch.4.023101} {\bibfield  {journal}
  {\bibinfo  {journal} {Phys. Rev. Res.}\ }\textbf {\bibinfo {volume} {4}},\
  \bibinfo {pages} {023101} (\bibinfo {year} {2022})}\BibitemShut {NoStop}%
\bibitem [{\citenamefont {Iemini}\ \emph {et~al.}(2018)\citenamefont {Iemini},
  \citenamefont {Russomanno}, \citenamefont {Keeling}, \citenamefont
  {Schir{\`o}}, \citenamefont {Dalmonte},\ and\ \citenamefont
  {Fazio}}]{iemini2018}%
  \BibitemOpen
  \bibfield  {author} {\bibinfo {author} {\bibfnamefont {F.}~\bibnamefont
  {Iemini}}, \bibinfo {author} {\bibfnamefont {A.}~\bibnamefont {Russomanno}},
  \bibinfo {author} {\bibfnamefont {J.}~\bibnamefont {Keeling}}, \bibinfo
  {author} {\bibfnamefont {M.}~\bibnamefont {Schir{\`o}}}, \bibinfo {author}
  {\bibfnamefont {M.}~\bibnamefont {Dalmonte}},\ and\ \bibinfo {author}
  {\bibfnamefont {R.}~\bibnamefont {Fazio}},\ }\bibfield  {title} {\bibinfo
  {title} {{Boundary time crystals}},\ }\href
  {https://doi.org/https://doi.org/10.1103/PhysRevLett.121.035301} {\bibfield
  {journal} {\bibinfo  {journal} {Phys. Rev. Lett.}\ }\textbf {\bibinfo
  {volume} {121}},\ \bibinfo {pages} {035301} (\bibinfo {year}
  {2018})}\BibitemShut {NoStop}%
\bibitem [{\citenamefont {Bu{\v c}a}\ \emph {et~al.}(2019)\citenamefont {Bu{\v
  c}a}, \citenamefont {Tindall},\ and\ \citenamefont {Jaksch}}]{buca2019}%
  \BibitemOpen
  \bibfield  {author} {\bibinfo {author} {\bibfnamefont {B.}~\bibnamefont
  {Bu{\v c}a}}, \bibinfo {author} {\bibfnamefont {J.}~\bibnamefont {Tindall}},\
  and\ \bibinfo {author} {\bibfnamefont {D.}~\bibnamefont {Jaksch}},\
  }\bibfield  {title} {\bibinfo {title} {{Non-stationary coherent quantum
  many-body dynamics through dissipation}},\ }\href
  {https://doi.org/10.1038/s41467-019-09757-y} {\bibfield  {journal} {\bibinfo
  {journal} {Nat. Commun.}\ }\textbf {\bibinfo {volume} {10}},\ \bibinfo
  {pages} {1730} (\bibinfo {year} {2019})}\BibitemShut {NoStop}%
\bibitem [{\citenamefont {Bu\ifmmode~\check{c}\else \v{c}\fi{}a}\ and\
  \citenamefont {Jaksch}(2019)}]{buca2019b}%
  \BibitemOpen
  \bibfield  {author} {\bibinfo {author} {\bibfnamefont {B.}~\bibnamefont
  {Bu\ifmmode~\check{c}\else \v{c}\fi{}a}}\ and\ \bibinfo {author}
  {\bibfnamefont {D.}~\bibnamefont {Jaksch}},\ }\bibfield  {title} {\bibinfo
  {title} {{Dissipation Induced Nonstationarity in a Quantum Gas}},\ }\href
  {https://doi.org/10.1103/PhysRevLett.123.260401} {\bibfield  {journal}
  {\bibinfo  {journal} {Phys. Rev. Lett.}\ }\textbf {\bibinfo {volume} {123}},\
  \bibinfo {pages} {260401} (\bibinfo {year} {2019})}\BibitemShut {NoStop}%
\bibitem [{\citenamefont {Chiacchio}\ and\ \citenamefont
  {Nunnenkamp}(2019)}]{chiacchio2019}%
  \BibitemOpen
  \bibfield  {author} {\bibinfo {author} {\bibfnamefont {E.~I.~R.}\
  \bibnamefont {Chiacchio}}\ and\ \bibinfo {author} {\bibfnamefont
  {A.}~\bibnamefont {Nunnenkamp}},\ }\bibfield  {title} {\bibinfo {title}
  {{Dissipation-Induced Instabilities of a Spinor Bose-Einstein Condensate
  Inside an Optical Cavity}},\ }\href
  {https://doi.org/10.1103/PhysRevLett.122.193605} {\bibfield  {journal}
  {\bibinfo  {journal} {Phys. Rev. Lett.}\ }\textbf {\bibinfo {volume} {122}},\
  \bibinfo {pages} {193605} (\bibinfo {year} {2019})}\BibitemShut {NoStop}%
\bibitem [{\citenamefont {Booker}\ \emph {et~al.}(2020)\citenamefont {Booker},
  \citenamefont {Buča},\ and\ \citenamefont {Jaksch}}]{booker2020}%
  \BibitemOpen
  \bibfield  {author} {\bibinfo {author} {\bibfnamefont {C.}~\bibnamefont
  {Booker}}, \bibinfo {author} {\bibfnamefont {B.}~\bibnamefont {Buča}},\ and\
  \bibinfo {author} {\bibfnamefont {D.}~\bibnamefont {Jaksch}},\ }\bibfield
  {title} {\bibinfo {title} {{Non-stationarity and dissipative time crystals:
  spectral properties and finite-size effects}},\ }\href
  {https://doi.org/10.1088/1367-2630/ababc4} {\bibfield  {journal} {\bibinfo
  {journal} {New J. Phys.}\ }\textbf {\bibinfo {volume} {22}},\ \bibinfo
  {pages} {085007} (\bibinfo {year} {2020})}\BibitemShut {NoStop}%
\bibitem [{\citenamefont {Wilczek}(2012)}]{wilczek2012}%
  \BibitemOpen
  \bibfield  {author} {\bibinfo {author} {\bibfnamefont {F.}~\bibnamefont
  {Wilczek}},\ }\bibfield  {title} {\bibinfo {title} {{Quantum time
  crystals}},\ }\href
  {https://doi.org/https://doi.org/10.1103/PhysRevLett.109.160401} {\bibfield
  {journal} {\bibinfo  {journal} {Phys. Rev. Lett.}\ }\textbf {\bibinfo
  {volume} {109}},\ \bibinfo {pages} {160401} (\bibinfo {year}
  {2012})}\BibitemShut {NoStop}%
\bibitem [{\citenamefont {Shapere}\ and\ \citenamefont
  {Wilczek}(2012)}]{shapere2012}%
  \BibitemOpen
  \bibfield  {author} {\bibinfo {author} {\bibfnamefont {A.}~\bibnamefont
  {Shapere}}\ and\ \bibinfo {author} {\bibfnamefont {F.}~\bibnamefont
  {Wilczek}},\ }\bibfield  {title} {\bibinfo {title} {{Classical Time
  Crystals}},\ }\href {https://doi.org/10.1103/PhysRevLett.109.160402}
  {\bibfield  {journal} {\bibinfo  {journal} {Phys. Rev. Lett.}\ }\textbf
  {\bibinfo {volume} {109}},\ \bibinfo {pages} {160402} (\bibinfo {year}
  {2012})}\BibitemShut {NoStop}%
\bibitem [{\citenamefont {Bruno}(2013)}]{bruno2013}%
  \BibitemOpen
  \bibfield  {author} {\bibinfo {author} {\bibfnamefont {P.}~\bibnamefont
  {Bruno}},\ }\bibfield  {title} {\bibinfo {title} {{Impossibility of
  Spontaneously Rotating Time Crystals: A No-Go Theorem}},\ }\href
  {https://doi.org/10.1103/PhysRevLett.111.070402} {\bibfield  {journal}
  {\bibinfo  {journal} {Phys. Rev. Lett.}\ }\textbf {\bibinfo {volume} {111}},\
  \bibinfo {pages} {070402} (\bibinfo {year} {2013})}\BibitemShut {NoStop}%
\bibitem [{\citenamefont {Watanabe}\ and\ \citenamefont
  {Oshikawa}(2015)}]{watanabe2015}%
  \BibitemOpen
  \bibfield  {author} {\bibinfo {author} {\bibfnamefont {H.}~\bibnamefont
  {Watanabe}}\ and\ \bibinfo {author} {\bibfnamefont {M.}~\bibnamefont
  {Oshikawa}},\ }\bibfield  {title} {\bibinfo {title} {{Absence of Quantum Time
  Crystals}},\ }\href {https://doi.org/10.1103/PhysRevLett.114.251603}
  {\bibfield  {journal} {\bibinfo  {journal} {Phys. Rev. Lett.}\ }\textbf
  {\bibinfo {volume} {114}},\ \bibinfo {pages} {251603} (\bibinfo {year}
  {2015})}\BibitemShut {NoStop}%
\bibitem [{\citenamefont {Sacha}(2015)}]{sacha2015}%
  \BibitemOpen
  \bibfield  {author} {\bibinfo {author} {\bibfnamefont {K.}~\bibnamefont
  {Sacha}},\ }\bibfield  {title} {\bibinfo {title} {{Modeling spontaneous
  breaking of time-translation symmetry}},\ }\href
  {https://doi.org/10.1103/PhysRevA.91.033617} {\bibfield  {journal} {\bibinfo
  {journal} {Phys. Rev. A}\ }\textbf {\bibinfo {volume} {91}},\ \bibinfo
  {pages} {033617} (\bibinfo {year} {2015})}\BibitemShut {NoStop}%
\bibitem [{\citenamefont {Khemani}\ \emph {et~al.}(2016)\citenamefont
  {Khemani}, \citenamefont {Lazarides}, \citenamefont {Moessner},\ and\
  \citenamefont {Sondhi}}]{khemani2016}%
  \BibitemOpen
  \bibfield  {author} {\bibinfo {author} {\bibfnamefont {V.}~\bibnamefont
  {Khemani}}, \bibinfo {author} {\bibfnamefont {A.}~\bibnamefont {Lazarides}},
  \bibinfo {author} {\bibfnamefont {R.}~\bibnamefont {Moessner}},\ and\
  \bibinfo {author} {\bibfnamefont {S.~L.}\ \bibnamefont {Sondhi}},\ }\bibfield
   {title} {\bibinfo {title} {{Phase Structure of Driven Quantum Systems}},\
  }\href {https://doi.org/10.1103/PhysRevLett.116.250401} {\bibfield  {journal}
  {\bibinfo  {journal} {Phys. Rev. Lett.}\ }\textbf {\bibinfo {volume} {116}},\
  \bibinfo {pages} {250401} (\bibinfo {year} {2016})}\BibitemShut {NoStop}%
\bibitem [{\citenamefont {Else}\ \emph {et~al.}(2016)\citenamefont {Else},
  \citenamefont {Bauer},\ and\ \citenamefont {Nayak}}]{else2016}%
  \BibitemOpen
  \bibfield  {author} {\bibinfo {author} {\bibfnamefont {D.~V.}\ \bibnamefont
  {Else}}, \bibinfo {author} {\bibfnamefont {B.}~\bibnamefont {Bauer}},\ and\
  \bibinfo {author} {\bibfnamefont {C.}~\bibnamefont {Nayak}},\ }\bibfield
  {title} {\bibinfo {title} {{Floquet time crystals}},\ }\href@noop {}
  {\bibfield  {journal} {\bibinfo  {journal} {Phys. Rev. Lett.}\ }\textbf
  {\bibinfo {volume} {117}},\ \bibinfo {pages} {090402} (\bibinfo {year}
  {2016})}\BibitemShut {NoStop}%
\bibitem [{\citenamefont {Yao}\ \emph {et~al.}(2017)\citenamefont {Yao},
  \citenamefont {Potter}, \citenamefont {Potirniche},\ and\ \citenamefont
  {Vishwanath}}]{yao2017}%
  \BibitemOpen
  \bibfield  {author} {\bibinfo {author} {\bibfnamefont {N.~Y.}\ \bibnamefont
  {Yao}}, \bibinfo {author} {\bibfnamefont {A.~C.}\ \bibnamefont {Potter}},
  \bibinfo {author} {\bibfnamefont {I.-D.}\ \bibnamefont {Potirniche}},\ and\
  \bibinfo {author} {\bibfnamefont {A.}~\bibnamefont {Vishwanath}},\ }\bibfield
   {title} {\bibinfo {title} {{Discrete time crystals: Rigidity, criticality,
  and realizations}},\ }\href
  {https://doi.org/https://doi.org/10.1103/PhysRevLett.118.030401} {\bibfield
  {journal} {\bibinfo  {journal} {Phys. Rev. Lett.}\ }\textbf {\bibinfo
  {volume} {118}},\ \bibinfo {pages} {030401} (\bibinfo {year}
  {2017})}\BibitemShut {NoStop}%
\bibitem [{\citenamefont {Sacha}\ and\ \citenamefont
  {Zakrzewski}(2017)}]{sacha2018}%
  \BibitemOpen
  \bibfield  {author} {\bibinfo {author} {\bibfnamefont {K.}~\bibnamefont
  {Sacha}}\ and\ \bibinfo {author} {\bibfnamefont {J.}~\bibnamefont
  {Zakrzewski}},\ }\bibfield  {title} {\bibinfo {title} {{Time crystals: a
  review}},\ }\href {https://doi.org/10.1088/1361-6633/aa8b38} {\bibfield
  {journal} {\bibinfo  {journal} {Rep. Prog. Phys.}\ }\textbf {\bibinfo
  {volume} {81}},\ \bibinfo {pages} {016401} (\bibinfo {year}
  {2017})}\BibitemShut {NoStop}%
\bibitem [{\citenamefont {Gambetta}\ \emph {et~al.}(2019)\citenamefont
  {Gambetta}, \citenamefont {Carollo}, \citenamefont {Marcuzzi}, \citenamefont
  {Garrahan},\ and\ \citenamefont {Lesanovsky}}]{gambetta2019}%
  \BibitemOpen
  \bibfield  {author} {\bibinfo {author} {\bibfnamefont {F.}~\bibnamefont
  {Gambetta}}, \bibinfo {author} {\bibfnamefont {F.}~\bibnamefont {Carollo}},
  \bibinfo {author} {\bibfnamefont {M.}~\bibnamefont {Marcuzzi}}, \bibinfo
  {author} {\bibfnamefont {J.}~\bibnamefont {Garrahan}},\ and\ \bibinfo
  {author} {\bibfnamefont {I.}~\bibnamefont {Lesanovsky}},\ }\bibfield  {title}
  {\bibinfo {title} {{Discrete time crystals in the absence of manifest
  symmetries or disorder in open quantum systems}},\ }\href
  {https://doi.org/https://doi.org/10.1103/PhysRevLett.122.015701} {\bibfield
  {journal} {\bibinfo  {journal} {Phys. Rev. Lett.}\ }\textbf {\bibinfo
  {volume} {122}},\ \bibinfo {pages} {015701} (\bibinfo {year}
  {2019})}\BibitemShut {NoStop}%
\bibitem [{\citenamefont {Else}\ \emph {et~al.}(2020)\citenamefont {Else},
  \citenamefont {Monroe}, \citenamefont {Nayak},\ and\ \citenamefont
  {Yao}}]{else2020}%
  \BibitemOpen
  \bibfield  {author} {\bibinfo {author} {\bibfnamefont {D.~V.}\ \bibnamefont
  {Else}}, \bibinfo {author} {\bibfnamefont {C.}~\bibnamefont {Monroe}},
  \bibinfo {author} {\bibfnamefont {C.}~\bibnamefont {Nayak}},\ and\ \bibinfo
  {author} {\bibfnamefont {N.~Y.}\ \bibnamefont {Yao}},\ }\bibfield  {title}
  {\bibinfo {title} {{Discrete time crystals}},\ }\href
  {https://doi.org/https://doi.org/10.1146/annurev-conmatphys-031119-050658}
  {\bibfield  {journal} {\bibinfo  {journal} {Annu. Rev. Condens. Matter
  Phys.}\ }\textbf {\bibinfo {volume} {11}},\ \bibinfo {pages} {467} (\bibinfo
  {year} {2020})}\BibitemShut {NoStop}%
\bibitem [{\citenamefont {Taheri}\ \emph {et~al.}(2022)\citenamefont {Taheri},
  \citenamefont {Matsko}, \citenamefont {Maleki},\ and\ \citenamefont
  {Sacha}}]{taheri2022}%
  \BibitemOpen
  \bibfield  {author} {\bibinfo {author} {\bibfnamefont {H.}~\bibnamefont
  {Taheri}}, \bibinfo {author} {\bibfnamefont {A.~B.}\ \bibnamefont {Matsko}},
  \bibinfo {author} {\bibfnamefont {L.}~\bibnamefont {Maleki}},\ and\ \bibinfo
  {author} {\bibfnamefont {K.}~\bibnamefont {Sacha}},\ }\bibfield  {title}
  {\bibinfo {title} {{All-optical dissipative discrete time crystals}},\ }\href
  {https://doi.org/10.1038/s41467-022-28462-x} {\bibfield  {journal} {\bibinfo
  {journal} {Nat. Commun.}\ }\textbf {\bibinfo {volume} {13}},\ \bibinfo
  {pages} {848} (\bibinfo {year} {2022})}\BibitemShut {NoStop}%
\bibitem [{\citenamefont {Kongkhambut}\ \emph {et~al.}(2022)\citenamefont
  {Kongkhambut}, \citenamefont {Skulte}, \citenamefont {Mathey}, \citenamefont
  {Cosme}, \citenamefont {Hemmerich},\ and\ \citenamefont
  {Keßler}}]{kongkhambut2022}%
  \BibitemOpen
  \bibfield  {author} {\bibinfo {author} {\bibfnamefont {P.}~\bibnamefont
  {Kongkhambut}}, \bibinfo {author} {\bibfnamefont {J.}~\bibnamefont {Skulte}},
  \bibinfo {author} {\bibfnamefont {L.}~\bibnamefont {Mathey}}, \bibinfo
  {author} {\bibfnamefont {J.~G.}\ \bibnamefont {Cosme}}, \bibinfo {author}
  {\bibfnamefont {A.}~\bibnamefont {Hemmerich}},\ and\ \bibinfo {author}
  {\bibfnamefont {H.}~\bibnamefont {Keßler}},\ }\bibfield  {title} {\bibinfo
  {title} {Observation of a continuous time crystal},\ }\href
  {https://doi.org/10.1126/science.abo3382} {\bibfield  {journal} {\bibinfo
  {journal} {Science}\ }\textbf {\bibinfo {volume} {377}},\ \bibinfo {pages}
  {670} (\bibinfo {year} {2022})}\BibitemShut {NoStop}%
\bibitem [{\citenamefont {Nie}\ and\ \citenamefont {Zheng}(2023)}]{nie2023}%
  \BibitemOpen
  \bibfield  {author} {\bibinfo {author} {\bibfnamefont {X.}~\bibnamefont
  {Nie}}\ and\ \bibinfo {author} {\bibfnamefont {W.}~\bibnamefont {Zheng}},\
  }\bibfield  {title} {\bibinfo {title} {Mode softening in time-crystalline
  transitions of open quantum systems},\ }\href
  {https://doi.org/10.1103/PhysRevA.107.033311} {\bibfield  {journal} {\bibinfo
   {journal} {Phys. Rev. A}\ }\textbf {\bibinfo {volume} {107}},\ \bibinfo
  {pages} {033311} (\bibinfo {year} {2023})}\BibitemShut {NoStop}%
\bibitem [{\citenamefont {S\'anchez Mu\~noz}\ \emph {et~al.}(2019)\citenamefont
  {S\'anchez Mu\~noz}, \citenamefont {Bu\ifmmode~\check{c}\else \v{c}\fi{}a},
  \citenamefont {Tindall}, \citenamefont {Gonz\'alez-Tudela}, \citenamefont
  {Jaksch},\ and\ \citenamefont {Porras}}]{sanchezmunoz2019}%
  \BibitemOpen
  \bibfield  {author} {\bibinfo {author} {\bibfnamefont {C.}~\bibnamefont
  {S\'anchez Mu\~noz}}, \bibinfo {author} {\bibfnamefont {B.}~\bibnamefont
  {Bu\ifmmode~\check{c}\else \v{c}\fi{}a}}, \bibinfo {author} {\bibfnamefont
  {J.}~\bibnamefont {Tindall}}, \bibinfo {author} {\bibfnamefont
  {A.}~\bibnamefont {Gonz\'alez-Tudela}}, \bibinfo {author} {\bibfnamefont
  {D.}~\bibnamefont {Jaksch}},\ and\ \bibinfo {author} {\bibfnamefont
  {D.}~\bibnamefont {Porras}},\ }\bibfield  {title} {\bibinfo {title}
  {{Symmetries and conservation laws in quantum trajectories: Dissipative
  freezing}},\ }\href {https://doi.org/10.1103/PhysRevA.100.042113} {\bibfield
  {journal} {\bibinfo  {journal} {Phys. Rev. A}\ }\textbf {\bibinfo {volume}
  {100}},\ \bibinfo {pages} {042113} (\bibinfo {year} {2019})}\BibitemShut
  {NoStop}%
\bibitem [{\citenamefont {Piccitto}\ \emph {et~al.}(2021)\citenamefont
  {Piccitto}, \citenamefont {Wauters}, \citenamefont {Nori},\ and\
  \citenamefont {Shammah}}]{piccitto2021}%
  \BibitemOpen
  \bibfield  {author} {\bibinfo {author} {\bibfnamefont {G.}~\bibnamefont
  {Piccitto}}, \bibinfo {author} {\bibfnamefont {M.}~\bibnamefont {Wauters}},
  \bibinfo {author} {\bibfnamefont {F.}~\bibnamefont {Nori}},\ and\ \bibinfo
  {author} {\bibfnamefont {N.}~\bibnamefont {Shammah}},\ }\bibfield  {title}
  {\bibinfo {title} {{Symmetries and conserved quantities of boundary time
  crystals in generalized spin models}},\ }\href
  {https://doi.org/10.1103/PhysRevB.104.014307} {\bibfield  {journal} {\bibinfo
   {journal} {Phys. Rev. B}\ }\textbf {\bibinfo {volume} {104}},\ \bibinfo
  {pages} {014307} (\bibinfo {year} {2021})}\BibitemShut {NoStop}%
\bibitem [{\citenamefont {Prazeres}\ \emph {et~al.}(2021)\citenamefont
  {Prazeres}, \citenamefont {Souza},\ and\ \citenamefont
  {Iemini}}]{prazeres2021}%
  \BibitemOpen
  \bibfield  {author} {\bibinfo {author} {\bibfnamefont {L.~F.~d.}\
  \bibnamefont {Prazeres}}, \bibinfo {author} {\bibfnamefont {L.~d.~S.}\
  \bibnamefont {Souza}},\ and\ \bibinfo {author} {\bibfnamefont
  {F.}~\bibnamefont {Iemini}},\ }\bibfield  {title} {\bibinfo {title}
  {{Boundary time crystals in collective $d$-level systems}},\ }\href
  {https://doi.org/10.1103/PhysRevB.103.184308} {\bibfield  {journal} {\bibinfo
   {journal} {Phys. Rev. B}\ }\textbf {\bibinfo {volume} {103}},\ \bibinfo
  {pages} {184308} (\bibinfo {year} {2021})}\BibitemShut {NoStop}%
\bibitem [{\citenamefont {Passarelli}\ \emph {et~al.}(2022)\citenamefont
  {Passarelli}, \citenamefont {Lucignano}, \citenamefont {Fazio},\ and\
  \citenamefont {Russomanno}}]{passarelli2022}%
  \BibitemOpen
  \bibfield  {author} {\bibinfo {author} {\bibfnamefont {G.}~\bibnamefont
  {Passarelli}}, \bibinfo {author} {\bibfnamefont {P.}~\bibnamefont
  {Lucignano}}, \bibinfo {author} {\bibfnamefont {R.}~\bibnamefont {Fazio}},\
  and\ \bibinfo {author} {\bibfnamefont {A.}~\bibnamefont {Russomanno}},\
  }\bibfield  {title} {\bibinfo {title} {Dissipative time crystals with
  long-range lindbladians},\ }\href
  {https://doi.org/10.1103/PhysRevB.106.224308} {\bibfield  {journal} {\bibinfo
   {journal} {Phys. Rev. B}\ }\textbf {\bibinfo {volume} {106}},\ \bibinfo
  {pages} {224308} (\bibinfo {year} {2022})}\BibitemShut {NoStop}%
\bibitem [{\citenamefont {Ferioli}\ \emph {et~al.}(2022)\citenamefont
  {Ferioli}, \citenamefont {Glicenstein}, \citenamefont {Ferrier-Barbut},\ and\
  \citenamefont {Browaeys}}]{ferioli2022}%
  \BibitemOpen
  \bibfield  {author} {\bibinfo {author} {\bibfnamefont {G.}~\bibnamefont
  {Ferioli}}, \bibinfo {author} {\bibfnamefont {A.}~\bibnamefont
  {Glicenstein}}, \bibinfo {author} {\bibfnamefont {I.}~\bibnamefont
  {Ferrier-Barbut}},\ and\ \bibinfo {author} {\bibfnamefont {A.}~\bibnamefont
  {Browaeys}},\ }\bibfield  {title} {\bibinfo {title} {{Observation of a
  non-equilibrium superradiant phase transition in free space}},\ }\href
  {https://doi.org/10.48550/ARXIV.2207.10361} {\bibfield  {journal} {\bibinfo
  {journal} {arXiv:2207.10361}\ } (\bibinfo {year} {2022})}\BibitemShut
  {NoStop}%
\bibitem [{\citenamefont {Chase}\ and\ \citenamefont
  {Geremia}(2008)}]{chase2008}%
  \BibitemOpen
  \bibfield  {author} {\bibinfo {author} {\bibfnamefont {B.~A.}\ \bibnamefont
  {Chase}}\ and\ \bibinfo {author} {\bibfnamefont {J.~M.}\ \bibnamefont
  {Geremia}},\ }\bibfield  {title} {\bibinfo {title} {{Collective processes of
  an ensemble of spin-1/2 particles}},\ }\href
  {https://doi.org/10.1103/PhysRevA.78.052101} {\bibfield  {journal} {\bibinfo
  {journal} {Phys. Rev. A}\ }\textbf {\bibinfo {volume} {78}},\ \bibinfo
  {pages} {052101} (\bibinfo {year} {2008})}\BibitemShut {NoStop}%
\bibitem [{\citenamefont {Baragiola}\ \emph {et~al.}(2010)\citenamefont
  {Baragiola}, \citenamefont {Chase},\ and\ \citenamefont
  {Geremia}}]{baragiola2010}%
  \BibitemOpen
  \bibfield  {author} {\bibinfo {author} {\bibfnamefont {B.~Q.}\ \bibnamefont
  {Baragiola}}, \bibinfo {author} {\bibfnamefont {B.~A.}\ \bibnamefont
  {Chase}},\ and\ \bibinfo {author} {\bibfnamefont {J.}~\bibnamefont
  {Geremia}},\ }\bibfield  {title} {\bibinfo {title} {{Collective uncertainty
  in partially polarized and partially decohered spin-1/2 systems}},\ }\href
  {https://doi.org/10.1103/PhysRevA.81.032104} {\bibfield  {journal} {\bibinfo
  {journal} {Phys. Rev. A}\ }\textbf {\bibinfo {volume} {81}},\ \bibinfo
  {pages} {032104} (\bibinfo {year} {2010})}\BibitemShut {NoStop}%
\bibitem [{\citenamefont {Kirton}\ and\ \citenamefont
  {Keeling}(2017)}]{kirton2017}%
  \BibitemOpen
  \bibfield  {author} {\bibinfo {author} {\bibfnamefont {P.}~\bibnamefont
  {Kirton}}\ and\ \bibinfo {author} {\bibfnamefont {J.}~\bibnamefont
  {Keeling}},\ }\bibfield  {title} {\bibinfo {title} {{Suppressing and
  Restoring the Dicke Superradiance Transition by Dephasing and Decay}},\
  }\href {https://doi.org/10.1103/PhysRevLett.118.123602} {\bibfield  {journal}
  {\bibinfo  {journal} {Phys. Rev. Lett.}\ }\textbf {\bibinfo {volume} {118}},\
  \bibinfo {pages} {123602} (\bibinfo {year} {2017})}\BibitemShut {NoStop}%
\bibitem [{\citenamefont {Kirton}\ and\ \citenamefont
  {Keeling}(2018)}]{kirton2018}%
  \BibitemOpen
  \bibfield  {author} {\bibinfo {author} {\bibfnamefont {P.}~\bibnamefont
  {Kirton}}\ and\ \bibinfo {author} {\bibfnamefont {J.}~\bibnamefont
  {Keeling}},\ }\bibfield  {title} {\bibinfo {title} {{Superradiant and lasing
  states in driven-dissipative Dicke models}},\ }\href
  {https://doi.org/10.1088/1367-2630/aaa11d} {\bibfield  {journal} {\bibinfo
  {journal} {New J. Phys.}\ }\textbf {\bibinfo {volume} {20}},\ \bibinfo
  {pages} {015009} (\bibinfo {year} {2018})}\BibitemShut {NoStop}%
\bibitem [{\citenamefont {Shammah}\ \emph {et~al.}(2018)\citenamefont
  {Shammah}, \citenamefont {Ahmed}, \citenamefont {Lambert}, \citenamefont
  {De~Liberato},\ and\ \citenamefont {Nori}}]{shammah2018}%
  \BibitemOpen
  \bibfield  {author} {\bibinfo {author} {\bibfnamefont {N.}~\bibnamefont
  {Shammah}}, \bibinfo {author} {\bibfnamefont {S.}~\bibnamefont {Ahmed}},
  \bibinfo {author} {\bibfnamefont {N.}~\bibnamefont {Lambert}}, \bibinfo
  {author} {\bibfnamefont {S.}~\bibnamefont {De~Liberato}},\ and\ \bibinfo
  {author} {\bibfnamefont {F.}~\bibnamefont {Nori}},\ }\bibfield  {title}
  {\bibinfo {title} {{Open quantum systems with local and collective incoherent
  processes: Efficient numerical simulations using permutational invariance}},\
  }\href {https://doi.org/10.1103/PhysRevA.98.063815} {\bibfield  {journal}
  {\bibinfo  {journal} {Phys. Rev. A}\ }\textbf {\bibinfo {volume} {98}},\
  \bibinfo {pages} {063815} (\bibinfo {year} {2018})}\BibitemShut {NoStop}%
\bibitem [{\citenamefont {Huybrechts}\ \emph {et~al.}(2020)\citenamefont
  {Huybrechts}, \citenamefont {Minganti}, \citenamefont {Nori}, \citenamefont
  {Wouters},\ and\ \citenamefont {Shammah}}]{huybrechts2020}%
  \BibitemOpen
  \bibfield  {author} {\bibinfo {author} {\bibfnamefont {D.}~\bibnamefont
  {Huybrechts}}, \bibinfo {author} {\bibfnamefont {F.}~\bibnamefont
  {Minganti}}, \bibinfo {author} {\bibfnamefont {F.}~\bibnamefont {Nori}},
  \bibinfo {author} {\bibfnamefont {M.}~\bibnamefont {Wouters}},\ and\ \bibinfo
  {author} {\bibfnamefont {N.}~\bibnamefont {Shammah}},\ }\bibfield  {title}
  {\bibinfo {title} {{Validity of mean-field theory in a dissipative critical
  system: Liouvillian gap, $\mathbb{PT}$-symmetric antigap, and permutational
  symmetry in the $\mathit{XYZ}$ model}},\ }\href
  {https://doi.org/10.1103/PhysRevB.101.214302} {\bibfield  {journal} {\bibinfo
   {journal} {Phys. Rev. B}\ }\textbf {\bibinfo {volume} {101}},\ \bibinfo
  {pages} {214302} (\bibinfo {year} {2020})}\BibitemShut {NoStop}%
\bibitem [{\citenamefont {Carmichael}(1980)}]{carmichael1980}%
  \BibitemOpen
  \bibfield  {author} {\bibinfo {author} {\bibfnamefont {H.~J.}\ \bibnamefont
  {Carmichael}},\ }\bibfield  {title} {\bibinfo {title} {{Analytical and
  numerical results for the steady state in cooperative resonance
  fluorescence}},\ }\href {https://doi.org/10.1088/0022-3700/13/18/009}
  {\bibfield  {journal} {\bibinfo  {journal} {J. Phys. B: Atom. Mol. Phys.}\
  }\textbf {\bibinfo {volume} {13}},\ \bibinfo {pages} {3551} (\bibinfo {year}
  {1980})}\BibitemShut {NoStop}%
\bibitem [{\citenamefont {Alicki}\ and\ \citenamefont
  {Messer}(1983)}]{alicki1983}%
  \BibitemOpen
  \bibfield  {author} {\bibinfo {author} {\bibfnamefont {R.}~\bibnamefont
  {Alicki}}\ and\ \bibinfo {author} {\bibfnamefont {J.}~\bibnamefont
  {Messer}},\ }\bibfield  {title} {\bibinfo {title} {{Nonlinear quantum
  dynamical semigroups for many-body open systems}},\ }\href
  {https://doi.org/10.1007/BF01012712} {\bibfield  {journal} {\bibinfo
  {journal} {J. Stat. Phys.}\ }\textbf {\bibinfo {volume} {32}},\ \bibinfo
  {pages} {299} (\bibinfo {year} {1983})}\BibitemShut {NoStop}%
\bibitem [{\citenamefont {Benatti}\ \emph {et~al.}(2018)\citenamefont
  {Benatti}, \citenamefont {Carollo}, \citenamefont {Floreanini},\ and\
  \citenamefont {Narnhofer}}]{benatti2018}%
  \BibitemOpen
  \bibfield  {author} {\bibinfo {author} {\bibfnamefont {F.}~\bibnamefont
  {Benatti}}, \bibinfo {author} {\bibfnamefont {F.}~\bibnamefont {Carollo}},
  \bibinfo {author} {\bibfnamefont {R.}~\bibnamefont {Floreanini}},\ and\
  \bibinfo {author} {\bibfnamefont {H.}~\bibnamefont {Narnhofer}},\ }\bibfield
  {title} {\bibinfo {title} {{Quantum spin chain dissipative mean-field
  dynamics}},\ }\href {https://dx.doi.org/10.1088/1751-8121/aacbdb} {\bibfield
  {journal} {\bibinfo  {journal} {J. Phys. A}\ }\textbf {\bibinfo {volume}
  {51}},\ \bibinfo {pages} {325001} (\bibinfo {year} {2018})}\BibitemShut
  {NoStop}%
\bibitem [{\citenamefont {Buonaiuto}\ \emph {et~al.}(2021)\citenamefont
  {Buonaiuto}, \citenamefont {Carollo}, \citenamefont {Olmos},\ and\
  \citenamefont {Lesanovsky}}]{buonaiuto2021}%
  \BibitemOpen
  \bibfield  {author} {\bibinfo {author} {\bibfnamefont {G.}~\bibnamefont
  {Buonaiuto}}, \bibinfo {author} {\bibfnamefont {F.}~\bibnamefont {Carollo}},
  \bibinfo {author} {\bibfnamefont {B.}~\bibnamefont {Olmos}},\ and\ \bibinfo
  {author} {\bibfnamefont {I.}~\bibnamefont {Lesanovsky}},\ }\bibfield  {title}
  {\bibinfo {title} {{Dynamical Phases and Quantum Correlations in an
  Emitter-Waveguide System with Feedback}},\ }\href
  {https://doi.org/10.1103/PhysRevLett.127.133601} {\bibfield  {journal}
  {\bibinfo  {journal} {Phys. Rev. Lett.}\ }\textbf {\bibinfo {volume} {127}},\
  \bibinfo {pages} {133601} (\bibinfo {year} {2021})}\BibitemShut {NoStop}%
\bibitem [{\citenamefont {Carollo}\ and\ \citenamefont
  {Lesanovsky}(2022)}]{carollo2022}%
  \BibitemOpen
  \bibfield  {author} {\bibinfo {author} {\bibfnamefont {F.}~\bibnamefont
  {Carollo}}\ and\ \bibinfo {author} {\bibfnamefont {I.}~\bibnamefont
  {Lesanovsky}},\ }\bibfield  {title} {\bibinfo {title} {{Exact solution of a
  boundary time-crystal phase transition: time-translation symmetry breaking
  and non-Markovian dynamics of correlations}},\ }\href
  {https://doi.org/https://doi.org/10.1103/PhysRevA.105.L040202} {\bibfield
  {journal} {\bibinfo  {journal} {Phys. Rev. A}\ }\textbf {\bibinfo {volume}
  {105}},\ \bibinfo {pages} {L040202} (\bibinfo {year} {2022})}\BibitemShut
  {NoStop}%
\bibitem [{\citenamefont {Boneberg}\ \emph {et~al.}(2022)\citenamefont
  {Boneberg}, \citenamefont {Lesanovsky},\ and\ \citenamefont
  {Carollo}}]{boneberg2022}%
  \BibitemOpen
  \bibfield  {author} {\bibinfo {author} {\bibfnamefont {M.}~\bibnamefont
  {Boneberg}}, \bibinfo {author} {\bibfnamefont {I.}~\bibnamefont
  {Lesanovsky}},\ and\ \bibinfo {author} {\bibfnamefont {F.}~\bibnamefont
  {Carollo}},\ }\bibfield  {title} {\bibinfo {title} {{Quantum fluctuations and
  correlations in open quantum Dicke models}},\ }\href
  {https://doi.org/10.1103/PhysRevA.106.012212} {\bibfield  {journal} {\bibinfo
   {journal} {Phys. Rev. A}\ }\textbf {\bibinfo {volume} {106}},\ \bibinfo
  {pages} {012212} (\bibinfo {year} {2022})}\BibitemShut {NoStop}%
\bibitem [{\citenamefont {Hannukainen}\ and\ \citenamefont
  {Larson}(2018)}]{hannukainen2018}%
  \BibitemOpen
  \bibfield  {author} {\bibinfo {author} {\bibfnamefont {J.}~\bibnamefont
  {Hannukainen}}\ and\ \bibinfo {author} {\bibfnamefont {J.}~\bibnamefont
  {Larson}},\ }\bibfield  {title} {\bibinfo {title} {{Dissipation-driven
  quantum phase transitions and symmetry breaking}},\ }\href
  {https://doi.org/10.1103/PhysRevA.98.042113} {\bibfield  {journal} {\bibinfo
  {journal} {Phys. Rev. A}\ }\textbf {\bibinfo {volume} {98}},\ \bibinfo
  {pages} {042113} (\bibinfo {year} {2018})}\BibitemShut {NoStop}%
\bibitem [{\citenamefont {Louren\ifmmode~\mbox{\c{c}}\else \c{c}\fi{}o}\ \emph
  {et~al.}(2022)\citenamefont {Louren\ifmmode~\mbox{\c{c}}\else \c{c}\fi{}o},
  \citenamefont {Prazeres}, \citenamefont {Maciel}, \citenamefont {Iemini},\
  and\ \citenamefont {Duzzioni}}]{lourenco2022}%
  \BibitemOpen
  \bibfield  {author} {\bibinfo {author} {\bibfnamefont {A.~C.}\ \bibnamefont
  {Louren\ifmmode~\mbox{\c{c}}\else \c{c}\fi{}o}}, \bibinfo {author}
  {\bibfnamefont {L.~F.~d.}\ \bibnamefont {Prazeres}}, \bibinfo {author}
  {\bibfnamefont {T.~O.}\ \bibnamefont {Maciel}}, \bibinfo {author}
  {\bibfnamefont {F.}~\bibnamefont {Iemini}},\ and\ \bibinfo {author}
  {\bibfnamefont {E.~I.}\ \bibnamefont {Duzzioni}},\ }\bibfield  {title}
  {\bibinfo {title} {{Genuine multipartite correlations in a boundary time
  crystal}},\ }\href {https://doi.org/10.1103/PhysRevB.105.134422} {\bibfield
  {journal} {\bibinfo  {journal} {Phys. Rev. B}\ }\textbf {\bibinfo {volume}
  {105}},\ \bibinfo {pages} {134422} (\bibinfo {year} {2022})}\BibitemShut
  {NoStop}%
\bibitem [{\citenamefont {Montenegro}\ \emph {et~al.}(2023)\citenamefont
  {Montenegro}, \citenamefont {Genoni}, \citenamefont {Bayat},\ and\
  \citenamefont {Paris}}]{montenegro2023}%
  \BibitemOpen
  \bibfield  {author} {\bibinfo {author} {\bibfnamefont {V.}~\bibnamefont
  {Montenegro}}, \bibinfo {author} {\bibfnamefont {M.~G.}\ \bibnamefont
  {Genoni}}, \bibinfo {author} {\bibfnamefont {A.}~\bibnamefont {Bayat}},\ and\
  \bibinfo {author} {\bibfnamefont {M.~G.~A.}\ \bibnamefont {Paris}},\
  }\bibfield  {title} {\bibinfo {title} {{Quantum-Enhanced Boundary Time
  Crystal Sensors}},\ }\href {https://doi.org/10.48550/ARXIV.2301.02103}
  {\bibfield  {journal} {\bibinfo  {journal} {arXiv:2301.02103}\ } (\bibinfo
  {year} {2023})}\BibitemShut {NoStop}%
\bibitem [{\citenamefont {Pavlov}\ \emph {et~al.}(2023)\citenamefont {Pavlov},
  \citenamefont {Porras},\ and\ \citenamefont {Ivanov}}]{pavlov2023}%
  \BibitemOpen
  \bibfield  {author} {\bibinfo {author} {\bibfnamefont {V.~P.}\ \bibnamefont
  {Pavlov}}, \bibinfo {author} {\bibfnamefont {D.}~\bibnamefont {Porras}},\
  and\ \bibinfo {author} {\bibfnamefont {P.~A.}\ \bibnamefont {Ivanov}},\
  }\bibfield  {title} {\bibinfo {title} {{Quantum metrology with critical
  driven-dissipative collective spin system}},\ }\href
  {https://doi.org/10.48550/ARXIV.2302.05216} {\bibfield  {journal} {\bibinfo
  {journal} {arXiv:2302.05216}\ } (\bibinfo {year} {2023})}\BibitemShut
  {NoStop}%
\bibitem [{\citenamefont {Cabot}\ \emph
  {et~al.}(2022{\natexlab{a}})\citenamefont {Cabot}, \citenamefont {Carollo},\
  and\ \citenamefont {Lesanovsky}}]{cabot2022}%
  \BibitemOpen
  \bibfield  {author} {\bibinfo {author} {\bibfnamefont {A.}~\bibnamefont
  {Cabot}}, \bibinfo {author} {\bibfnamefont {F.}~\bibnamefont {Carollo}},\
  and\ \bibinfo {author} {\bibfnamefont {I.}~\bibnamefont {Lesanovsky}},\
  }\bibfield  {title} {\bibinfo {title} {{Metastable discrete time-crystal
  resonances in a dissipative central spin system}},\ }\href
  {https://doi.org/10.1103/PhysRevB.106.134311} {\bibfield  {journal} {\bibinfo
   {journal} {Phys. Rev. B}\ }\textbf {\bibinfo {volume} {106}},\ \bibinfo
  {pages} {134311} (\bibinfo {year} {2022}{\natexlab{a}})}\BibitemShut
  {NoStop}%
\bibitem [{\citenamefont {Tavis}\ and\ \citenamefont
  {Cummings}(1967)}]{tavis1967}%
  \BibitemOpen
  \bibfield  {author} {\bibinfo {author} {\bibfnamefont {M.}~\bibnamefont
  {Tavis}}\ and\ \bibinfo {author} {\bibfnamefont {F.}~\bibnamefont
  {Cummings}},\ }\bibfield  {title} {\bibinfo {title} {{The exact solution of N
  two level systems interacting with a single mode, quantized radiation
  field}},\ }\href
  {https://doi.org/https://doi.org/10.1016/0375-9601(67)90957-7} {\bibfield
  {journal} {\bibinfo  {journal} {Phys. Lett. A}\ }\textbf {\bibinfo {volume}
  {25}},\ \bibinfo {pages} {714} (\bibinfo {year} {1967})}\BibitemShut
  {NoStop}%
\bibitem [{\citenamefont {Tavis}\ and\ \citenamefont
  {Cummings}(1969)}]{tavis1969}%
  \BibitemOpen
  \bibfield  {author} {\bibinfo {author} {\bibfnamefont {M.}~\bibnamefont
  {Tavis}}\ and\ \bibinfo {author} {\bibfnamefont {F.~W.}\ \bibnamefont
  {Cummings}},\ }\bibfield  {title} {\bibinfo {title} {{Approximate Solutions
  for an $N$-Molecule-Radiation-Field Hamiltonian}},\ }\href
  {https://doi.org/10.1103/PhysRev.188.692} {\bibfield  {journal} {\bibinfo
  {journal} {Phys. Rev.}\ }\textbf {\bibinfo {volume} {188}},\ \bibinfo {pages}
  {692} (\bibinfo {year} {1969})}\BibitemShut {NoStop}%
\bibitem [{\citenamefont {Carollo}\ \emph {et~al.}(2020)\citenamefont
  {Carollo}, \citenamefont {Brandner},\ and\ \citenamefont
  {Lesanovsky}}]{carollo2020}%
  \BibitemOpen
  \bibfield  {author} {\bibinfo {author} {\bibfnamefont {F.}~\bibnamefont
  {Carollo}}, \bibinfo {author} {\bibfnamefont {K.}~\bibnamefont {Brandner}},\
  and\ \bibinfo {author} {\bibfnamefont {I.}~\bibnamefont {Lesanovsky}},\
  }\bibfield  {title} {\bibinfo {title} {{Nonequilibrium Many-Body Quantum
  Engine Driven by Time-Translation Symmetry Breaking}},\ }\href
  {https://doi.org/10.1103/PhysRevLett.125.240602} {\bibfield  {journal}
  {\bibinfo  {journal} {Phys. Rev. Lett.}\ }\textbf {\bibinfo {volume} {125}},\
  \bibinfo {pages} {240602} (\bibinfo {year} {2020})}\BibitemShut {NoStop}%
\bibitem [{\citenamefont {Paulino}\ \emph {et~al.}(2022)\citenamefont
  {Paulino}, \citenamefont {Lesanovsky},\ and\ \citenamefont
  {Carollo}}]{paulino2022}%
  \BibitemOpen
  \bibfield  {author} {\bibinfo {author} {\bibfnamefont {P.~J.}\ \bibnamefont
  {Paulino}}, \bibinfo {author} {\bibfnamefont {I.}~\bibnamefont
  {Lesanovsky}},\ and\ \bibinfo {author} {\bibfnamefont {F.}~\bibnamefont
  {Carollo}},\ }\bibfield  {title} {\bibinfo {title} {{Nonequilibrium
  thermodynamics and power generation in open quantum optomechanical
  systems}},\ }\href {https://doi.org/10.48550/ARXIV.2212.10194} {\bibfield
  {journal} {\bibinfo  {journal} {arXiv:2212.10194}\ } (\bibinfo {year}
  {2022})}\BibitemShut {NoStop}%
\bibitem [{\citenamefont {Ritsch}\ \emph {et~al.}(2013)\citenamefont {Ritsch},
  \citenamefont {Domokos}, \citenamefont {Brennecke},\ and\ \citenamefont
  {Esslinger}}]{ritsch2013}%
  \BibitemOpen
  \bibfield  {author} {\bibinfo {author} {\bibfnamefont {H.}~\bibnamefont
  {Ritsch}}, \bibinfo {author} {\bibfnamefont {P.}~\bibnamefont {Domokos}},
  \bibinfo {author} {\bibfnamefont {F.}~\bibnamefont {Brennecke}},\ and\
  \bibinfo {author} {\bibfnamefont {T.}~\bibnamefont {Esslinger}},\ }\bibfield
  {title} {\bibinfo {title} {{Cold atoms in cavity-generated dynamical optical
  potentials}},\ }\href {https://doi.org/10.1103/RevModPhys.85.553} {\bibfield
  {journal} {\bibinfo  {journal} {Rev. Mod. Phys.}\ }\textbf {\bibinfo {volume}
  {85}},\ \bibinfo {pages} {553} (\bibinfo {year} {2013})}\BibitemShut
  {NoStop}%
\bibitem [{\citenamefont {Feng}\ \emph {et~al.}(2015)\citenamefont {Feng},
  \citenamefont {Zhong}, \citenamefont {Liu}, \citenamefont {Yan},
  \citenamefont {Yang}, \citenamefont {Twamley},\ and\ \citenamefont
  {Wang}}]{feng2015}%
  \BibitemOpen
  \bibfield  {author} {\bibinfo {author} {\bibfnamefont {M.}~\bibnamefont
  {Feng}}, \bibinfo {author} {\bibfnamefont {Y.~P.}\ \bibnamefont {Zhong}},
  \bibinfo {author} {\bibfnamefont {T.}~\bibnamefont {Liu}}, \bibinfo {author}
  {\bibfnamefont {L.~L.}\ \bibnamefont {Yan}}, \bibinfo {author} {\bibfnamefont
  {W.~L.}\ \bibnamefont {Yang}}, \bibinfo {author} {\bibfnamefont
  {J.}~\bibnamefont {Twamley}},\ and\ \bibinfo {author} {\bibfnamefont
  {H.}~\bibnamefont {Wang}},\ }\bibfield  {title} {\bibinfo {title} {{Exploring
  the quantum critical behaviour in a driven Tavis--Cummings circuit}},\ }\href
  {https://doi.org/10.1038/ncomms8111} {\bibfield  {journal} {\bibinfo
  {journal} {Nat. Commun.}\ }\textbf {\bibinfo {volume} {6}},\ \bibinfo {pages}
  {7111} (\bibinfo {year} {2015})}\BibitemShut {NoStop}%
\bibitem [{\citenamefont {Piazza}\ and\ \citenamefont
  {Ritsch}(2015)}]{piazza2015}%
  \BibitemOpen
  \bibfield  {author} {\bibinfo {author} {\bibfnamefont {F.}~\bibnamefont
  {Piazza}}\ and\ \bibinfo {author} {\bibfnamefont {H.}~\bibnamefont
  {Ritsch}},\ }\bibfield  {title} {\bibinfo {title} {Self-ordered limit cycles,
  chaos, and phase slippage with a superfluid inside an optical resonator},\
  }\href {https://doi.org/10.1103/PhysRevLett.115.163601} {\bibfield  {journal}
  {\bibinfo  {journal} {Phys. Rev. Lett.}\ }\textbf {\bibinfo {volume} {115}},\
  \bibinfo {pages} {163601} (\bibinfo {year} {2015})}\BibitemShut {NoStop}%
\bibitem [{\citenamefont {Norcia}\ \emph {et~al.}(2016)\citenamefont {Norcia},
  \citenamefont {Winchester}, \citenamefont {Cline},\ and\ \citenamefont
  {Thompson}}]{norcia2016}%
  \BibitemOpen
  \bibfield  {author} {\bibinfo {author} {\bibfnamefont {M.~A.}\ \bibnamefont
  {Norcia}}, \bibinfo {author} {\bibfnamefont {M.~N.}\ \bibnamefont
  {Winchester}}, \bibinfo {author} {\bibfnamefont {J.~R.~K.}\ \bibnamefont
  {Cline}},\ and\ \bibinfo {author} {\bibfnamefont {J.~K.}\ \bibnamefont
  {Thompson}},\ }\bibfield  {title} {\bibinfo {title} {{Superradiance on the
  millihertz linewidth strontium clock transition}},\ }\href
  {https://doi.org/10.1126/sciadv.1601231} {\bibfield  {journal} {\bibinfo
  {journal} {Sci. Adv.}\ }\textbf {\bibinfo {volume} {2}},\ \bibinfo {pages}
  {e1601231} (\bibinfo {year} {2016})}\BibitemShut {NoStop}%
\bibitem [{\citenamefont {Rose}\ \emph {et~al.}(2017)\citenamefont {Rose},
  \citenamefont {Tyryshkin}, \citenamefont {Riemann}, \citenamefont
  {Abrosimov}, \citenamefont {Becker}, \citenamefont {Pohl}, \citenamefont
  {Thewalt}, \citenamefont {Itoh},\ and\ \citenamefont {Lyon}}]{rose2017}%
  \BibitemOpen
  \bibfield  {author} {\bibinfo {author} {\bibfnamefont {B.~C.}\ \bibnamefont
  {Rose}}, \bibinfo {author} {\bibfnamefont {A.~M.}\ \bibnamefont {Tyryshkin}},
  \bibinfo {author} {\bibfnamefont {H.}~\bibnamefont {Riemann}}, \bibinfo
  {author} {\bibfnamefont {N.~V.}\ \bibnamefont {Abrosimov}}, \bibinfo {author}
  {\bibfnamefont {P.}~\bibnamefont {Becker}}, \bibinfo {author} {\bibfnamefont
  {H.-J.}\ \bibnamefont {Pohl}}, \bibinfo {author} {\bibfnamefont {M.~L.~W.}\
  \bibnamefont {Thewalt}}, \bibinfo {author} {\bibfnamefont {K.~M.}\
  \bibnamefont {Itoh}},\ and\ \bibinfo {author} {\bibfnamefont {S.~A.}\
  \bibnamefont {Lyon}},\ }\bibfield  {title} {\bibinfo {title} {{Coherent Rabi
  Dynamics of a Superradiant Spin Ensemble in a Microwave Cavity}},\ }\href
  {https://doi.org/10.1103/PhysRevX.7.031002} {\bibfield  {journal} {\bibinfo
  {journal} {Phys. Rev. X}\ }\textbf {\bibinfo {volume} {7}},\ \bibinfo {pages}
  {031002} (\bibinfo {year} {2017})}\BibitemShut {NoStop}%
\bibitem [{\citenamefont {Norcia}\ \emph {et~al.}(2018)\citenamefont {Norcia},
  \citenamefont {Lewis-Swan}, \citenamefont {Cline}, \citenamefont {Zhu},
  \citenamefont {Rey},\ and\ \citenamefont {Thompson}}]{norcia2018}%
  \BibitemOpen
  \bibfield  {author} {\bibinfo {author} {\bibfnamefont {M.~A.}\ \bibnamefont
  {Norcia}}, \bibinfo {author} {\bibfnamefont {R.~J.}\ \bibnamefont
  {Lewis-Swan}}, \bibinfo {author} {\bibfnamefont {J.~R.~K.}\ \bibnamefont
  {Cline}}, \bibinfo {author} {\bibfnamefont {B.}~\bibnamefont {Zhu}}, \bibinfo
  {author} {\bibfnamefont {A.~M.}\ \bibnamefont {Rey}},\ and\ \bibinfo {author}
  {\bibfnamefont {J.~K.}\ \bibnamefont {Thompson}},\ }\bibfield  {title}
  {\bibinfo {title} {{Cavity-mediated collective spin-exchange interactions in
  a strontium superradiant laser}},\ }\href
  {https://doi.org/10.1126/science.aar3102} {\bibfield  {journal} {\bibinfo
  {journal} {Science}\ }\textbf {\bibinfo {volume} {361}},\ \bibinfo {pages}
  {259} (\bibinfo {year} {2018})}\BibitemShut {NoStop}%
\bibitem [{\citenamefont {Dogra}\ \emph {et~al.}(2019)\citenamefont {Dogra},
  \citenamefont {Landini}, \citenamefont {Kroeger}, \citenamefont {Hruby},
  \citenamefont {Donner},\ and\ \citenamefont {Esslinger}}]{dogra2019}%
  \BibitemOpen
  \bibfield  {author} {\bibinfo {author} {\bibfnamefont {N.}~\bibnamefont
  {Dogra}}, \bibinfo {author} {\bibfnamefont {M.}~\bibnamefont {Landini}},
  \bibinfo {author} {\bibfnamefont {K.}~\bibnamefont {Kroeger}}, \bibinfo
  {author} {\bibfnamefont {L.}~\bibnamefont {Hruby}}, \bibinfo {author}
  {\bibfnamefont {T.}~\bibnamefont {Donner}},\ and\ \bibinfo {author}
  {\bibfnamefont {T.}~\bibnamefont {Esslinger}},\ }\bibfield  {title} {\bibinfo
  {title} {{Dissipation-induced structural instability and chiral dynamics in a
  quantum gas}},\ }\href {https://doi.org/10.1126/science.aaw4465} {\bibfield
  {journal} {\bibinfo  {journal} {Science}\ }\textbf {\bibinfo {volume}
  {366}},\ \bibinfo {pages} {1496} (\bibinfo {year} {2019})}\BibitemShut
  {NoStop}%
\bibitem [{\citenamefont {Sch\"affer}\ \emph {et~al.}(2020)\citenamefont
  {Sch\"affer}, \citenamefont {Tang}, \citenamefont {Henriksen}, \citenamefont
  {J\o{}rgensen}, \citenamefont {Christensen},\ and\ \citenamefont
  {Thomsen}}]{schaffer2020}%
  \BibitemOpen
  \bibfield  {author} {\bibinfo {author} {\bibfnamefont {S.~A.}\ \bibnamefont
  {Sch\"affer}}, \bibinfo {author} {\bibfnamefont {M.}~\bibnamefont {Tang}},
  \bibinfo {author} {\bibfnamefont {M.~R.}\ \bibnamefont {Henriksen}}, \bibinfo
  {author} {\bibfnamefont {A.~A.}\ \bibnamefont {J\o{}rgensen}}, \bibinfo
  {author} {\bibfnamefont {B.~T.~R.}\ \bibnamefont {Christensen}},\ and\
  \bibinfo {author} {\bibfnamefont {J.~W.}\ \bibnamefont {Thomsen}},\
  }\bibfield  {title} {\bibinfo {title} {{Lasing on a narrow transition in a
  cold thermal strontium ensemble}},\ }\href
  {https://doi.org/10.1103/PhysRevA.101.013819} {\bibfield  {journal} {\bibinfo
   {journal} {Phys. Rev. A}\ }\textbf {\bibinfo {volume} {101}},\ \bibinfo
  {pages} {013819} (\bibinfo {year} {2020})}\BibitemShut {NoStop}%
\bibitem [{\citenamefont {Mivehvar}\ \emph {et~al.}(2021)\citenamefont
  {Mivehvar}, \citenamefont {Piazza}, \citenamefont {Donner},\ and\
  \citenamefont {Ritsch}}]{mivehvar2021}%
  \BibitemOpen
  \bibfield  {author} {\bibinfo {author} {\bibfnamefont {F.}~\bibnamefont
  {Mivehvar}}, \bibinfo {author} {\bibfnamefont {F.}~\bibnamefont {Piazza}},
  \bibinfo {author} {\bibfnamefont {T.}~\bibnamefont {Donner}},\ and\ \bibinfo
  {author} {\bibfnamefont {H.}~\bibnamefont {Ritsch}},\ }\bibfield  {title}
  {\bibinfo {title} {{Cavity QED with quantum gases: new paradigms in many-body
  physics}},\ }\href {https://doi.org/10.1080/00018732.2021.1969727} {\bibfield
   {journal} {\bibinfo  {journal} {Adv. Phys.}\ }\textbf {\bibinfo {volume}
  {70}},\ \bibinfo {pages} {1} (\bibinfo {year} {2021})}\BibitemShut {NoStop}%
\bibitem [{\citenamefont {Agarwal}\ \emph {et~al.}(1997)\citenamefont
  {Agarwal}, \citenamefont {Puri},\ and\ \citenamefont {Singh}}]{agarwal1997}%
  \BibitemOpen
  \bibfield  {author} {\bibinfo {author} {\bibfnamefont {G.~S.}\ \bibnamefont
  {Agarwal}}, \bibinfo {author} {\bibfnamefont {R.~R.}\ \bibnamefont {Puri}},\
  and\ \bibinfo {author} {\bibfnamefont {R.~P.}\ \bibnamefont {Singh}},\
  }\bibfield  {title} {\bibinfo {title} {{Atomic Schr\"odinger cat states}},\
  }\href {https://doi.org/10.1103/PhysRevA.56.2249} {\bibfield  {journal}
  {\bibinfo  {journal} {Phys. Rev. A}\ }\textbf {\bibinfo {volume} {56}},\
  \bibinfo {pages} {2249} (\bibinfo {year} {1997})}\BibitemShut {NoStop}%
\bibitem [{\citenamefont {Gopalakrishnan}\ \emph {et~al.}(2011)\citenamefont
  {Gopalakrishnan}, \citenamefont {Lev},\ and\ \citenamefont
  {Goldbart}}]{gopalakrishnan2011}%
  \BibitemOpen
  \bibfield  {author} {\bibinfo {author} {\bibfnamefont {S.}~\bibnamefont
  {Gopalakrishnan}}, \bibinfo {author} {\bibfnamefont {B.~L.}\ \bibnamefont
  {Lev}},\ and\ \bibinfo {author} {\bibfnamefont {P.~M.}\ \bibnamefont
  {Goldbart}},\ }\bibfield  {title} {\bibinfo {title} {{Frustration and
  Glassiness in Spin Models with Cavity-Mediated Interactions}},\ }\href
  {https://doi.org/10.1103/PhysRevLett.107.277201} {\bibfield  {journal}
  {\bibinfo  {journal} {Phys. Rev. Lett.}\ }\textbf {\bibinfo {volume} {107}},\
  \bibinfo {pages} {277201} (\bibinfo {year} {2011})}\BibitemShut {NoStop}%
\bibitem [{\citenamefont {Damanet}\ \emph {et~al.}(2019)\citenamefont
  {Damanet}, \citenamefont {Daley},\ and\ \citenamefont
  {Keeling}}]{damanet2019}%
  \BibitemOpen
  \bibfield  {author} {\bibinfo {author} {\bibfnamefont {F.}~\bibnamefont
  {Damanet}}, \bibinfo {author} {\bibfnamefont {A.~J.}\ \bibnamefont {Daley}},\
  and\ \bibinfo {author} {\bibfnamefont {J.}~\bibnamefont {Keeling}},\
  }\bibfield  {title} {\bibinfo {title} {{Atom-only descriptions of the
  driven-dissipative Dicke model}},\ }\href
  {https://doi.org/10.1103/PhysRevA.99.033845} {\bibfield  {journal} {\bibinfo
  {journal} {Phys. Rev. A}\ }\textbf {\bibinfo {volume} {99}},\ \bibinfo
  {pages} {033845} (\bibinfo {year} {2019})}\BibitemShut {NoStop}%
\bibitem [{\citenamefont {Carollo}\ and\ \citenamefont
  {Lesanovsky}(2021)}]{carollo2021}%
  \BibitemOpen
  \bibfield  {author} {\bibinfo {author} {\bibfnamefont {F.}~\bibnamefont
  {Carollo}}\ and\ \bibinfo {author} {\bibfnamefont {I.}~\bibnamefont
  {Lesanovsky}},\ }\bibfield  {title} {\bibinfo {title} {{Exactness of
  Mean-Field Equations for Open Dicke Models with an Application to Pattern
  Retrieval Dynamics}},\ }\href
  {https://doi.org/10.1103/PhysRevLett.126.230601} {\bibfield  {journal}
  {\bibinfo  {journal} {Phys. Rev. Lett.}\ }\textbf {\bibinfo {volume} {126}},\
  \bibinfo {pages} {230601} (\bibinfo {year} {2021})}\BibitemShut {NoStop}%
\bibitem [{\citenamefont {Lindblad}(1976)}]{lindblad1976}%
  \BibitemOpen
  \bibfield  {author} {\bibinfo {author} {\bibfnamefont {G.}~\bibnamefont
  {Lindblad}},\ }\bibfield  {title} {\bibinfo {title} {{On the generators of
  quantum dynamical semigroups}},\ }\href {https://doi.org/10.1007/BF01608499}
  {\bibfield  {journal} {\bibinfo  {journal} {Commun. Math. Phys.}\ }\textbf
  {\bibinfo {volume} {48}},\ \bibinfo {pages} {119} (\bibinfo {year}
  {1976})}\BibitemShut {NoStop}%
\bibitem [{\citenamefont {Gorini}\ \emph {et~al.}(1976)\citenamefont {Gorini},
  \citenamefont {Kossakowski},\ and\ \citenamefont {Sudarshan}}]{gorini1976}%
  \BibitemOpen
  \bibfield  {author} {\bibinfo {author} {\bibfnamefont {V.}~\bibnamefont
  {Gorini}}, \bibinfo {author} {\bibfnamefont {A.}~\bibnamefont
  {Kossakowski}},\ and\ \bibinfo {author} {\bibfnamefont {E.~C.~G.}\
  \bibnamefont {Sudarshan}},\ }\bibfield  {title} {\bibinfo {title}
  {{Completely positive dynamical semigroups of N‐level systems}},\ }\href
  {https://doi.org/10.1063/1.522979} {\bibfield  {journal} {\bibinfo  {journal}
  {J. Math. Phys.}\ }\textbf {\bibinfo {volume} {17}},\ \bibinfo {pages} {821}
  (\bibinfo {year} {1976})}\BibitemShut {NoStop}%
\bibitem [{\citenamefont {Bogoliubov}\ \emph {et~al.}(1996)\citenamefont
  {Bogoliubov}, \citenamefont {Bullough},\ and\ \citenamefont
  {Timonen}}]{bogoliubov1996}%
  \BibitemOpen
  \bibfield  {author} {\bibinfo {author} {\bibfnamefont {N.~M.}\ \bibnamefont
  {Bogoliubov}}, \bibinfo {author} {\bibfnamefont {R.~K.}\ \bibnamefont
  {Bullough}},\ and\ \bibinfo {author} {\bibfnamefont {J.}~\bibnamefont
  {Timonen}},\ }\bibfield  {title} {\bibinfo {title} {{Exact solution of
  generalized Tavis - Cummings models in quantum optics}},\ }\href
  {https://doi.org/10.1088/0305-4470/29/19/015} {\bibfield  {journal} {\bibinfo
   {journal} {J. Phys. A Math. Gen.}\ }\textbf {\bibinfo {volume} {29}},\
  \bibinfo {pages} {6305} (\bibinfo {year} {1996})}\BibitemShut {NoStop}%
\bibitem [{\citenamefont {López}\ \emph {et~al.}(2007)\citenamefont {López},
  \citenamefont {Lastra}, \citenamefont {Romero},\ and\ \citenamefont
  {Retamal}}]{lopez2007}%
  \BibitemOpen
  \bibfield  {author} {\bibinfo {author} {\bibfnamefont {C.~E.}\ \bibnamefont
  {López}}, \bibinfo {author} {\bibfnamefont {F.}~\bibnamefont {Lastra}},
  \bibinfo {author} {\bibfnamefont {G.}~\bibnamefont {Romero}},\ and\ \bibinfo
  {author} {\bibfnamefont {J.~C.}\ \bibnamefont {Retamal}},\ }\bibfield
  {title} {\bibinfo {title} {{Concurrence in the inhomogeneous Tavis-Cummings
  model}},\ }\href {https://doi.org/10.1088/1742-6596/84/1/012013} {\bibfield
  {journal} {\bibinfo  {journal} {J. Phys.: Conf. Ser.}\ }\textbf {\bibinfo
  {volume} {84}},\ \bibinfo {pages} {012013} (\bibinfo {year}
  {2007})}\BibitemShut {NoStop}%
\bibitem [{\citenamefont {Wood}\ \emph {et~al.}(2014)\citenamefont {Wood},
  \citenamefont {Borneman},\ and\ \citenamefont {Cory}}]{wood2014}%
  \BibitemOpen
  \bibfield  {author} {\bibinfo {author} {\bibfnamefont {C.~J.}\ \bibnamefont
  {Wood}}, \bibinfo {author} {\bibfnamefont {T.~W.}\ \bibnamefont {Borneman}},\
  and\ \bibinfo {author} {\bibfnamefont {D.~G.}\ \bibnamefont {Cory}},\
  }\bibfield  {title} {\bibinfo {title} {{Cavity Cooling of an Ensemble Spin
  System}},\ }\href {https://doi.org/10.1103/PhysRevLett.112.050501} {\bibfield
   {journal} {\bibinfo  {journal} {Phys. Rev. Lett.}\ }\textbf {\bibinfo
  {volume} {112}},\ \bibinfo {pages} {050501} (\bibinfo {year}
  {2014})}\BibitemShut {NoStop}%
\bibitem [{\citenamefont {Genway}\ \emph {et~al.}(2014)\citenamefont {Genway},
  \citenamefont {Li}, \citenamefont {Ates}, \citenamefont {Lanyon},\ and\
  \citenamefont {Lesanovsky}}]{genway2014}%
  \BibitemOpen
  \bibfield  {author} {\bibinfo {author} {\bibfnamefont {S.}~\bibnamefont
  {Genway}}, \bibinfo {author} {\bibfnamefont {W.}~\bibnamefont {Li}}, \bibinfo
  {author} {\bibfnamefont {C.}~\bibnamefont {Ates}}, \bibinfo {author}
  {\bibfnamefont {B.~P.}\ \bibnamefont {Lanyon}},\ and\ \bibinfo {author}
  {\bibfnamefont {I.}~\bibnamefont {Lesanovsky}},\ }\bibfield  {title}
  {\bibinfo {title} {{Generalized Dicke Nonequilibrium Dynamics in Trapped
  Ions}},\ }\href {https://doi.org/10.1103/PhysRevLett.112.023603} {\bibfield
  {journal} {\bibinfo  {journal} {Phys. Rev. Lett.}\ }\textbf {\bibinfo
  {volume} {112}},\ \bibinfo {pages} {023603} (\bibinfo {year}
  {2014})}\BibitemShut {NoStop}%
\bibitem [{\citenamefont {Lamata}(2017)}]{lamata2017}%
  \BibitemOpen
  \bibfield  {author} {\bibinfo {author} {\bibfnamefont {L.}~\bibnamefont
  {Lamata}},\ }\bibfield  {title} {\bibinfo {title} {{Digital-analog quantum
  simulation of generalized Dicke models with superconducting circuits}},\
  }\href {https://doi.org/10.1038/srep43768} {\bibfield  {journal} {\bibinfo
  {journal} {Sci. Rep.}\ }\textbf {\bibinfo {volume} {7}},\ \bibinfo {pages}
  {43768} (\bibinfo {year} {2017})}\BibitemShut {NoStop}%
\bibitem [{\citenamefont {Trivedi}\ \emph {et~al.}(2019)\citenamefont
  {Trivedi}, \citenamefont {Radulaski}, \citenamefont {Fischer}, \citenamefont
  {Fan},\ and\ \citenamefont {Vu\ifmmode \check{c}\else
  \v{c}\fi{}kovi\ifmmode~\acute{c}\else \'{c}\fi{}}}]{trivedi2019}%
  \BibitemOpen
  \bibfield  {author} {\bibinfo {author} {\bibfnamefont {R.}~\bibnamefont
  {Trivedi}}, \bibinfo {author} {\bibfnamefont {M.}~\bibnamefont {Radulaski}},
  \bibinfo {author} {\bibfnamefont {K.~A.}\ \bibnamefont {Fischer}}, \bibinfo
  {author} {\bibfnamefont {S.}~\bibnamefont {Fan}},\ and\ \bibinfo {author}
  {\bibfnamefont {J.}~\bibnamefont {Vu\ifmmode \check{c}\else
  \v{c}\fi{}kovi\ifmmode~\acute{c}\else \'{c}\fi{}}},\ }\bibfield  {title}
  {\bibinfo {title} {{Photon Blockade in Weakly Driven Cavity Quantum
  Electrodynamics Systems with Many Emitters}},\ }\href
  {https://doi.org/10.1103/PhysRevLett.122.243602} {\bibfield  {journal}
  {\bibinfo  {journal} {Phys. Rev. Lett.}\ }\textbf {\bibinfo {volume} {122}},\
  \bibinfo {pages} {243602} (\bibinfo {year} {2019})}\BibitemShut {NoStop}%
\bibitem [{\citenamefont {Andolina}\ \emph {et~al.}(2019)\citenamefont
  {Andolina}, \citenamefont {Keck}, \citenamefont {Mari}, \citenamefont
  {Campisi}, \citenamefont {Giovannetti},\ and\ \citenamefont
  {Polini}}]{andolina2019}%
  \BibitemOpen
  \bibfield  {author} {\bibinfo {author} {\bibfnamefont {G.~M.}\ \bibnamefont
  {Andolina}}, \bibinfo {author} {\bibfnamefont {M.}~\bibnamefont {Keck}},
  \bibinfo {author} {\bibfnamefont {A.}~\bibnamefont {Mari}}, \bibinfo {author}
  {\bibfnamefont {M.}~\bibnamefont {Campisi}}, \bibinfo {author} {\bibfnamefont
  {V.}~\bibnamefont {Giovannetti}},\ and\ \bibinfo {author} {\bibfnamefont
  {M.}~\bibnamefont {Polini}},\ }\bibfield  {title} {\bibinfo {title}
  {{Extractable Work, the Role of Correlations, and Asymptotic Freedom in
  Quantum Batteries}},\ }\href {https://doi.org/10.1103/PhysRevLett.122.047702}
  {\bibfield  {journal} {\bibinfo  {journal} {Phys. Rev. Lett.}\ }\textbf
  {\bibinfo {volume} {122}},\ \bibinfo {pages} {047702} (\bibinfo {year}
  {2019})}\BibitemShut {NoStop}%
\bibitem [{\citenamefont {Shapiro}\ \emph {et~al.}(2020)\citenamefont
  {Shapiro}, \citenamefont {Pogosov},\ and\ \citenamefont
  {Lozovik}}]{shapiro2020}%
  \BibitemOpen
  \bibfield  {author} {\bibinfo {author} {\bibfnamefont {D.~S.}\ \bibnamefont
  {Shapiro}}, \bibinfo {author} {\bibfnamefont {W.~V.}\ \bibnamefont
  {Pogosov}},\ and\ \bibinfo {author} {\bibfnamefont {Y.~E.}\ \bibnamefont
  {Lozovik}},\ }\bibfield  {title} {\bibinfo {title} {{Universal fluctuations
  and squeezing in a generalized Dicke model near the superradiant phase
  transition}},\ }\href {https://doi.org/10.1103/PhysRevA.102.023703}
  {\bibfield  {journal} {\bibinfo  {journal} {Phys. Rev. A}\ }\textbf {\bibinfo
  {volume} {102}},\ \bibinfo {pages} {023703} (\bibinfo {year}
  {2020})}\BibitemShut {NoStop}%
\bibitem [{\citenamefont {Cuartas}\ and\ \citenamefont
  {Vinck-Posada}(2021)}]{restrepo2021}%
  \BibitemOpen
  \bibfield  {author} {\bibinfo {author} {\bibfnamefont {J.~R.}\ \bibnamefont
  {Cuartas}}\ and\ \bibinfo {author} {\bibfnamefont {H.}~\bibnamefont
  {Vinck-Posada}},\ }\bibfield  {title} {\bibinfo {title} {{Uncover quantumness
  in the crossover from coherent to quantum-correlated phases via photon
  statistics and entanglement in the Tavis–Cummings model}},\ }\href
  {https://doi.org/https://doi.org/10.1016/j.ijleo.2021.167672} {\bibfield
  {journal} {\bibinfo  {journal} {Optik}\ }\textbf {\bibinfo {volume} {245}},\
  \bibinfo {pages} {167672} (\bibinfo {year} {2021})}\BibitemShut {NoStop}%
\bibitem [{\citenamefont {Baum}\ \emph {et~al.}(2022)\citenamefont {Baum},
  \citenamefont {Broman}, \citenamefont {Clarke}, \citenamefont {Costa},
  \citenamefont {Mucciaccio}, \citenamefont {Yue}, \citenamefont {Zhang},
  \citenamefont {Norman}, \citenamefont {Patton}, \citenamefont {Radulaski},\
  and\ \citenamefont {Scalettar}}]{baum2022}%
  \BibitemOpen
  \bibfield  {author} {\bibinfo {author} {\bibfnamefont {E.}~\bibnamefont
  {Baum}}, \bibinfo {author} {\bibfnamefont {A.}~\bibnamefont {Broman}},
  \bibinfo {author} {\bibfnamefont {T.}~\bibnamefont {Clarke}}, \bibinfo
  {author} {\bibfnamefont {N.~C.}\ \bibnamefont {Costa}}, \bibinfo {author}
  {\bibfnamefont {J.}~\bibnamefont {Mucciaccio}}, \bibinfo {author}
  {\bibfnamefont {A.}~\bibnamefont {Yue}}, \bibinfo {author} {\bibfnamefont
  {Y.}~\bibnamefont {Zhang}}, \bibinfo {author} {\bibfnamefont
  {V.}~\bibnamefont {Norman}}, \bibinfo {author} {\bibfnamefont
  {J.}~\bibnamefont {Patton}}, \bibinfo {author} {\bibfnamefont
  {M.}~\bibnamefont {Radulaski}},\ and\ \bibinfo {author} {\bibfnamefont
  {R.~T.}\ \bibnamefont {Scalettar}},\ }\bibfield  {title} {\bibinfo {title}
  {{Effect of emitters on quantum state transfer in coupled cavity arrays}},\
  }\href {https://doi.org/10.1103/PhysRevB.105.195429} {\bibfield  {journal}
  {\bibinfo  {journal} {Phys. Rev. B}\ }\textbf {\bibinfo {volume} {105}},\
  \bibinfo {pages} {195429} (\bibinfo {year} {2022})}\BibitemShut {NoStop}%
\bibitem [{\citenamefont {Blaha}\ \emph {et~al.}(2022)\citenamefont {Blaha},
  \citenamefont {Johnson}, \citenamefont {Rauschenbeutel},\ and\ \citenamefont
  {Volz}}]{blaha2022}%
  \BibitemOpen
  \bibfield  {author} {\bibinfo {author} {\bibfnamefont {M.}~\bibnamefont
  {Blaha}}, \bibinfo {author} {\bibfnamefont {A.}~\bibnamefont {Johnson}},
  \bibinfo {author} {\bibfnamefont {A.}~\bibnamefont {Rauschenbeutel}},\ and\
  \bibinfo {author} {\bibfnamefont {J.}~\bibnamefont {Volz}},\ }\bibfield
  {title} {\bibinfo {title} {{Beyond the Tavis-Cummings model: Revisiting
  cavity QED with ensembles of quantum emitters}},\ }\href
  {https://doi.org/10.1103/PhysRevA.105.013719} {\bibfield  {journal} {\bibinfo
   {journal} {Phys. Rev. A}\ }\textbf {\bibinfo {volume} {105}},\ \bibinfo
  {pages} {013719} (\bibinfo {year} {2022})}\BibitemShut {NoStop}%
\bibitem [{\citenamefont {Valencia-Tortora}\ \emph {et~al.}(2022)\citenamefont
  {Valencia-Tortora}, \citenamefont {Kelly}, \citenamefont {Donner},
  \citenamefont {Morigi}, \citenamefont {Fazio},\ and\ \citenamefont
  {Marino}}]{valenciatortora2022}%
  \BibitemOpen
  \bibfield  {author} {\bibinfo {author} {\bibfnamefont {R.~J.}\ \bibnamefont
  {Valencia-Tortora}}, \bibinfo {author} {\bibfnamefont {S.~P.}\ \bibnamefont
  {Kelly}}, \bibinfo {author} {\bibfnamefont {T.}~\bibnamefont {Donner}},
  \bibinfo {author} {\bibfnamefont {G.}~\bibnamefont {Morigi}}, \bibinfo
  {author} {\bibfnamefont {R.}~\bibnamefont {Fazio}},\ and\ \bibinfo {author}
  {\bibfnamefont {J.}~\bibnamefont {Marino}},\ }\bibfield  {title} {\bibinfo
  {title} {{Crafting the dynamical structure of synchronization by harnessing
  bosonic multi-level cavity QED}},\ }\href
  {https://doi.org/10.48550/arXiv.2210.14224} {\bibfield  {journal} {\bibinfo
  {journal} {arXiv:2210.14224}\ } (\bibinfo {year} {2022})}\BibitemShut
  {NoStop}%
\bibitem [{\citenamefont {Kelly}\ \emph {et~al.}(2022)\citenamefont {Kelly},
  \citenamefont {Thompson}, \citenamefont {Rey},\ and\ \citenamefont
  {Marino}}]{kelly2022}%
  \BibitemOpen
  \bibfield  {author} {\bibinfo {author} {\bibfnamefont {S.~P.}\ \bibnamefont
  {Kelly}}, \bibinfo {author} {\bibfnamefont {J.~K.}\ \bibnamefont {Thompson}},
  \bibinfo {author} {\bibfnamefont {A.~M.}\ \bibnamefont {Rey}},\ and\ \bibinfo
  {author} {\bibfnamefont {J.}~\bibnamefont {Marino}},\ }\bibfield  {title}
  {\bibinfo {title} {{Resonant light enhances phase coherence in a cavity QED
  simulator of fermionic superfluidity}},\ }\href
  {https://doi.org/10.1103/PhysRevResearch.4.L042032} {\bibfield  {journal}
  {\bibinfo  {journal} {Phys. Rev. Res.}\ }\textbf {\bibinfo {volume} {4}},\
  \bibinfo {pages} {L042032} (\bibinfo {year} {2022})}\BibitemShut {NoStop}%
\bibitem [{\citenamefont {Stitely}\ \emph {et~al.}(2023)\citenamefont
  {Stitely}, \citenamefont {Finger}, \citenamefont {Rosa-Medina}, \citenamefont
  {Ferri}, \citenamefont {Donner}, \citenamefont {Esslinger}, \citenamefont
  {Parkins},\ and\ \citenamefont {Krauskopf}}]{stitely2023}%
  \BibitemOpen
  \bibfield  {author} {\bibinfo {author} {\bibfnamefont {K.}~\bibnamefont
  {Stitely}}, \bibinfo {author} {\bibfnamefont {F.}~\bibnamefont {Finger}},
  \bibinfo {author} {\bibfnamefont {R.}~\bibnamefont {Rosa-Medina}}, \bibinfo
  {author} {\bibfnamefont {F.}~\bibnamefont {Ferri}}, \bibinfo {author}
  {\bibfnamefont {T.}~\bibnamefont {Donner}}, \bibinfo {author} {\bibfnamefont
  {T.}~\bibnamefont {Esslinger}}, \bibinfo {author} {\bibfnamefont
  {S.}~\bibnamefont {Parkins}},\ and\ \bibinfo {author} {\bibfnamefont
  {B.}~\bibnamefont {Krauskopf}},\ }\bibfield  {title} {\bibinfo {title}
  {{Quantum Fluctuation Dynamics of Dispersive Superradiant Pulses in a Hybrid
  Light-Matter System}},\ }\href {https://arxiv.org/abs/2302.08078} {\bibfield
  {journal} {\bibinfo  {journal} {arXiv:2302.08078}\ } (\bibinfo {year}
  {2023})}\BibitemShut {NoStop}%
\bibitem [{\citenamefont {Hebenstreit}\ \emph {et~al.}(2017)\citenamefont
  {Hebenstreit}, \citenamefont {Kraus}, \citenamefont {Ostermann},\ and\
  \citenamefont {Ritsch}}]{hebenstreit2017}%
  \BibitemOpen
  \bibfield  {author} {\bibinfo {author} {\bibfnamefont {M.}~\bibnamefont
  {Hebenstreit}}, \bibinfo {author} {\bibfnamefont {B.}~\bibnamefont {Kraus}},
  \bibinfo {author} {\bibfnamefont {L.}~\bibnamefont {Ostermann}},\ and\
  \bibinfo {author} {\bibfnamefont {H.}~\bibnamefont {Ritsch}},\ }\bibfield
  {title} {\bibinfo {title} {{Subradiance via Entanglement in Atoms with
  Several Independent Decay Channels}},\ }\href
  {https://doi.org/10.1103/PhysRevLett.118.143602} {\bibfield  {journal}
  {\bibinfo  {journal} {Phys. Rev. Lett.}\ }\textbf {\bibinfo {volume} {118}},\
  \bibinfo {pages} {143602} (\bibinfo {year} {2017})}\BibitemShut {NoStop}%
\bibitem [{\citenamefont {Plankensteiner}\ \emph {et~al.}(2017)\citenamefont
  {Plankensteiner}, \citenamefont {Sommer}, \citenamefont {Ritsch},\ and\
  \citenamefont {Genes}}]{plankensteiner2017}%
  \BibitemOpen
  \bibfield  {author} {\bibinfo {author} {\bibfnamefont {D.}~\bibnamefont
  {Plankensteiner}}, \bibinfo {author} {\bibfnamefont {C.}~\bibnamefont
  {Sommer}}, \bibinfo {author} {\bibfnamefont {H.}~\bibnamefont {Ritsch}},\
  and\ \bibinfo {author} {\bibfnamefont {C.}~\bibnamefont {Genes}},\ }\bibfield
   {title} {\bibinfo {title} {{Cavity Antiresonance Spectroscopy of Dipole
  Coupled Subradiant Arrays}},\ }\href
  {https://doi.org/10.1103/PhysRevLett.119.093601} {\bibfield  {journal}
  {\bibinfo  {journal} {Phys. Rev. Lett.}\ }\textbf {\bibinfo {volume} {119}},\
  \bibinfo {pages} {093601} (\bibinfo {year} {2017})}\BibitemShut {NoStop}%
\bibitem [{\citenamefont {Plankensteiner}\ \emph {et~al.}(2019)\citenamefont
  {Plankensteiner}, \citenamefont {Sommer}, \citenamefont {Reitz},
  \citenamefont {Ritsch},\ and\ \citenamefont {Genes}}]{plankensteiner2019}%
  \BibitemOpen
  \bibfield  {author} {\bibinfo {author} {\bibfnamefont {D.}~\bibnamefont
  {Plankensteiner}}, \bibinfo {author} {\bibfnamefont {C.}~\bibnamefont
  {Sommer}}, \bibinfo {author} {\bibfnamefont {M.}~\bibnamefont {Reitz}},
  \bibinfo {author} {\bibfnamefont {H.}~\bibnamefont {Ritsch}},\ and\ \bibinfo
  {author} {\bibfnamefont {C.}~\bibnamefont {Genes}},\ }\bibfield  {title}
  {\bibinfo {title} {{Enhanced collective Purcell effect of coupled quantum
  emitter systems}},\ }\href {https://doi.org/10.1103/PhysRevA.99.043843}
  {\bibfield  {journal} {\bibinfo  {journal} {Phys. Rev. A}\ }\textbf {\bibinfo
  {volume} {99}},\ \bibinfo {pages} {043843} (\bibinfo {year}
  {2019})}\BibitemShut {NoStop}%
\bibitem [{\citenamefont {Reitz}\ \emph {et~al.}(2022)\citenamefont {Reitz},
  \citenamefont {Sommer},\ and\ \citenamefont {Genes}}]{reitz2022}%
  \BibitemOpen
  \bibfield  {author} {\bibinfo {author} {\bibfnamefont {M.}~\bibnamefont
  {Reitz}}, \bibinfo {author} {\bibfnamefont {C.}~\bibnamefont {Sommer}},\ and\
  \bibinfo {author} {\bibfnamefont {C.}~\bibnamefont {Genes}},\ }\bibfield
  {title} {\bibinfo {title} {{Cooperative Quantum Phenomena in Light-Matter
  Platforms}},\ }\href {https://doi.org/10.1103/PRXQuantum.3.010201} {\bibfield
   {journal} {\bibinfo  {journal} {PRX Quantum}\ }\textbf {\bibinfo {volume}
  {3}},\ \bibinfo {pages} {010201} (\bibinfo {year} {2022})}\BibitemShut
  {NoStop}%
\bibitem [{\citenamefont {Dimer}\ \emph {et~al.}(2007)\citenamefont {Dimer},
  \citenamefont {Estienne}, \citenamefont {Parkins},\ and\ \citenamefont
  {Carmichael}}]{dimer2007}%
  \BibitemOpen
  \bibfield  {author} {\bibinfo {author} {\bibfnamefont {F.}~\bibnamefont
  {Dimer}}, \bibinfo {author} {\bibfnamefont {B.}~\bibnamefont {Estienne}},
  \bibinfo {author} {\bibfnamefont {A.~S.}\ \bibnamefont {Parkins}},\ and\
  \bibinfo {author} {\bibfnamefont {H.~J.}\ \bibnamefont {Carmichael}},\
  }\bibfield  {title} {\bibinfo {title} {Proposed realization of the
  dicke-model quantum phase transition in an optical cavity qed system},\
  }\href {https://doi.org/10.1103/PhysRevA.75.013804} {\bibfield  {journal}
  {\bibinfo  {journal} {Phys. Rev. A}\ }\textbf {\bibinfo {volume} {75}},\
  \bibinfo {pages} {013804} (\bibinfo {year} {2007})}\BibitemShut {NoStop}%
\bibitem [{\citenamefont {Zhou}\ \emph {et~al.}(2023)\citenamefont {Zhou},
  \citenamefont {Deng},\ and\ \citenamefont {Tan}}]{zhou2023}%
  \BibitemOpen
  \bibfield  {author} {\bibinfo {author} {\bibfnamefont {S.}~\bibnamefont
  {Zhou}}, \bibinfo {author} {\bibfnamefont {W.}~\bibnamefont {Deng}},\ and\
  \bibinfo {author} {\bibfnamefont {H.}~\bibnamefont {Tan}},\ }\bibfield
  {title} {\bibinfo {title} {Robust entanglement and steering in open dicke
  models with individual atomic spontaneous emission and dephasing},\ }\href
  {https://doi.org/10.1364/OE.480191} {\bibfield  {journal} {\bibinfo
  {journal} {Opt. Express}\ }\textbf {\bibinfo {volume} {31}},\ \bibinfo
  {pages} {8548} (\bibinfo {year} {2023})}\BibitemShut {NoStop}%
\bibitem [{\citenamefont {Tessier}\ \emph {et~al.}(2003)\citenamefont
  {Tessier}, \citenamefont {Deutsch}, \citenamefont {Delgado},\ and\
  \citenamefont {Fuentes-Guridi}}]{tessier2003}%
  \BibitemOpen
  \bibfield  {author} {\bibinfo {author} {\bibfnamefont {T.~E.}\ \bibnamefont
  {Tessier}}, \bibinfo {author} {\bibfnamefont {I.~H.}\ \bibnamefont
  {Deutsch}}, \bibinfo {author} {\bibfnamefont {A.}~\bibnamefont {Delgado}},\
  and\ \bibinfo {author} {\bibfnamefont {I.}~\bibnamefont {Fuentes-Guridi}},\
  }\bibfield  {title} {\bibinfo {title} {{Entanglement sharing in the two-atom
  Tavis-Cummings model}},\ }\href {https://doi.org/10.1103/PhysRevA.68.062316}
  {\bibfield  {journal} {\bibinfo  {journal} {Phys. Rev. A}\ }\textbf {\bibinfo
  {volume} {68}},\ \bibinfo {pages} {062316} (\bibinfo {year}
  {2003})}\BibitemShut {NoStop}%
\bibitem [{\citenamefont {Retzker}\ \emph {et~al.}(2007)\citenamefont
  {Retzker}, \citenamefont {Solano},\ and\ \citenamefont
  {Reznik}}]{retzker2007}%
  \BibitemOpen
  \bibfield  {author} {\bibinfo {author} {\bibfnamefont {A.}~\bibnamefont
  {Retzker}}, \bibinfo {author} {\bibfnamefont {E.}~\bibnamefont {Solano}},\
  and\ \bibinfo {author} {\bibfnamefont {B.}~\bibnamefont {Reznik}},\
  }\bibfield  {title} {\bibinfo {title} {{Tavis-Cummings model and collective
  multiqubit entanglement in trapped ions}},\ }\href
  {https://doi.org/10.1103/PhysRevA.75.022312} {\bibfield  {journal} {\bibinfo
  {journal} {Phys. Rev. A}\ }\textbf {\bibinfo {volume} {75}},\ \bibinfo
  {pages} {022312} (\bibinfo {year} {2007})}\BibitemShut {NoStop}%
\bibitem [{\citenamefont {Guo}\ and\ \citenamefont {Song}(2008)}]{Guo2008}%
  \BibitemOpen
  \bibfield  {author} {\bibinfo {author} {\bibfnamefont {J.-L.}\ \bibnamefont
  {Guo}}\ and\ \bibinfo {author} {\bibfnamefont {H.-S.}\ \bibnamefont {Song}},\
  }\bibfield  {title} {\bibinfo {title} {{Dynamics of pairwise entanglement
  between two Tavis–Cummings atoms}},\ }\href
  {https://doi.org/10.1088/1751-8113/41/8/085302} {\bibfield  {journal}
  {\bibinfo  {journal} {J. Phys. A: Math. Theor.}\ }\textbf {\bibinfo {volume}
  {41}},\ \bibinfo {pages} {085302} (\bibinfo {year} {2008})}\BibitemShut
  {NoStop}%
\bibitem [{\citenamefont {Zhang}\ and\ \citenamefont {Chen}(2009)}]{zhang2009}%
  \BibitemOpen
  \bibfield  {author} {\bibinfo {author} {\bibfnamefont {J.-S.}\ \bibnamefont
  {Zhang}}\ and\ \bibinfo {author} {\bibfnamefont {A.-X.}\ \bibnamefont
  {Chen}},\ }\bibfield  {title} {\bibinfo {title} {{Entanglement dynamics in
  the three-atom Tavis-Cummings model}},\ }\href
  {https://doi.org/10.1142/S0219749909005638} {\bibfield  {journal} {\bibinfo
  {journal} {Int. J. Quantum Inf.}\ }\textbf {\bibinfo {volume} {07}},\
  \bibinfo {pages} {1001} (\bibinfo {year} {2009})}\BibitemShut {NoStop}%
\bibitem [{\citenamefont {Man}\ \emph {et~al.}(2009)\citenamefont {Man},
  \citenamefont {Xia},\ and\ \citenamefont {An}}]{man2009}%
  \BibitemOpen
  \bibfield  {author} {\bibinfo {author} {\bibfnamefont {Z.~X.}\ \bibnamefont
  {Man}}, \bibinfo {author} {\bibfnamefont {Y.~J.}\ \bibnamefont {Xia}},\ and\
  \bibinfo {author} {\bibfnamefont {N.~B.}\ \bibnamefont {An}},\ }\bibfield
  {title} {\bibinfo {title} {{Entanglement dynamics for the double
  Tavis-Cummings model}},\ }\href {https://doi.org/10.1140/epjd/e2009-00095-7}
  {\bibfield  {journal} {\bibinfo  {journal} {Eur. Phys. J. D}\ }\textbf
  {\bibinfo {volume} {53}},\ \bibinfo {pages} {229} (\bibinfo {year}
  {2009})}\BibitemShut {NoStop}%
\bibitem [{\citenamefont {Youssef}\ \emph {et~al.}(2010)\citenamefont
  {Youssef}, \citenamefont {Metwally},\ and\ \citenamefont
  {Obada}}]{youssef2010}%
  \BibitemOpen
  \bibfield  {author} {\bibinfo {author} {\bibfnamefont {M.}~\bibnamefont
  {Youssef}}, \bibinfo {author} {\bibfnamefont {N.}~\bibnamefont {Metwally}},\
  and\ \bibinfo {author} {\bibfnamefont {A.-S.~F.}\ \bibnamefont {Obada}},\
  }\bibfield  {title} {\bibinfo {title} {{Some entanglement features of a
  three-atom Tavis–Cummings model: a cooperative case}},\ }\href
  {https://doi.org/10.1088/0953-4075/43/9/095501} {\bibfield  {journal}
  {\bibinfo  {journal} {J. Phys. B: At. Mol. Opt. Phys.}\ }\textbf {\bibinfo
  {volume} {43}},\ \bibinfo {pages} {095501} (\bibinfo {year}
  {2010})}\BibitemShut {NoStop}%
\bibitem [{\citenamefont {Mohamed}(2012)}]{mohamed2012}%
  \BibitemOpen
  \bibfield  {author} {\bibinfo {author} {\bibfnamefont {A.-B.}\ \bibnamefont
  {Mohamed}},\ }\bibfield  {title} {\bibinfo {title} {{Non-local correlation
  and quantum discord in two atoms in the non-degenerate model}},\ }\href
  {https://doi.org/https://doi.org/10.1016/j.aop.2012.08.003} {\bibfield
  {journal} {\bibinfo  {journal} {Ann. Phys. (N. Y.)}\ }\textbf {\bibinfo
  {volume} {327}},\ \bibinfo {pages} {3130} (\bibinfo {year}
  {2012})}\BibitemShut {NoStop}%
\bibitem [{\citenamefont {Hu}\ and\ \citenamefont {Xu}(2013)}]{hu2013}%
  \BibitemOpen
  \bibfield  {author} {\bibinfo {author} {\bibfnamefont {Z.-D.}\ \bibnamefont
  {Hu}}\ and\ \bibinfo {author} {\bibfnamefont {J.-B.}\ \bibnamefont {Xu}},\
  }\bibfield  {title} {\bibinfo {title} {{Control of quantum information flow
  and quantum correlations in the two-atom Tavis–Cummings model}},\ }\href
  {https://doi.org/10.1088/1751-8113/46/15/155303} {\bibfield  {journal}
  {\bibinfo  {journal} {J. Phys. A: Math. Theor.}\ }\textbf {\bibinfo {volume}
  {46}},\ \bibinfo {pages} {155303} (\bibinfo {year} {2013})}\BibitemShut
  {NoStop}%
\bibitem [{\citenamefont {Torres}(2014)}]{torres2014}%
  \BibitemOpen
  \bibfield  {author} {\bibinfo {author} {\bibfnamefont {J.~M.}\ \bibnamefont
  {Torres}},\ }\bibfield  {title} {\bibinfo {title} {{Closed-form solution of
  Lindblad master equations without gain}},\ }\href
  {https://doi.org/10.1103/PhysRevA.89.052133} {\bibfield  {journal} {\bibinfo
  {journal} {Phys. Rev. A}\ }\textbf {\bibinfo {volume} {89}},\ \bibinfo
  {pages} {052133} (\bibinfo {year} {2014})}\BibitemShut {NoStop}%
\bibitem [{\citenamefont {Fan}\ and\ \citenamefont {Zhang}(2014)}]{fan2014}%
  \BibitemOpen
  \bibfield  {author} {\bibinfo {author} {\bibfnamefont {K.-M.}\ \bibnamefont
  {Fan}}\ and\ \bibinfo {author} {\bibfnamefont {G.-F.}\ \bibnamefont
  {Zhang}},\ }\bibfield  {title} {\bibinfo {title} {{Geometric quantum discord
  and entanglement between two atoms in Tavis-Cummings model with dipole-dipole
  interaction under intrinsic decoherence}},\ }\href
  {https://doi.org/10.1140/epjd/e2014-50145-0} {\bibfield  {journal} {\bibinfo
  {journal} {Eur. Phys. J. D}\ }\textbf {\bibinfo {volume} {68}},\ \bibinfo
  {pages} {163} (\bibinfo {year} {2014})}\BibitemShut {NoStop}%
\bibitem [{\citenamefont {Restrepo}\ and\ \citenamefont
  {Rodríguez}(2016)}]{restrepo2016}%
  \BibitemOpen
  \bibfield  {author} {\bibinfo {author} {\bibfnamefont {J.}~\bibnamefont
  {Restrepo}}\ and\ \bibinfo {author} {\bibfnamefont {B.~A.}\ \bibnamefont
  {Rodríguez}},\ }\bibfield  {title} {\bibinfo {title} {{Dynamics of
  entanglement and quantum discord in the {Tavis–Cummings} model}},\ }\href
  {https://doi.org/10.1088/0953-4075/49/12/125502} {\bibfield  {journal}
  {\bibinfo  {journal} {J. Phys. B: At. Mol. Opt. Phys.}\ }\textbf {\bibinfo
  {volume} {49}},\ \bibinfo {pages} {125502} (\bibinfo {year}
  {2016})}\BibitemShut {NoStop}%
\bibitem [{\citenamefont {Rundle}\ and\ \citenamefont
  {Everitt}(2021)}]{rundle2021}%
  \BibitemOpen
  \bibfield  {author} {\bibinfo {author} {\bibfnamefont {R.~P.}\ \bibnamefont
  {Rundle}}\ and\ \bibinfo {author} {\bibfnamefont {M.~J.}\ \bibnamefont
  {Everitt}},\ }\bibfield  {title} {\bibinfo {title} {{An informationally
  complete Wigner function for the Tavis--Cummings model}},\ }\href
  {https://doi.org/10.1007/s10825-021-01777-6} {\bibfield  {journal} {\bibinfo
  {journal} {J. Comput. Electron.}\ }\textbf {\bibinfo {volume} {20}},\
  \bibinfo {pages} {2180} (\bibinfo {year} {2021})}\BibitemShut {NoStop}%
\bibitem [{\citenamefont {Carnio}\ \emph {et~al.}(2021)\citenamefont {Carnio},
  \citenamefont {Buchleitner},\ and\ \citenamefont {Schlawin}}]{carnio2021}%
  \BibitemOpen
  \bibfield  {author} {\bibinfo {author} {\bibfnamefont {E.~G.}\ \bibnamefont
  {Carnio}}, \bibinfo {author} {\bibfnamefont {A.}~\bibnamefont
  {Buchleitner}},\ and\ \bibinfo {author} {\bibfnamefont {F.}~\bibnamefont
  {Schlawin}},\ }\bibfield  {title} {\bibinfo {title} {{Optimization of
  selective two-photon absorption in cavity polaritons}},\ }\href
  {https://doi.org/10.1063/5.0049863} {\bibfield  {journal} {\bibinfo
  {journal} {J. Chem. Phys.}\ }\textbf {\bibinfo {volume} {154}},\ \bibinfo
  {pages} {214114} (\bibinfo {year} {2021})}\BibitemShut {NoStop}%
\bibitem [{\citenamefont {{Strocchi}}(2021)}]{strocchi2005}%
  \BibitemOpen
  \bibfield  {author} {\bibinfo {author} {\bibfnamefont {F.}~\bibnamefont
  {{Strocchi}}},\ }\href {https://doi.org/10.1007/978-3-662-62166-0} {\emph
  {\bibinfo {title} {{Symmetry Breaking}}}}\ (\bibinfo  {publisher} {Springer
  Berlin, Heidelberg},\ \bibinfo {year} {2021})\BibitemShut {NoStop}%
\bibitem [{\citenamefont {Fiorelli}\ \emph {et~al.}(2023)\citenamefont
  {Fiorelli}, \citenamefont {M\"uller}, \citenamefont {Lesanovsky},\ and\
  \citenamefont {Carollo}}]{fiorelli2023}%
  \BibitemOpen
  \bibfield  {author} {\bibinfo {author} {\bibfnamefont {E.}~\bibnamefont
  {Fiorelli}}, \bibinfo {author} {\bibfnamefont {M.}~\bibnamefont {M\"uller}},
  \bibinfo {author} {\bibfnamefont {I.}~\bibnamefont {Lesanovsky}},\ and\
  \bibinfo {author} {\bibfnamefont {F.}~\bibnamefont {Carollo}},\ }\bibfield
  {title} {\bibinfo {title} {{Mean-field dynamics of open quantum systems with
  collective operator-valued rates: validity and application}},\ }\href
  {https://doi.org/10.48550/ARXIV.2302.04155} {\bibfield  {journal} {\bibinfo
  {journal} {arXiv:2302.04155}\ } (\bibinfo {year} {2023})}\BibitemShut
  {NoStop}%
\bibitem [{\citenamefont {{Strogatz}}(2018)}]{strogatz}%
  \BibitemOpen
  \bibfield  {author} {\bibinfo {author} {\bibfnamefont {S.}~\bibnamefont
  {{Strogatz}}},\ }\href@noop {} {\emph {\bibinfo {title} {{Nonlinear Dynamics
  and Chaos: With Applications to Physics, Biology, Chemistry, and Engineering
  }}}}\ (\bibinfo  {publisher} {CRC Press},\ \bibinfo {year}
  {2018})\BibitemShut {NoStop}%
\bibitem [{\citenamefont {Cabot}\ \emph
  {et~al.}(2022{\natexlab{b}})\citenamefont {Cabot}, \citenamefont {Muhle},
  \citenamefont {Carollo},\ and\ \citenamefont
  {Lesanovsky}}]{cabot2022quantum}%
  \BibitemOpen
  \bibfield  {author} {\bibinfo {author} {\bibfnamefont {A.}~\bibnamefont
  {Cabot}}, \bibinfo {author} {\bibfnamefont {L.~S.}\ \bibnamefont {Muhle}},
  \bibinfo {author} {\bibfnamefont {F.}~\bibnamefont {Carollo}},\ and\ \bibinfo
  {author} {\bibfnamefont {I.}~\bibnamefont {Lesanovsky}},\ }\bibfield  {title}
  {\bibinfo {title} {{Quantum trajectories of dissipative time-crystals}},\
  }\href {https://doi.org/10.48550/arXiv.2212.06460} {\bibfield  {journal}
  {\bibinfo  {journal} {arXiv:2212.06460}\ } (\bibinfo {year}
  {2022}{\natexlab{b}})}\BibitemShut {NoStop}%
\bibitem [{\citenamefont {Adesso}\ and\ \citenamefont
  {Datta}(2010)}]{adesso2010}%
  \BibitemOpen
  \bibfield  {author} {\bibinfo {author} {\bibfnamefont {G.}~\bibnamefont
  {Adesso}}\ and\ \bibinfo {author} {\bibfnamefont {A.}~\bibnamefont {Datta}},\
  }\bibfield  {title} {\bibinfo {title} {{Quantum versus Classical Correlations
  in Gaussian States}},\ }\href
  {https://doi.org/10.1103/PhysRevLett.105.030501} {\bibfield  {journal}
  {\bibinfo  {journal} {Phys. Rev. Lett.}\ }\textbf {\bibinfo {volume} {105}},\
  \bibinfo {pages} {030501} (\bibinfo {year} {2010})}\BibitemShut {NoStop}%
\bibitem [{\citenamefont {Giorda}\ and\ \citenamefont
  {Paris}(2010)}]{giorda2010}%
  \BibitemOpen
  \bibfield  {author} {\bibinfo {author} {\bibfnamefont {P.}~\bibnamefont
  {Giorda}}\ and\ \bibinfo {author} {\bibfnamefont {M.~G.~A.}\ \bibnamefont
  {Paris}},\ }\bibfield  {title} {\bibinfo {title} {{Gaussian Quantum
  Discord}},\ }\href {https://doi.org/10.1103/PhysRevLett.105.020503}
  {\bibfield  {journal} {\bibinfo  {journal} {Phys. Rev. Lett.}\ }\textbf
  {\bibinfo {volume} {105}},\ \bibinfo {pages} {020503} (\bibinfo {year}
  {2010})}\BibitemShut {NoStop}%
\bibitem [{\citenamefont {Simon}(2000)}]{simon2000}%
  \BibitemOpen
  \bibfield  {author} {\bibinfo {author} {\bibfnamefont {R.}~\bibnamefont
  {Simon}},\ }\bibfield  {title} {\bibinfo {title} {{Peres-Horodecki
  Separability Criterion for Continuous Variable Systems}},\ }\href
  {https://doi.org/10.1103/PhysRevLett.84.2726} {\bibfield  {journal} {\bibinfo
   {journal} {Phys. Rev. Lett.}\ }\textbf {\bibinfo {volume} {84}},\ \bibinfo
  {pages} {2726} (\bibinfo {year} {2000})}\BibitemShut {NoStop}%
\bibitem [{\citenamefont {Adesso}\ \emph {et~al.}(2014)\citenamefont {Adesso},
  \citenamefont {Ragy},\ and\ \citenamefont {Lee}}]{adesso2014}%
  \BibitemOpen
  \bibfield  {author} {\bibinfo {author} {\bibfnamefont {G.}~\bibnamefont
  {Adesso}}, \bibinfo {author} {\bibfnamefont {S.}~\bibnamefont {Ragy}},\ and\
  \bibinfo {author} {\bibfnamefont {A.~R.}\ \bibnamefont {Lee}},\ }\bibfield
  {title} {\bibinfo {title} {{Continuous Variable Quantum Information: Gaussian
  States and Beyond}},\ }\href {https://doi.org/10.1142/S1230161214400010}
  {\bibfield  {journal} {\bibinfo  {journal} {Open Syst. Inf. Dyn.}\ }\textbf
  {\bibinfo {volume} {21}},\ \bibinfo {pages} {1440001} (\bibinfo {year}
  {2014})}\BibitemShut {NoStop}%
\bibitem [{\citenamefont {Carr}\ \emph {et~al.}(2013)\citenamefont {Carr},
  \citenamefont {Ritter}, \citenamefont {Wade}, \citenamefont {Adams},\ and\
  \citenamefont {Weatherill}}]{carr2013}%
  \BibitemOpen
  \bibfield  {author} {\bibinfo {author} {\bibfnamefont {C.}~\bibnamefont
  {Carr}}, \bibinfo {author} {\bibfnamefont {R.}~\bibnamefont {Ritter}},
  \bibinfo {author} {\bibfnamefont {C.~G.}\ \bibnamefont {Wade}}, \bibinfo
  {author} {\bibfnamefont {C.~S.}\ \bibnamefont {Adams}},\ and\ \bibinfo
  {author} {\bibfnamefont {K.~J.}\ \bibnamefont {Weatherill}},\ }\bibfield
  {title} {\bibinfo {title} {{Nonequilibrium Phase Transition in a Dilute
  Rydberg Ensemble}},\ }\href {https://doi.org/10.1103/PhysRevLett.111.113901}
  {\bibfield  {journal} {\bibinfo  {journal} {Phys. Rev. Lett.}\ }\textbf
  {\bibinfo {volume} {111}},\ \bibinfo {pages} {113901} (\bibinfo {year}
  {2013})}\BibitemShut {NoStop}%
\bibitem [{\citenamefont {Marcuzzi}\ \emph {et~al.}(2014)\citenamefont
  {Marcuzzi}, \citenamefont {Levi}, \citenamefont {Diehl}, \citenamefont
  {Garrahan},\ and\ \citenamefont {Lesanovsky}}]{marcuzzi2014}%
  \BibitemOpen
  \bibfield  {author} {\bibinfo {author} {\bibfnamefont {M.}~\bibnamefont
  {Marcuzzi}}, \bibinfo {author} {\bibfnamefont {E.}~\bibnamefont {Levi}},
  \bibinfo {author} {\bibfnamefont {S.}~\bibnamefont {Diehl}}, \bibinfo
  {author} {\bibfnamefont {J.~P.}\ \bibnamefont {Garrahan}},\ and\ \bibinfo
  {author} {\bibfnamefont {I.}~\bibnamefont {Lesanovsky}},\ }\bibfield  {title}
  {\bibinfo {title} {{Universal Nonequilibrium Properties of Dissipative
  Rydberg Gases}},\ }\href {https://doi.org/10.1103/PhysRevLett.113.210401}
  {\bibfield  {journal} {\bibinfo  {journal} {Phys. Rev. Lett.}\ }\textbf
  {\bibinfo {volume} {113}},\ \bibinfo {pages} {210401} (\bibinfo {year}
  {2014})}\BibitemShut {NoStop}%
\bibitem [{SM()}]{SM}%
  \BibitemOpen
  \href@noop {} {\bibinfo {title} {Animations of the different bifurcations
  occurring in our model}},\ \bibinfo {howpublished}
  {\url{https://github.com/RobertMattes/SM_animations}}\BibitemShut {NoStop}%
\bibitem [{\citenamefont {Lanford}\ and\ \citenamefont
  {Ruelle}(1969)}]{landford1969}%
  \BibitemOpen
  \bibfield  {author} {\bibinfo {author} {\bibfnamefont {O.~E.}\ \bibnamefont
  {Lanford}}\ and\ \bibinfo {author} {\bibfnamefont {D.}~\bibnamefont
  {Ruelle}},\ }\bibfield  {title} {\bibinfo {title} {{Observables at infinity
  and states with short range correlations in statistical mechanics}},\ }\href
  {https://doi.org/10.1007/BF01645487} {\bibfield  {journal} {\bibinfo
  {journal} {Commun. Math. Phys.}\ }\textbf {\bibinfo {volume} {13}},\ \bibinfo
  {pages} {194} (\bibinfo {year} {1969})}\BibitemShut {NoStop}%
\bibitem [{\citenamefont {Goderis}\ \emph {et~al.}(1989)\citenamefont
  {Goderis}, \citenamefont {Verbeure},\ and\ \citenamefont
  {Vets}}]{goderis1989}%
  \BibitemOpen
  \bibfield  {author} {\bibinfo {author} {\bibfnamefont {D.}~\bibnamefont
  {Goderis}}, \bibinfo {author} {\bibfnamefont {A.}~\bibnamefont {Verbeure}},\
  and\ \bibinfo {author} {\bibfnamefont {P.}~\bibnamefont {Vets}},\ }\bibfield
  {title} {\bibinfo {title} {{Non-commutative central limits}},\ }\href
  {https://doi.org/10.1007/BF00341282} {\bibfield  {journal} {\bibinfo
  {journal} {Probab. Theory Relat. Fields}\ }\textbf {\bibinfo {volume} {82}},\
  \bibinfo {pages} {527} (\bibinfo {year} {1989})}\BibitemShut {NoStop}%
\bibitem [{\citenamefont {Goderis}\ \emph {et~al.}(1990)\citenamefont
  {Goderis}, \citenamefont {Verbeure},\ and\ \citenamefont
  {Vets}}]{goderis1990}%
  \BibitemOpen
  \bibfield  {author} {\bibinfo {author} {\bibfnamefont {D.}~\bibnamefont
  {Goderis}}, \bibinfo {author} {\bibfnamefont {A.}~\bibnamefont {Verbeure}},\
  and\ \bibinfo {author} {\bibfnamefont {P.}~\bibnamefont {Vets}},\ }\bibfield
  {title} {\bibinfo {title} {{Dynamics of fluctuations for quantum lattice
  systems}},\ }\href {https://doi.org/10.1007/BF02096872} {\bibfield  {journal}
  {\bibinfo  {journal} {Comm. Math. Phys.}\ }\textbf {\bibinfo {volume}
  {128}},\ \bibinfo {pages} {533} (\bibinfo {year} {1990})}\BibitemShut
  {NoStop}%
\bibitem [{\citenamefont {Narnhofer}\ and\ \citenamefont
  {Thirring}(2002)}]{narnhofer2002}%
  \BibitemOpen
  \bibfield  {author} {\bibinfo {author} {\bibfnamefont {H.}~\bibnamefont
  {Narnhofer}}\ and\ \bibinfo {author} {\bibfnamefont {W.}~\bibnamefont
  {Thirring}},\ }\bibfield  {title} {\bibinfo {title} {{Entanglement of
  mesoscopic systems}},\ }\href {https://doi.org/10.1103/PhysRevA.66.052304}
  {\bibfield  {journal} {\bibinfo  {journal} {Phys. Rev. A}\ }\textbf {\bibinfo
  {volume} {66}},\ \bibinfo {pages} {052304} (\bibinfo {year}
  {2002})}\BibitemShut {NoStop}%
\bibitem [{\citenamefont {Benatti}\ \emph {et~al.}(2014)\citenamefont
  {Benatti}, \citenamefont {Carollo},\ and\ \citenamefont
  {Floreanini}}]{benatti2014}%
  \BibitemOpen
  \bibfield  {author} {\bibinfo {author} {\bibfnamefont {F.}~\bibnamefont
  {Benatti}}, \bibinfo {author} {\bibfnamefont {F.}~\bibnamefont {Carollo}},\
  and\ \bibinfo {author} {\bibfnamefont {R.}~\bibnamefont {Floreanini}},\
  }\bibfield  {title} {\bibinfo {title} {{Environment induced entanglement in
  many-body mesoscopic systems}},\ }\href
  {https://doi.org/https://doi.org/10.1016/j.physleta.2014.04.034} {\bibfield
  {journal} {\bibinfo  {journal} {Phys. Lett. A}\ }\textbf {\bibinfo {volume}
  {378}},\ \bibinfo {pages} {1700} (\bibinfo {year} {2014})}\BibitemShut
  {NoStop}%
\bibitem [{\citenamefont {Benatti}\ \emph
  {et~al.}(2016{\natexlab{a}})\citenamefont {Benatti}, \citenamefont {Carollo},
  \citenamefont {Floreanini},\ and\ \citenamefont {Narnhofer}}]{benatti2016}%
  \BibitemOpen
  \bibfield  {author} {\bibinfo {author} {\bibfnamefont {F.}~\bibnamefont
  {Benatti}}, \bibinfo {author} {\bibfnamefont {F.}~\bibnamefont {Carollo}},
  \bibinfo {author} {\bibfnamefont {R.}~\bibnamefont {Floreanini}},\ and\
  \bibinfo {author} {\bibfnamefont {H.}~\bibnamefont {Narnhofer}},\ }\bibfield
  {title} {\bibinfo {title} {{Non-markovian mesoscopic dissipative dynamics of
  open quantum spin chains}},\ }\href
  {https://doi.org/https://doi.org/10.1016/j.physleta.2015.10.062} {\bibfield
  {journal} {\bibinfo  {journal} {Phys. Lett. A}\ }\textbf {\bibinfo {volume}
  {380}},\ \bibinfo {pages} {381} (\bibinfo {year}
  {2016}{\natexlab{a}})}\BibitemShut {NoStop}%
\bibitem [{\citenamefont {Benatti}\ \emph
  {et~al.}(2016{\natexlab{b}})\citenamefont {Benatti}, \citenamefont
  {Carollo},\ and\ \citenamefont {Floreanini}}]{benatti2016b}%
  \BibitemOpen
  \bibfield  {author} {\bibinfo {author} {\bibfnamefont {F.}~\bibnamefont
  {Benatti}}, \bibinfo {author} {\bibfnamefont {F.}~\bibnamefont {Carollo}},\
  and\ \bibinfo {author} {\bibfnamefont {R.}~\bibnamefont {Floreanini}},\
  }\bibfield  {title} {\bibinfo {title} {{Dissipative entanglement of quantum
  spin fluctuations}},\ }\href {https://doi.org/10.1063/1.4954072} {\bibfield
  {journal} {\bibinfo  {journal} {J. Math. Phys.}\ }\textbf {\bibinfo {volume}
  {57}},\ \bibinfo {pages} {062208} (\bibinfo {year}
  {2016}{\natexlab{b}})}\BibitemShut {NoStop}%
\bibitem [{\citenamefont {Heinosaari}\ \emph {et~al.}(2010)\citenamefont
  {Heinosaari}, \citenamefont {Holevo},\ and\ \citenamefont
  {Wolf}}]{heinosaari2010}%
  \BibitemOpen
  \bibfield  {author} {\bibinfo {author} {\bibfnamefont {T.}~\bibnamefont
  {Heinosaari}}, \bibinfo {author} {\bibfnamefont {A.~S.}\ \bibnamefont
  {Holevo}},\ and\ \bibinfo {author} {\bibfnamefont {M.~M.}\ \bibnamefont
  {Wolf}},\ }\bibfield  {title} {\bibinfo {title} {{The Semigroup Structure of
  Gaussian Channels}},\ }\href@noop {} {\bibfield  {journal} {\bibinfo
  {journal} {Quantum Info. Comput.}\ }\textbf {\bibinfo {volume} {10}},\
  \bibinfo {pages} {619–635} (\bibinfo {year} {2010})}\BibitemShut {NoStop}%
\bibitem [{\citenamefont {Henderson}\ and\ \citenamefont
  {Vedral}(2001)}]{henderson2001}%
  \BibitemOpen
  \bibfield  {author} {\bibinfo {author} {\bibfnamefont {L.}~\bibnamefont
  {Henderson}}\ and\ \bibinfo {author} {\bibfnamefont {V.}~\bibnamefont
  {Vedral}},\ }\bibfield  {title} {\bibinfo {title} {{Classical, quantum and
  total correlations}},\ }\href {https://doi.org/10.1088/0305-4470/34/35/315}
  {\bibfield  {journal} {\bibinfo  {journal} {J. Phys. A: Math. Gen.}\ }\textbf
  {\bibinfo {volume} {34}},\ \bibinfo {pages} {6899} (\bibinfo {year}
  {2001})}\BibitemShut {NoStop}%
\bibitem [{\citenamefont {Ollivier}\ and\ \citenamefont
  {Zurek}(2001)}]{ollivier2001}%
  \BibitemOpen
  \bibfield  {author} {\bibinfo {author} {\bibfnamefont {H.}~\bibnamefont
  {Ollivier}}\ and\ \bibinfo {author} {\bibfnamefont {W.~H.}\ \bibnamefont
  {Zurek}},\ }\bibfield  {title} {\bibinfo {title} {{Quantum Discord: A Measure
  of the Quantumness of Correlations}},\ }\href
  {https://doi.org/10.1103/PhysRevLett.88.017901} {\bibfield  {journal}
  {\bibinfo  {journal} {Phys. Rev. Lett.}\ }\textbf {\bibinfo {volume} {88}},\
  \bibinfo {pages} {017901} (\bibinfo {year} {2001})}\BibitemShut {NoStop}%
\bibitem [{\citenamefont {Isar}(2014)}]{isar2014}%
  \BibitemOpen
  \bibfield  {author} {\bibinfo {author} {\bibfnamefont {A.}~\bibnamefont
  {Isar}},\ }\bibfield  {title} {\bibinfo {title} {{Quantum Discord and
  Classical Correlations of Two Bosonic Modes in the Two-Reservoir Model}},\
  }\href {https://doi.org/10.1007/s10946-014-9401-z} {\bibfield  {journal}
  {\bibinfo  {journal} {J. Russ. Laser Res.}\ }\textbf {\bibinfo {volume}
  {35}},\ \bibinfo {pages} {62} (\bibinfo {year} {2014})}\BibitemShut {NoStop}%
\bibitem [{\citenamefont {Chan}\ \emph {et~al.}(2015)\citenamefont {Chan},
  \citenamefont {Lee},\ and\ \citenamefont {Gopalakrishnan}}]{chan2015}%
  \BibitemOpen
  \bibfield  {author} {\bibinfo {author} {\bibfnamefont {C.-K.}\ \bibnamefont
  {Chan}}, \bibinfo {author} {\bibfnamefont {T.~E.}\ \bibnamefont {Lee}},\ and\
  \bibinfo {author} {\bibfnamefont {S.}~\bibnamefont {Gopalakrishnan}},\
  }\bibfield  {title} {\bibinfo {title} {Limit-cycle phase in
  driven-dissipative spin systems},\ }\href
  {https://doi.org/10.1103/PhysRevA.91.051601} {\bibfield  {journal} {\bibinfo
  {journal} {Phys. Rev. A}\ }\textbf {\bibinfo {volume} {91}},\ \bibinfo
  {pages} {051601} (\bibinfo {year} {2015})}\BibitemShut {NoStop}%
\bibitem [{\citenamefont {Navarrete-Benlloch}\ \emph
  {et~al.}(2017)\citenamefont {Navarrete-Benlloch}, \citenamefont {Weiss},
  \citenamefont {Walter},\ and\ \citenamefont {de~Valc\'arcel}}]{benlloch2017}%
  \BibitemOpen
  \bibfield  {author} {\bibinfo {author} {\bibfnamefont {C.}~\bibnamefont
  {Navarrete-Benlloch}}, \bibinfo {author} {\bibfnamefont {T.}~\bibnamefont
  {Weiss}}, \bibinfo {author} {\bibfnamefont {S.}~\bibnamefont {Walter}},\ and\
  \bibinfo {author} {\bibfnamefont {G.~J.}\ \bibnamefont {de~Valc\'arcel}},\
  }\bibfield  {title} {\bibinfo {title} {General linearized theory of quantum
  fluctuations around arbitrary limit cycles},\ }\href
  {https://doi.org/10.1103/PhysRevLett.119.133601} {\bibfield  {journal}
  {\bibinfo  {journal} {Phys. Rev. Lett.}\ }\textbf {\bibinfo {volume} {119}},\
  \bibinfo {pages} {133601} (\bibinfo {year} {2017})}\BibitemShut {NoStop}%
\end{thebibliography}%
\end{document}